\pdfoutput=1
\documentclass[a4paper, 11pt]{article}
\usepackage{jheppub, amssymb, amsmath, graphicx, feynmp-auto, hyperref, 
color, multirow, tabularx, arydshln, slashed, orcidlink} 

\usepackage[T1]{fontenc}
\definecolor{nicered}{rgb}{.7,.1,.1}
\definecolor{nicegreen}{rgb}{.1,.5,.1}
\definecolor{darkblue}{rgb}{0,0,.5}
\hypersetup{colorlinks=true, urlcolor=darkblue, linkcolor=nicered, citecolor=nicegreen}
\definecolor{lightgray}{gray}{0.6}

\renewcommand\d{{\mathrm d}}

\newcommand\GeV{\text{GeV}}
\newcommand\TeV{\text{TeV}}
\newcommand\e{\epsilon}

\renewcommand\b{\beta}
\renewcommand\t{\theta}

\title{Long-lived Left-Right signals at the FCC-ee}

\author[a]{\hspace*{-3pt} Benjamin Fuks \orcidlink{0000-0002-0041-0566},}
\emailAdd{fuks@lpthe.jussieu.fr}
\affiliation[a]{\normalsize \it Laboratoire de Physique Th\'eorique et 
Hautes Energies (LPTHE),
UMR 7589, Sorbonne Universit\'e \& CNRS, 4 place Jussieu, 75252 Paris 
Cedex 05, France}

\author[b]{Jonathan Kriewald \orcidlink{0000-0002-3313-115X},}
\emailAdd{jonathan.kriewald@ijs.si}
\affiliation[b]{\normalsize \it  Jo\v zef Stefan Institute, 
Jamova 39, 1000 Ljubljana, Slovenia}

\author[c, b]{Miha Nemev\v{s}ek \orcidlink{0000-0003-1110-342X}}
\emailAdd{miha.nemevsek@ijs.si}
\affiliation[c]{\normalsize \it  Faculty of Mathematics and Physics, 
University of Ljubljana, Jadranska 19, 1000 Ljubljana, Slovenia}

\author[d, e]{and Fabrizio Nesti \orcidlink{0000-0003-4508-8141}}
\emailAdd{fabrizio.nesti@aquila.infn.it}
\affiliation[d]{\normalsize \it 
Dipartimento di Scienze Fisiche e Chimiche, Universit\`a dell'Aquila, 
via Vetoio, I-67100, L'Aquila, Italy}
\affiliation[e]{\normalsize \it INFN, Laboratori Nazionali del Gran Sasso,
I-67100 Assergi (AQ), Italy}

\date{\today}

\abstract{
We give an extensive discussion of the displaced signals of heavy Majorana neutrino production at future electron-positron colliders operating at various proposed energies in the context of the Left-Right symmetric model.
A comprehensive collection of channels is taken into account, ranging from those featuring $W$ and $W_R$ mediation to those induced by scalar mixing and gauge/scalar boson fusion, with connections to the mechanism of neutrino mass origin.
The emerging signatures feature possibly multiple displaced heavy neutrinos that are in some cases accompanied by prompt activity and forward leptons.
We derive the corresponding total production rates and differential distributions, which allow us to differentiate the channels and have analytical estimates of the signal yield.
We then develop realistic estimates of the selection efficiencies using a 
dedicated vertexing algorithm which establishes the displaced decay positions 
and supplies a reliable proxy for reconstructing the full four-momenta of
long-lived particles.
This allows to determine the realistic reaches in the parameter space of the Left-Right symmetric model across the various channels, and we show that these can strongly surpass the LHC ones, demonstrating that future lepton colliders are sensitive to left-right symmetry breaking scales in the deep multi-TeV regime.
}

\keywords{Neutrino mass origin, Majorana neutrinos, Collider physics, 
Left-Right symmetry, Extended Higgs bosons, Long-lived particles, 
Heavy Neutral Leptons.}

\begin{document}

\maketitle
\flushbottom

%
%
\section{Introduction} \label{sec:Intro}
Determining the origin of neutrino masses and uncovering whether neutrinos are
Majorana~\cite{Majorana:1937vz} or Dirac~\cite{Dirac:1928hu} particles remain 
among the foremost open questions in particle physics. 
The Higgs mechanism indeed successfully explains the masses of the charged fermions, 
the gauge bosons and the Higgs boson within the Standard Model (SM)~\cite{Higgs:1964pj, 
Englert:1964et, Guralnik:1964eu, Weinberg:1967tq}, and it has been thoroughly tested 
at the LHC~\cite{ATLAS:2022vkf, CMS:2022dwd}. 
However, it cannot account for neutrino masses, which clearly point to physics 
beyond the SM. 
In this context, the possible Majorana nature of neutrinos would imply lepton-number 
violation (LNV) and the existence of new degrees of freedom, such as the heavy 
neutral leptons often predicted in seesaw or left-right symmetric frameworks~\cite{
Pati:1974yy, Mohapatra:1974gc, Senjanovic:1975rk, Mohapatra:1977mj, Gell-Mann:1979vob, 
Glashow:1979nm, Sawada:1979dis, Minkowski:1977sc, Mohapatra:1979ia}.

A future high-luminosity circular electron-positron collider like the FCC-ee~\cite{
FCC:2018byv, FCC:2018evy, Aslanides:2019gew}, operating at multiple centre-of-mass 
energies $\sqrt{s}$ ranging from the $Z$ pole to the top-anti-top threshold and 
including a Higgs factory mode~\cite{Blondel:2024mry}, offers an ideal environment 
to explore these questions with unprecedented precision. 
The especially clean experimental environment of such a collider, with negligible 
soft hadronic backgrounds compared to hadron machines, indeed enables the detection 
of soft or significantly displaced signatures characteristic of long-lived particles 
and typical of several neutrino mass models~\cite{Alimena:2019zri, Blondel:2022qqo}. 
Moreover, as a global flagship endeavour jointly supported by international communities 
across the world~\cite{FCC:2025lpp, CEPCStudyGroup:2018rmc, CEPCStudyGroup:2018ghi, 
CEPCPhysicsStudyGroup:2022uwl, Antusch:2025lpm, Maltoni:2022bqs}, the FCC-ee is 
envisioned as a successor to the High-Luminosity LHC era~\cite{Benedikt:2020ejr} and 
a step towards future high-energy hadron colliders~\cite{FCC:2018vvp}. 
Consequently, the FCC-ee provides an excellent laboratory to probe new physics scenarios 
motivated by neutrino mass generation mechanisms and LNV processes~\cite{Cai:2017mow, 
Abdullahi:2022jlv}.

%
%
One of the most compelling theoretical frameworks for generating neutrino masses consists 
of the seesaw mechanism~\cite{Gell-Mann:1979vob, Glashow:1979nm, Sawada:1979dis, Minkowski:1977sc,
Mohapatra:1979ia}, which naturally connects the smallness of light neutrino masses with 
the existence of heavier states. 
Such a mechanism often emerges in theories addressing the apparent maximal parity violation 
of weak interactions, thereby providing further motivation for physics beyond the Standard 
Model (BSM).
A particularly elegant realisation arises in Left-Right (LR) symmetric theories~\cite{Pati:1974yy,
Mohapatra:1974gc}, where parity is a fundamental symmetry of the Lagrangian that is spontaneously 
broken at low energies~\cite{Senjanovic:1975rk}. 
In the minimal incarnation of such a setup, called the minimal Left-Right Symmetric Model 
(LRSM)~\cite{Minkowski:1977sc, Mohapatra:1979ia}, a dynamical origin for the mass of the heavy 
neutrino $N$ is provided through the vacuum expectation value of a right-handed scalar triplet 
$\Delta_R$, whose neutral component may mix with the SM-like Higgs boson $h$. 

%
%
To date, the LRSM has been confronted with a wide range of experimental constraints, spanning 
indirect constraints from flavour observables to direct collider searches with or without 
manifest LNV. 
%
Early bounds from kaon mixing~\cite{Beall:1981ze} have already placed the scale of right-handed 
interactions beyond the reach of the Tevatron, but the LHC has since pushed this frontier 
considerably~\cite{Nemevsek:2011hz}. 
The most recent searches from the ATLAS and CMS collaborations~\cite{CMS:2021dzb, ATLAS:2023cjo} 
now yield limits that are comparable to, or even stronger than those obtained from flavour 
data~\cite{Bertolini:2014sua}, while the precise constraints depend on the specific realisation 
of parity~\cite{Maiezza:2010ic} and how the strong CP problem is 
addressed~\cite{Maiezza:2014ala, Bertolini:2019out, Bertolini:2020hjc}. 
The bounds from collider searches generally focus on characteristic LRSM final states and include, 
most notably, the Keung-Senjanović (KS) process~\cite{Keung:1983uu}, where a heavy Majorana neutrino 
is produced via a charged current exchange $pp \to \ell N$, and then decays as $N \to \ell jj$. 
In addition, when the heavy neutrino mass satisfies $m_N > m_t$, third-generation quarks can 
appear in the final state through an $N \to \ell t b$ decay~\cite{Frank:2023epx}. 
In both cases, depending on the mass hierarchy and the lifetime of the heavy neutrino, 
current searches probe resolved, displaced~\cite{Helo:2013esa, Nemevsek:2018bbt, 
Cottin:2018kmq, Cottin:2019drg} or merged~\cite{Mitra:2016kov, Nemevsek:2018bbt} topologies.
For heavier neutrinos $m_N > M_{W_R}$, direct $W_R$ searches in purely hadronic final states 
become dominant.
These target the production of both di-jet or $tb$ resonances~\cite{ATLAS:2019fgd, CMS:2019gwf}, 
as well as $tb$ pairs~\cite{CMS:2023gte, ATLAS:2023ibb}. 
Current lower bounds on the $W_R$-boson mass of $M_{W_R} \gtrsim 4.5~\mathrm{TeV}$ have however been established. 
At the opposite extreme, if the heavy neutrino $N$ is very light and its decay products escape detection, the constraints stemming from the searches for the decay $W_R \to \ell\nu$~\cite{CMS:2022krd, CMS:2022ncp, ATLAS:2019lsy} apply, 
typically constraining $M_{W_R} \gtrsim 6\text{--}7~\mathrm{TeV}$. 
The last regime is particularly motivated by a connection to dark matter~\cite{Bezrukov:2009th, 
Nemevsek:2012cd}. 
However, in this context, generic dilution constraints~\cite{Nemevsek:2022anh} typically push the LR symmetry breaking scale to very high values~\cite{Nemevsek:2023yjl}.

All the above-mentioned limits can vary significantly across model realisations. 
For instance, alternative LR constructions~\cite{Ma:1986we, Babu:1987kp, Frank:2005rb,
Ma:2010us} with different LR fermionic pairings can partially evade current bounds, thereby 
keeping sizeable regions of parameter space open for exploration, while additionally including 
dark matter candidates~\cite{Frank:2019nid, Frank:2022tbm}, realising baryogenesis through 
leptogenesis~\cite{Frank:2020odd}, addressing vacuum stability~\cite{Frank:2021ekj} or 
featuring increased flavour violation in the fermion sector~\cite{Frank:2023fkc, Frank:2024bss}.

To investigate the collider phenomenology of heavy Majorana neutrinos at the FCC-ee, we focus 
on the LRSM framework in which the heavy neutrinos $N$ couple to SM particles via both gauge 
and scalar interactions. 
This subsequently yields a variety of production and decay processes, which we systematically 
examine. 
More precisely, we focus on all gauge-mediated and scalar-mediated production channels that 
can give rise to LNV final states at the FCC-ee ($s$-channel, $t/u$-channel as well as fusion 
processes).
Such final states can reveal the Majorana nature of the heavy neutrinos, and in the
case of $\Delta_R$ production, allow us to measure the heavy neutrino Yukawa coupling, which would
establish the spontaneous origin of their mass.
Our objective is to quantify the FCC-ee sensitivity to the associated signatures, 
reconstruct the kinematic features distinguishing the different channels, and compare 
the expected reach with that of complementary searches at other facilities.

The foundation of our analysis is to collectively extend previous collider studies 
of heavy neutrinos in Higgs~\cite{Maiezza:2015lza}, $\Delta \cong 
\Delta_R$~\cite{Nemevsek:2016enw, BhupalDev:2018vpr} and gauge~\cite{Huitu:1997vh,Barry:2012ga,Urquia-Calderon:2023dkf}
signatures to the environment of the FCC-ee.
We show that their interplay yields striking signals with substantial sensitivity to the 
LR symmetry-breaking scale. 
We remind in particular that the scalar $\Delta$ plays a key role as the \emph{Majorana Higgs}
boson~\cite{Nemevsek:2016enw} (giving rise to the Majorana masses of the neutral leptons), 
with production and decay processes potentially leading to distinctive \emph{beautiful} 
and \emph{displaced} final states containing $N$ pairs and $b$-quarks~\cite{Fuks:2025jrn}. 
In practice, we update initial searches for heavy neutral leptons conducted during the LEP era 
by the ALEPH~\cite{ALEPH:1989tsb}, L3~\cite{L3:1990aqk, L3:1992xaz, L3:2001zfe}, 
DELPHI~\cite{DELPHI:1996qcc}, and CHARM II~\cite{CHARMII:1994jjr} collaborations.
We include the pair production of heavy neutrinos via $Z$ exchange, as analysed in both 
Dirac and Majorana scenarios in~\cite{Ma:1989jpa, Maalampi:1991fx, Djouadi:1993pe} (with even 
an early excess of events interpreted in the context of $B\!-\!L$ gauge extensions of the 
SM~\cite{Rosner:1984ic}). 
On the other hand, early theoretical studies within the LRSM~\cite{Gluza:1993gf, Gluza:1995js, 
Gluza:1996bz} examined both the $N\nu$ and $NN$ production channels at lepton colliders, while  
more recent works have revisited these processes in the context of future facilities. 
Sterile-neutrino searches at the FCC-ee~\cite{Blondel:2014bra, Antusch:2016vyf, Antusch:2016ejd,
Bellagamba:2025xpd} and effective-operator approaches~\cite{Barducci:2022hll, Bolton:2025tqw} 
have been found to provide complementary perspectives, whereas explorations at muon colliders 
could yield significant sensitivity improvements~\cite{Li:2023lkl, Li:2023tbx, Kwok:2023dck, 
Jiang:2023mte, deLima:2024ohf, Dev:2023nha, Bandyopadhyay:2024gyg, Dehghani:2025xkd}.

To carry out our study, we employ the implementation of the LRSM at next-to-leading 
order in QCD~\cite{Kriewald:2024cgr} within \textsc{FeynRules}~\cite{Christensen:2008py, 
Christensen:2009jx, Alloul:2013bka}, and we use the corresponding UFO~\cite{Degrande:2011ua, 
Darme:2023jdn} model\footnote{We use the updated version of the model file, \texttt{mlrsm-1.2}, 
publicly available at \url{https://github.com/FeynRules/FeynRules}.} to generate signal 
events with \textsc{MadGraph5\_aMC@NLO}~\cite{Alwall:2014bza}. 
This implementation includes all the relevant mixings in the scalar, gauge and fermion 
sectors that govern the dominant production and decay modes of the heavy neutrinos $N$ 
and the scalar triplet $\Delta$. 
We compute the relevant production and decay rates for final state systems with $n = 2, 3$ 
and 4 particles, and for processes mediated by the SM gauge bosons $V = W, Z$, their right-handed 
counterparts $W_R, Z_{LR}$, and the scalar states $h$ and $\Delta$. 
At low $\sqrt{s}$, resonant production at the $Z$ pole dominates; away from it, associated 
scalar channels, such as $V \Delta$ production and the pair production of heavy neutrinos 
via $t/u$-channel $W_R$ exchange, become significant. 
At even higher $\sqrt{s}$, vector-boson fusion (VBF) and scalar-boson fusion (SBF) may 
also contribute, although these are typically subdominant.
Most of these processes feature displaced signatures due to the long-lived nature of 
the heavy neutrino and may also exhibit prompt activity from the associated gauge bosons 
or charged leptons in the final state.
This subsequently enables the efficient suppression of the SM backgrounds, provided that 
the detector geometry and the vertexing efficiencies are well understood.
We find that distinguishing between the different production mechanisms is possible using 
momentum variables, lepton flavours, transverse displacement, and angular distributions, 
which we compute analytically and validate through dedicated Monte Carlo simulations.
%
%
A key methodological improvement in this work is the refined treatment of displaced vertices. 
Previous studies within the FCC-ee's IDEA detector~\cite{IDEAStudyGroup:2025gbt} of 
searches for heavy neutrinos with muon final states~\cite{Bellagamba:2025xpd} and 
searches for dark scalar $s$ decays in $h\to ss, s\to b\bar b$~\cite{Ripellino:2024iem} 
relied on the vertex fitting code of~\cite{Bedeschi:2024uaf}, implemented in \textsc{Delphes}~3~\cite{deFavereau:2013fsa} 
and the full \textsc{FCCAnalyses} software stack.
Here, we rely on the new graph-based vertexing algorithms introduced in~\cite{Kriewald:2025eiy}, 
enabling the full kinematical reconstruction of displaced long-lived particle decays.
%
%
We employ this new framework to fully exploit the IDEA tracking capabilities, 
showcasing its performance within the LRSM and demonstrating the extraction of 
physical parameters across multiple production channels.

%
We find that the FCC-ee can probe right-handed gauge boson masses $M_{W_R}$ deep 
in the multi-TeV range, with the precise reach depending on the production channel 
under consideration. 
Furthermore, the scalar triplet mass $m_\Delta$ typically lies below the $WW$ threshold, 
$m_\Delta < 2 M_W \simeq 160~\mathrm{GeV}$, such that sizeable production rates can be expected. 
Finally, the sensitivity to the heavy neutrino mass $m_N$ extends, in principle, up to the 
kinematic endpoint $m_N \simeq \sqrt{s}/2$, although it is, in practice, constrained by the 
tracking volume and geometry of the detector. 
When a significant fraction of signal events occurs beyond the inner tracker with a very 
large displacement, reaching the end of the drift chamber and/or at large pseudo-rapidities, 
vertex reconstruction indeed becomes limited by the geometric coverage of the detector, 
leading to a drop in reconstruction efficiency. 
%
%
Nevertheless, for relatively light neutrinos with $m_N \sim \mathcal{O}(10~\mathrm{GeV})$, 
the expected FCC-ee reach exceeds that of current direct searches at the LHC.
The combination of a clean experimental environment, high luminosity, and excellent 
vertexing precision makes the FCC-ee an unparalleled probe of Majorana neutrinos and 
lepton-number violation in the LRSM framework.

%
The rest of this paper is structured as follows.
We first discuss in Section~\ref{sec:LRSM} the main features of the LRSM relevant for the 
FCC-ee study at hand, in particular, the mass scales and couplings of heavy neutrinos to 
gauge and scalar bosons. 
We then present analytical computations of decay rates, branching ratios, and lifetimes in 
Section~\ref{subsec:DecaysBrs}, before moving on to production cross sections in 
Section~\ref{subsec:Xsecs}. 
We continue with a discussion on the expected signal sensitivities in 
Section~\ref{sec:SigSens} and detail possible kinematic reconstruction strategies 
in Section~\ref{sec:KinRecon}. 
Finally, we conclude with an outlook in Section~\ref{sec:Conclusion}, while 
two appendices are dedicated to technical details related to phase-space integrals (Appendix~\ref{app:PhaseSpace}), and proposals for benchmarks for the FCC and CEPC colliders (Appendix~\ref{app:FCC_CEPC_benchmarks}).

%
%
\section{The minimal Left-Right Symmetric Model} \label{sec:LRSM}


The minimal Left-Right Symmetric Model is based on a parity-symmetric gauge 
group $G_{LR} = SU(3)_c \otimes SU(2)_L \otimes SU(2)_R \otimes U(1)_{B-L}$, 
with a discrete symmetry that swaps the left and right fermions as well as 
the two $SU(2)_{L,R}$ gauge groups.
These are thus assigned the same gauge coupling $g_L = g_R\equiv g_w$,  while 
we denote the gauge coupling of $U(1)_{B-L}$ by $g_{B-L}$. 
The discrete and gauge symmetries are spontaneously broken in the scalar sector, 
which encompasses a bi-doublet 
$\phi = (\mathbf{1}, \mathbf{2}, \mathbf{2}, \mathbf{0})$ as well as two triplets
$\Delta_L = (\mathbf{1}, \mathbf{3}, \mathbf{1}, \mathbf{2})$ and
$\Delta_R = (\mathbf{1}, \mathbf{1}, \mathbf{3}, \mathbf{2})$ under $G_{LR}$,
leaving unbroken only the electric charge generator
\begin{equation}
  Q = T_{3 L} + T_{3 R} + \frac{B-L}{2} \, . 
\end{equation}
This follows from the vacuum structure of the scalar multiplets,
\begin{align} \label{eq:ScalarVevs}
  \langle \phi \rangle &= \begin{pmatrix} v_1 & 0 \\ 0 & -e^{i \alpha} v_2 \end{pmatrix} \, ,
  &
  \langle \Delta_{L,R} \rangle &= \begin{pmatrix} 0 & 0 \\ v_{L,R} & 0 \end{pmatrix} \, ,
\end{align}
where the vacuum expectation values $v_1$ and $v_2$, for which we can generically assume $v_2 < v_1$, can be traded with the parameters
\begin{align}
  v^2 &= v_1^2 + v_2^2 \equiv (174\,\GeV)^2& &\text{and} & 
  0 \leq t_\b\equiv \tan \beta &= \frac{v_2}{v_1} < 1 \, .
\end{align}
The vacuum expectation values $v_{1,2}$ and $v_{L,R}$ also lead to the generation 
of the masses of all the states in the model. 
In addition, due to phenomenological constraints, we must have $v_L \lesssim 1\,\GeV$ 
and $v_R \gtrsim 1\,\TeV$ so that it is convenient to introduce the small parameter 
\begin{align}
  \epsilon &= \frac{v}{v_R} \, ,
\end{align}
which describes the hierarchy between the $SU(2)_L$ and $SU(2)_R$ breaking scales 
as well as the decoupling limit of the left-right symmetric phase for $v_R \to \infty$. 
In practice, we can safely assume $\epsilon \lesssim 0.03$ such that expansions 
in $\epsilon$ converge swiftly.

\medskip

\paragraph{Gauge sector.}
The $W_L$ and $W_R$ charged gauge boson states have different masses of the order 
of $v$ and $v_R$, respectively
\begin{align} \label{eq:MWs}
  M_{W_L} &\simeq \frac{g_w \,v}{\sqrt 2} \, ,
  & 
  M_{W_R} &\simeq g_w\, v_R   \, ,
\end{align}
and are mixed by the unitary rotation 
\begin{align} \label{eq:UW}
  U_W &=
  \begin{pmatrix}
    c_\xi & s_\xi\, e^{-i \alpha}
    \\[1ex] 
    - s_\xi\, e^{i \alpha}  & c_\xi
  \end{pmatrix} \, , \qquad \text{with}\qquad s_\xi \simeq \frac{\epsilon^2}{2} 
  s_{2 \beta} \simeq \frac{M_{W_L}^2}{M_{W_R}^2} s_{2 \beta} \, .
\end{align}
In our notation, $c_\xi=\cos\xi$, $s_\xi=\sin\xi$  and $s_{2\beta}\equiv \sin2\beta$. 
The neutral gauge boson sector features a massless photon, the SM-like $Z$ boson,
and the new heavier $Z_{LR}$ gauge boson. 
Defining the weak mixing angle as the ratio of the $Z$ and $W_L$ masses at leading 
order in $\e$, $c_w \equiv \cos\theta_w = M_{W_L}/M_{Z} \big|_{\mathcal O(\epsilon^0)}$, 
the $Z_{LR}$ mass is predicted as 
\begin{align} \label{eqn:MZRM_{W_R}}
  M_{Z_{LR}} &\simeq M_{W_R} \sqrt{\frac{2c_w^2}{c_{2w}}}\simeq 1.67 \, M_{W_R}\, ,
\end{align}
with $c_{2w} \equiv \cos2\theta_w$. 
The mixing matrix associated with these neutral gauge bosons, namely the real 
orthogonal matrix $O_Z$ transitioning from the neutral gauge basis to the related 
mass basis, will thus be relevant up to second order in $\e$,
\begin{align}\label{eq:OZ}
  O_Z = &
  \begin{pmatrix}
    s_w & -c_w & 0 
    \\
    s_w & s_w\, t_w & \!\!-\frac{\sqrt{c_{2 w}}}{c_w} 
    \\
    \sqrt{c_{2 w}} & \sqrt{c_{2 w}}\, t_w & t_w 
  \end{pmatrix} 
  + \frac{\epsilon^2}{4}
  {\renewcommand{\arraystretch}{1.6}
    \begin{pmatrix}
    0 & 0 & \frac{c_{2 w}^{3/2}}{c_w^3} 
    \\
    0 & -\frac{c_{2 w}^2}{c_w^5} & -\frac{c_{2 w}^{3/2} t_w^2}{c_w^3} 
    \\
    0 & \frac{c_{2 w}^{3/2} t_w}{c_w^4} & -\frac{c_{2 w}^2 t_w}{c_w^4} 
  \end{pmatrix}} + \mathcal{O}(\e^3)\,,
\end{align}
where $t_w$ and $s_w$ are the tangent and sine of the weak mixing angle. 
For more details, such as the related would-be Goldstone bosons and ghost 
fields, as well as the gauge fixing Lagrangian, we refer to the complete 
analysis in~\cite{Kriewald:2024cgr}.  

\medskip

\paragraph{Scalar sector.}
In the $v_L \to 0$ limit, the doubly-charged boson $\Delta_R^{++}$ and the 
$\Delta_L$ triplet of states do not mix with any other scalar, and the related 
mass matrices are correspondingly diagonal. 
From the three remaining singly-charged gauge eigenstates $\Phi_{1}^+$, $\Phi_{2}^+$, 
and $\Delta_R^+$, only one massive singly charged state $H^+$ persists after 
the breaking of the model's gauge symmetry. 
The diagonalisation of the neutral scalar mass matrix, on the other hand, 
yields the four physical mass eigenstates $h$, $\Delta$, $H$ and $A$, with the 
first one corresponding to the SM-like Higgs boson.

All masses and mixings have been calculated explicitly in terms of the Lagrangian 
parameters in~\cite{Kriewald:2024cgr}, and the obtained expressions have been inverted 
to yield couplings in terms of a chosen set of physical inputs. 
Thus, in the following, we express all quantities in terms of physical masses and 
mixing angles.
In particular, we take $m_H$, $m_A$, $m_\Delta$ (plus $m_h = 125\,\GeV$) 
as independent inputs, as well as $m_{\Delta_R^{++}}$ and $m_{\Delta_L^0}$. 
We recall that if $m_H$ and $m_A$ can be considered independent, they have to be 
quite degenerate because their splitting originates from electroweak symmetry breaking and we want to avoid non-perturbative issues. 
From flavour constraints, they must lie at or beyond  $\sim 20\,\TeV$, so we set
\vspace*{-1ex}
\begin{align}
  m_H\simeq m_A \simeq 20\,\TeV\,.
\end{align}
Then the other three states are determined by the sum rules
\begin{align} \label{eq:sumrules}
  m^2_{H^+} \!-\! m^2_{A} &= m_{\Delta_L^{++}}^2 \!-\! m_{\Delta_L^+}^2 = 
  m_{\Delta_L^+}^2 \!-\! m_{\Delta_L^0}^2 = \e^2 c_{2\b}^2\, m_{A}^2\sim 
  {\cal O}(500\,\GeV)^2 \left(\frac{5\,\TeV}{M_{W_R}}\right)^2 \, ,
\end{align}
with as usual $c_{2\beta}\equiv\cos2\beta$. 
For low $M_{W_R}$ values, the large splitting of the $\Delta_L$ triplet implies 
violations of the electroweak precision tests~\cite{Maiezza:2016bzp}, unless the 
whole multiplet is as heavy as approximately $1\,\TeV$. 
This bound is relaxed for heavier $M_{W_R}$, down to \emph{circa} $m_{\Delta_L^{++}} 
\gtrsim 500 \,\GeV$ as stemming from direct searches. 
Furthermore, the orthogonal rotations that diagonalise the neutral sector depend 
on three physical mixing angles $\theta$, $\eta$ and $\phi$ that are respectively associated with the $h$-$\Delta$, $H$-$\Delta$, and $h$-$H$ 
mixings~\cite{Kriewald:2024cgr}. 
The first angle has to be small due to Higgs decay constraints, such that only the 
$\eta$ angle may be of $\mathcal{O}(1)$, provided that $\Delta$ and $H$ are nearly 
degenerate~\cite{Kriewald:2024cgr}. 
We do not consider such scenarios in this work, since we are interested in
a fairly light $\Delta$.
In the following, we keep $\eta=\phi=0$ and adopt various choices for $\theta$.
\begin{table}\renewcommand{\arraystretch}{2.4}
  \centering \resizebox{.9\linewidth}{!}{\begin{tabular}{c c || c c}
  Diagram & Feynman rule & Diagram & Feynman rule\\ \hline %
  \hspace{0.35cm}\parbox{10\unitlength}{\begin{fmffile}{SSS1}\vspace{0.35cm}
    \begin{fmfgraph*}(52,39)
    \fmfleft{i1}
    \fmfright{o1,o2}
    \fmf{dashes,tension=2}{i1,v1}
    \fmf{scalar}{v1,o1}
    \fmf{scalar}{o2,v1}
    \fmfdot{v1}
    \fmfv{label=$h$}{i1}
    \fmfv{label=$\Delta_R^{++}$, label.angle=0}{o1}
    \fmfv{label=$\Delta_R^{--}$, label.angle=0}{o2}
  \end{fmfgraph*}\end{fmffile}}\hspace{2cm}
  &  $i \sqrt{2} g_w \left[\frac{\epsilon \, m_A^2 c_{2\b} ^2}{M_{W_R}}-
  \frac{\theta \, m_{\Delta_R^{\smash{++}}}^2}{M_{W_R}}\right]$ & \hspace{.4cm}
  \parbox{10\unitlength}{\begin{fmffile}{SSS2}\vspace{0.35cm}
    \begin{fmfgraph*}(52,39)
    \fmfleft{i1}
    \fmfright{o1,o2}
    \fmf{dashes,tension=2}{i1,v1}
    \fmf{scalar}{v1,o1}
    \fmf{scalar}{o2,v1}
    \fmfdot{v1}
    \fmfv{label=$\Delta$}{i1}
    \fmfv{label=$\Delta_R^{++}$, label.angle=0}{o1}
    \fmfv{label=$\Delta_R^{--}$, label.angle=0}{o2}
  \end{fmfgraph*}\end{fmffile}}\hspace{2cm}
  & $i \sqrt{2} g_w \left[\frac{\theta \, \epsilon \,m_A^2 c_{2\b} ^2}{M_{W_R}}+
  \frac{m_{\Delta }^2+ 2\, m_{\Delta_R^{\smash{++}}}^2}{2 M_{W_R}}\right]$\\[.75cm]
  \hspace{0.35cm}\parbox{10\unitlength}{\begin{fmffile}{SSS3}\vspace{0.25cm}
    \begin{fmfgraph*}(52,39)
    \fmfleft{i1}
    \fmfright{o1,o2}
    \fmf{dashes,tension=2}{i1,v1}
    \fmf{scalar}{v1,o1}
    \fmf{scalar}{o2,v1}
    \fmfdot{v1}
    \fmfv{label=$h$}{i1}
    \fmfv{label=$\Delta_L^{++}$, label.angle=0}{o1}
    \fmfv{label=$\Delta_L^{--}$, label.angle=0}{o2}
  \end{fmfgraph*}\end{fmffile}}\hspace{2cm}
  & $i \sqrt{2} g_w \left[\frac{\epsilon\,  m_A^2 \left(c_\alpha^2s^2_{2\b}-1\right)
  }{ M_{W_R}}-\frac{\theta\,  m_{\Delta_L}^2}{M_{W_R}}\right]$ & \hspace{.4cm}
  \parbox{10\unitlength}{\begin{fmffile}{SSS4}\vspace{0.25cm}
    \begin{fmfgraph*}(52,39)
    \fmfleft{i1}
    \fmfright{o1,o2}
    \fmf{dashes,tension=2}{i1,v1}
    \fmf{scalar}{v1,o1}
    \fmf{scalar}{o2,v1}
    \fmfdot{v1}
    \fmfv{label=$\Delta$}{i1}
    \fmfv{label=$\Delta_L^{++}$, label.angle=0}{o1}
    \fmfv{label=$\Delta_L^{--}$, label.angle=0}{o2}
  \end{fmfgraph*}\end{fmffile}}\hspace{2cm}
  & $i \sqrt{2} g_w \left[\frac{\theta\,  \epsilon\,  m_A^2 \left(c_\alpha^2s^2_{2\b}-
  1\right)}{ M_{W_R}}+\frac{m_{\Delta }^2+ 2\, m_{\Delta_L}^2}{2 M_{W_R}}\right]$\\[.75cm]
\hline
&&\\[-6.5ex]
\hspace{0.35cm}\parbox{10\unitlength}{\begin{fmffile}{SSS5}\vspace{0.25cm}
    \begin{fmfgraph*}(52,39)
    \fmfleft{i1}
    \fmfright{o1,o2}
    \fmf{dashes,tension=2}{i1,v1}
    \fmf{dashes}{v1,o1}
    \fmf{dashes}{o2,v1}
    \fmfdot{v1}
    \fmfv{label=$h$}{i1}
    \fmfv{label=$h$, label.angle=0}{o1}
    \fmfv{label=$h$, label.angle=0}{o2}
  \end{fmfgraph*}\end{fmffile}}\hspace{2cm}
  & $\frac{i g_w}{M_{W}} \frac{3}{2} m_h^2 \left(1-\frac32 \theta \right) $ 
  & \hspace{.4cm}
  \parbox{10\unitlength}{\begin{fmffile}{SSS6}\vspace{0.25cm}
    \begin{fmfgraph*}(52,39)
    \fmfleft{i1}
    \fmfright{o1,o2}
    \fmf{dashes,tension=2}{i1,v1}
    \fmf{dashes}{v1,o1}
    \fmf{dashes}{o2,v1}
    \fmfdot{v1}
    \fmfv{label=$\Delta$}{i1}
    \fmfv{label=$h$, label.angle=0}{o1}
    \fmfv{label=$h$, label.angle=0}{o2}
  \end{fmfgraph*}\end{fmffile}}\hspace{2cm}
  & $\frac{i g_w}{ M_{W}}\,\theta \left[\frac{3}{2}  m_h^2+\frac{1}{2} 
  m_{\Delta }^2 (1+\theta  \epsilon)\right]$\\[.75cm]
\hspace{0.35cm}\parbox{10\unitlength}{\begin{fmffile}{SSS7}\vspace{0.25cm}
    \begin{fmfgraph*}(52,39)
    \fmfleft{i1}
    \fmfright{o1,o2}
    \fmf{dashes,tension=2}{i1,v1}
    \fmf{dashes}{v1,o1}
    \fmf{dashes}{o2,v1}
    \fmfdot{v1}
    \fmfv{label=$\Delta$}{i1}
    \fmfv{label=$\Delta$, label.angle=0}{o1}
    \fmfv{label=$\Delta$, label.angle=0}{o2} \end{fmfgraph*}\end{fmffile}}\hspace{2cm}
  &$\frac{i g_w }{M_{W_R}}\frac{3}{\sqrt{2}}m_{\Delta }^2$
  &\hspace{0.4cm} \parbox{10\unitlength}{\begin{fmffile}{SSS8}\vspace{0.25cm}
    \begin{fmfgraph*}(52,39)
    \fmfleft{i1}
    \fmfright{o1,o2}
    \fmf{dashes,tension=2}{i1,v1}
    \fmf{dashes}{v1,o1}
    \fmf{dashes}{o2,v1}
    \fmfdot{v1}
    \fmfv{label=$\Delta$}{i1}
    \fmfv{label=$\Delta$, label.angle=0}{o1}
    \fmfv{label=$h$, label.angle=0}{o2}
  \end{fmfgraph*}\end{fmffile}}\hspace{2cm}
  & $\frac{i g_w}{ M_{W}} \,\theta \left[\frac{3}{2} m_h^2 \theta + m_{\Delta }^2 \left(\theta-\epsilon\right)\right]$ \\[.75cm]
 \end{tabular}}
 \caption{Relevant trilinear scalar vertices of the $h$ and $\Delta$ states at 
 first order in $\theta$ and $\epsilon$. 
 For couplings involving $\Delta_{L,R}^{\pm\pm}$, the leading contribution is 
 governed by the $m_A^2 \sim (20~\mathrm{TeV})^2$ term. 
 \label{tab:SSS}}
\end{table}

\begin{table}\renewcommand{\arraystretch}{2.4}
  \centering \resizebox{\linewidth}{!}{\begin{tabular}{c c || c c}
  Diagram & Feynman rule & Diagram & Feynman rule\\ \hline %
  \hspace{0.35cm}\parbox{10\unitlength}{\begin{fmffile}{SVV1}\vspace{0.35cm}
    \begin{fmfgraph*}(52,39)
    \fmfleft{i1}
    \fmfright{o1,o2}
    \fmf{dashes,tension=2}{i1,v1}
    \fmf{wiggly}{v1,o1}
    \fmf{wiggly}{o2,v1}
    \fmf{phantom_arrow}{v1,o1}
    \fmf{phantom_arrow}{o2,v1}
    \fmfdot{v1}
    \fmfv{label=$h$}{i1}
    \fmfv{label=$W^+$, label.angle=0}{o1}
    \fmfv{label=$W^-$, label.angle=0}{o2}
  \end{fmfgraph*}\end{fmffile}}\hspace{2cm}
  &  $i g_w M_W \text{O}_{N,11}\, \eta^{\mu\nu} \approx i g_w M_W \, 
  \eta^{\mu\nu} $ & \hspace{.4cm}
  \parbox{10\unitlength}{\begin{fmffile}{SVV2}\vspace{0.35cm}
    \begin{fmfgraph*}(52,39)
    \fmfleft{i1}
    \fmfright{o1,o2}
    \fmf{dashes,tension=2}{i1,v1}
    \fmf{wiggly}{v1,o1}
    \fmf{wiggly}{o2,v1}
    \fmf{phantom_arrow}{v1,o1}
    \fmf{phantom_arrow}{o2,v1}
    \fmfdot{v1}
    \fmfv{label=$\Delta$}{i1}
    \fmfv{label=$W^+$, label.angle=0}{o1}
    \fmfv{label=$W^-$, label.angle=0}{o2}
  \end{fmfgraph*}\end{fmffile}}\hspace{2cm}
  & $i g_w M_W \text{O}_{N,12}\, \eta^{\mu\nu} \approx i g_w M_W \, 
  \t\, \eta^{\mu\nu}$\\[.75cm]
  \hspace{0.35cm}\parbox{10\unitlength}{\begin{fmffile}{SVV3}\vspace{0.25cm}
    \begin{fmfgraph*}(52,39)
    \fmfleft{i1}
    \fmfright{o1,o2}
    \fmf{dashes,tension=2}{i1,v1}
    \fmf{wiggly}{v1,o1}
    \fmf{wiggly}{o2,v1}
    \fmf{phantom_arrow}{v1,o1}
    \fmf{phantom_arrow}{o2,v1}
    \fmfdot{v1}
    \fmfv{label=$h$}{i1}
    \fmfv{label=$W_R^+$, label.angle=0}{o1}
    \fmfv{label=$W^-$, label.angle=0}{o2}
  \end{fmfgraph*}\end{fmffile}}\hspace{2cm}
  & $\begin{aligned}&g_w M_W\! \left[i e^{i \alpha } s_{2\b} \text{O}_{N,11}\!+\!c_{2 \beta} 
  \left(\text{O}_{N,41}\!-\!i \text{O}_{N,31}\right)\right] \eta^{\mu\nu} 
  \\ & \qquad \approx i g_w M_W e^{i \alpha } s_{2\b}\, 
  \eta^{\mu\nu}\end{aligned}$ & \hspace{.4cm}
  \parbox{10\unitlength}{\begin{fmffile}{SVV4}\vspace{0.25cm}
    \begin{fmfgraph*}(52,39)
    \fmfleft{i1}
    \fmfright{o1,o2}
    \fmf{dashes,tension=2}{i1,v1}
    \fmf{wiggly}{v1,o1}
    \fmf{wiggly}{o2,v1}
    \fmf{phantom_arrow}{v1,o1}
    \fmf{phantom_arrow}{o2,v1}
    \fmfdot{v1}
    \fmfv{label=$\Delta$}{i1}
    \fmfv{label=$W_R^+$, label.angle=0}{o1}
    \fmfv{label=$W^-$, label.angle=0}{o2}
  \end{fmfgraph*}\end{fmffile}}\hspace{2cm}
  & $\begin{aligned}&g_w M_W \! \left[i e^{i \alpha } s_{2\b} \text{O}_{N,12}\!+
  \!c_{2 \beta} \left(\text{O}_{N,42}\!-\!i \text{O}_{N,32}\right)\right] \eta^{\mu\nu}
  \\& \qquad  \approx i g_w M_W  e^{i \alpha } s_{2\b}\, \t\, \eta^{\mu\nu}\end{aligned}$\\[.75cm]
  \hspace{0.35cm}\parbox{10\unitlength}{\begin{fmffile}{SVV5}\vspace{0.25cm}
    \begin{fmfgraph*}(52,39)
    \fmfleft{i1}
    \fmfright{o1,o2}
    \fmf{dashes,tension=2}{i1,v1}
    \fmf{wiggly}{v1,o1}
    \fmf{wiggly}{o2,v1}
    \fmf{phantom_arrow}{v1,o1}
    \fmf{phantom_arrow}{o2,v1}
    \fmfdot{v1}
    \fmfv{label=$h$}{i1}
    \fmfv{label=$W_R^+$, label.angle=0}{o1}
    \fmfv{label=$W_R^-$, label.angle=0}{o2}
  \end{fmfgraph*}\end{fmffile}}\hspace{2cm}
  & $\begin{aligned}&i g_w \left(\sqrt{2} M_{W_R}\text{O}_{N,21}
   +M_W \text{O}_{N,11}\right) \eta^{\mu\nu}\\ & \qquad  \approx i g_w \sqrt{2}  
   M_{W_R}(\e/2-\theta)\, \eta^{\mu\nu}\end{aligned}$ & \hspace{.4cm}
  \parbox{10\unitlength}{\begin{fmffile}{SVV6}\vspace{0.25cm}
    \begin{fmfgraph*}(52,39)
    \fmfleft{i1}
    \fmfright{o1,o2}
    \fmf{dashes,tension=2}{i1,v1}
    \fmf{wiggly}{v1,o1}
    \fmf{wiggly}{o2,v1}
    \fmf{phantom_arrow}{v1,o1}
    \fmf{phantom_arrow}{o2,v1}
    \fmfdot{v1}
    \fmfv{label=$\Delta$}{i1}
    \fmfv{label=$W_R^+$, label.angle=0}{o1}
    \fmfv{label=$W_R^-$, label.angle=0}{o2}
  \end{fmfgraph*}\end{fmffile}}\hspace{2cm}
  & $\begin{aligned}&i g_w \left(\sqrt{2} M_{W_R}\text{O}_{N,22}
   +M_W \text{O}_{N,12}\right) \eta^{\mu\nu} \\ & \qquad \approx i g_w \sqrt{2} 
   M_{W_R}\, \eta^{\mu\nu} \end{aligned}$\\[.75cm]
 \end{tabular}}
 \caption{Relevant scalar-vector-vector vertices involving only charged vector bosons, 
 shown at leading order in $\epsilon$ and $\theta$. 
 In the first line the factor $O_{N,1 1}\simeq \cos\theta \simeq 1$ is omitted in 
 the right-hand side of the Feynman rule. 
 However, deviations from unity reduces the Higgs coupling to the SM $W$-boson and 
 underlies the experimental bounds imposed on $\theta$. \label{tab:SVV}}
 \end{table}

Tables~\ref{tab:SSS} and \ref{tab:SVV} collect the trilinear couplings of the 
$h$ and $\Delta$ states to other scalars and vectors that are relevant in our study. 
The full expressions are cumbersome, primarily due to the presence of
elements of the neutral-scalar mixing matrix $O_N$. 
However, they simplify considerably with the scalar mixing angles chosen above, 
allowing us to retain only the leading-order terms in $\epsilon$ and $\theta$. 
In the four trilinear couplings of the $h$ and $\Delta$ states with the 
doubly-charged scalars $\Delta_{L,R}^{\pm\pm}$ (first part of Table~\ref{tab:SSS}),
the first term in each line, proportional to the large pseudoscalar mass squared $m_A^2$, 
dominates over the second term for the small values of $m_\Delta$ and $m_{\Delta^{++}}$ 
considered here. 
As a result, the $\Delta$ couplings are suppressed by $\theta$ relative to the 
corresponding $h$ couplings. The expressions are given here at tree 
level and can receive loop corrections;
these may be particularly relevant for the $hhh$ and $h\Delta\Delta$ interactions, 
as discussed in detail in~\cite{Nemevsek:2018bbt}, but a full loop treatment is 
beyond the scope of the present work. 
Finally, the suppression of the $\Delta WW$ couplings shown in Table~\ref{tab:SVV} is especially important for the vector boson fusion processes considered below.

\medskip

\paragraph{Fermion sector.} 
The LRSM fermions are arranged in LR symmetric multiplets transforming under 
$SU(2)_L\times SU(2)_R\times U(1)_{B-L}$ as 
\begin{align}
Q^\prime_{L i} &= \begin{pmatrix}   u_L^\prime 
  \\
  d_L^\prime\end{pmatrix}_i     \sim \left(\mathbf{2}, \mathbf{1}, \mathbf{\frac{1}{3}} \right)  ,
  &Q^\prime_{R i} = \begin{pmatrix} u_R^\prime 
  \\ 
  d_R^\prime\end{pmatrix}_i     \sim \left(\mathbf{1}, \mathbf{2}, \mathbf{\frac{1}{3}} \right)  ,
  \\[.7ex]
  L^\prime_{L i} &= \begin{pmatrix} \nu_L^\prime 
  \\ 
  \ell_L^\prime\end{pmatrix}_i  \sim \left(\mathbf{2}, \mathbf{1}, -\mathbf{1} \right)  ,
  &L^\prime_{R i} = \begin{pmatrix} \nu_R^\prime 
  \\
  \ell_R^\prime\end{pmatrix}_i  \sim \left(\mathbf{1}, \mathbf{2}, -\mathbf{1} \right)  ,
\end{align}
where $i=1,2,3$ denotes the flavour index. In addition to the usual gauge interactions, 
these fermions couple to the scalar fields through the Yukawa Lagrangians 
$\mathcal L_Y^q$ and $\mathcal L_Y^\ell$ given by
\begin{align}
  \mathcal L_Y^q &= \bar Q_L^\prime \left( Y_q \, \phi + \tilde Y_q \, 
  \tilde \phi \right) Q_R^\prime + \text{H.c.} \, ,
  \\[1ex]
  \begin{split}
  \mathcal L_Y^\ell &= \bar L_L^\prime \left( Y_\ell \, \phi + \tilde Y_\ell \, 
  \tilde \phi \right) L_R^\prime + {}
  \bar L_L^{\prime c} i \sigma_2 \Delta_L Y_\Delta L_L^\prime +
  \bar L_R^{\prime c} i \sigma_2 \Delta_R Y_\Delta L_R^\prime + \text{H.c.} \,,
  \end{split}
\end{align}
where $Y_{q,\ell}$ and $\tilde Y_{q,\ell}$ are the conventional $3\times 3$ Yukawa matrices 
and $Y_\Delta$ is a symmetric matrix. 
Moreover, in this expression $\tilde\phi = \sigma_2 \phi^* \sigma_2$ denotes the 
field dual to $\phi$ (with $\sigma_2$ being the second Pauli matrix).
After the spontaneous breaking of the two $SU(2)$ groups, these interactions 
give rise to all fermion mass matrices whose diagonalisation yields the physical 
quark and lepton states, as well as the light and heavy neutrino states $\nu$ and $N$. 
Specifically, this implies that the mass of $N$ is generated spontaneously and
is tied to the breaking of $SU(2)_R \times U(1)_{B-L}$ through $v_R$ and 
$Y_\Delta$, as shown below in~\eqref{eq:MnuSeesaw}.

In the quark sector, the charged current interactions involve the left-handed and right-handed CKM matrices $V_{L,R}^{\text{CKM}}$. Importantly, $V_R^{\text{CKM}}$ is not a free parameter: it is predicted by the model~\cite{Maiezza:2010ic} and has been shown to lie very close to $V_L^{\text{CKM}}$~\cite{Senjanovic:2014pva, Senjanovic:2015yea}. In the following we therefore set $V_R^{\text{CKM}} = V_L^{\text{CKM}}$. In the lepton sector, the charged-lepton and Dirac neutrino mass matrices are determined by the electroweak vacuum expectation value $v$. The Majorana masses of the heavy neutrinos are instead of $\mathcal{O}(v_R)$, while the light neutrino masses arise from a combination of type-I and type-II 
seesaw contributions, scaling as $\mathcal{O}(v^2/v_R)$ and $\mathcal{O}(v_L)$
\begin{align} \label{eq:MnuSeesaw}
  M_N & \simeq v_R Y_\Delta \, , & M_\nu &\simeq \frac{v_L}{v_R}M_N -M_DM_N^{-1}M_D^T \,.
\end{align} 
Remarkably, the Dirac mass matrix $M_D$ is also not a free parameter but is predicted by the left-right symmetry~\cite{Nemevsek:2012iq}. We can therefore fully determine all Yukawa couplings in terms of physical inputs. We take these to be the masses of the three light and three heavy neutrinos and their respective PMNS mixing matrices $U_\nu$ and $U_N$. We consequently treat $U_\nu$ and $U_N$ as unitary.
In practice, we work with $v_L \simeq 0$, \textit{i.e}.\ in the regime
of type I seesaw dominance, where we take $U_{N,ij} \simeq \delta_{ij}$
and compute $M_D$ (and the subsequent mixing effects) from Eq.~\eqref{eq:MnuSeesaw}.
The mixing between heavy and light states is of order $10^{-6}$ in 
this low-scale seesaw, and thus has a negligible impact on the 
processes relevant to our study except for the regions of 
parameter space where $M_{W_R} \gtrsim 50 \text{ TeV}$.

\begin{table}\renewcommand{\arraystretch}{2.4}
  \centering \resizebox{.9\linewidth}{!}{\begin{tabular}{c c || c c}
  Diagram & Feynman rule & Diagram & Feynman rule\\ \hline %
  \hspace{0.35cm}\parbox{10\unitlength}{\begin{fmffile}{SFF1}\vspace{0.35cm}
    \begin{fmfgraph*}(52,39)
    \fmfleft{i1}
    \fmfright{o1,o2}
    \fmf{dashes,tension=2}{i1,v1}
    \fmf{plain}{v1,o1}
    \fmf{plain}{o2,v1}
    \fmfdot{v1}
    \fmfv{label=$h$}{i1}
    \fmfv{label=$N_j$, label.angle=0}{o1}
    \fmfv{label=$N_i$, label.angle=0}{o2}
  \end{fmfgraph*}\end{fmffile}}\hspace{2cm}
  & $\begin{aligned}& \\[-.3cm] & -\frac{i O_{N, 21} U_{N, ki}^* U_{N, kj} Y_{\Delta, k}}{\sqrt{2}} \\ &\qquad \quad \approx \frac{i g_w M_{N_i}}{\sqrt{2} M_{W_R}}\delta _{ij}\t \end{aligned}$ & \hspace{.4cm}
  \parbox{10\unitlength}{\begin{fmffile}{SFF2}\vspace{0.35cm}
    \begin{fmfgraph*}(52,39)
    \fmfleft{i1}
    \fmfright{o1,o2}
    \fmf{dashes,tension=2}{i1,v1}
    \fmf{plain}{v1,o1}
    \fmf{plain}{o2,v1}
    \fmfdot{v1}
    \fmfv{label=$\Delta$}{i1}
    \fmfv{label=$N_j$, label.angle=0}{o1}
    \fmfv{label=$N_i$, label.angle=0}{o2}
  \end{fmfgraph*}\end{fmffile}}\hspace{2cm}
  & $\begin{aligned}& \\[-.3cm] & -\frac{i O_{N,2,2} U_{N, ki}^*   U_{N, kj} Y_{\Delta, k}}{\sqrt{2}} \\ &\qquad\quad  \approx -\frac{i  g_w M_{N_i}}{\sqrt{2} M_{W_R}}\delta _{ij}\end{aligned}$ \\[.75cm]
 \end{tabular}}
  \caption{Scalar-$NN$ vertices at leading order in $\theta$, neglecting subleading scalar mixings and Dirac heavy-light neutrino mixing. In this limit, the couplings are flavour diagonal and the $\Delta$ scalar is responsible for generating heavy neutrino masses via the Yukawa coupling $Y_{\Delta,i}=g_w\,M_{N_i}/M_{W_R}$.\label{tab:SNN}}
 \end{table}

In Table~\ref{tab:SNN}, we collect the couplings of the $h$ and $\Delta$ 
scalars to light and heavy neutrinos that are relevant for their decays 
into lepton-number-violating final states.
Throughout, we retain only the leading terms in the scalar mixing angle 
$\theta$ and neglect the tiny Dirac heavy-light mixing.
As expected, the SM-like Higgs boson $h$ couples to $NN$ pairs only through 
its admixture with the $\Delta$ state, which is directly controlled by~$\theta$.

In Table~\ref{tab:zff}, we collect the neutral-current couplings relevant
for the production and decay of the $Z$ and $Z_{LR}$ bosons into fermions, 
including the heavy neutrinos $N$. 
All expressions are shown at leading order in $\epsilon$ and are decomposed 
into vector and axial components. 
The standard quark and charged-lepton neutral currents appear together with 
their analogues for the heavier $Z_{LR}$ boson, and we can additionally observe 
that the neutrino sector exhibits a characteristic pattern. 
The $Z_{LR}$ state couples predominantly to the heavy state $N$, while its 
couplings to light neutrinos $\nu$ are instead suppressed by $s_w$.
This corresponds to the expected behaviour in the limit $s_w\!\to\!0$ in 
which electroweak symmetry breaking would not induce any mixing between the 
two gauge sectors. 
Conversely, the coupling of the SM-like $Z$ boson to an $NN$ pair vanishes 
at leading order and is generated only via the left-right-induced $Z$-$Z_{LR}$ 
mixing of order $\epsilon^2$ depicted by Eq.~\eqref{eq:OZ}. 
This $\epsilon^2$ contribution is therefore retained explicitly in the 
$ZNN$ entry of the table.

Finally, we report in Table~\ref{tab:WLL} the associated leptonic charged currents.
The mass eigenstates of the charged gauge bosons mediate interactions of both chiralities due to the left-right mixing encoded in Eq.~\eqref{eq:UW}. 
We omit from the discussion the analogous expressions holding for the quark sector that depend on the $V_{L,R}^{\text{CKM}}$ matrices.
\begin{table}\renewcommand{\arraystretch}{1.4}
  \centering \resizebox{.8\linewidth}{!}{\begin{tabular}{c c || c c}
  Diagram & Feynman rule & Diagram & Feynman rule\\ \hline %
  \hspace{0.35cm}\parbox{10\unitlength}{\begin{fmffile}{FFV1}\vspace{0.35cm}
    \begin{fmfgraph*}(52,39)
    \fmfleft{i1}
    \fmfright{o1,o2}
    \fmf{wiggly,tension=2}{i1,v1}
    \fmf{fermion}{v1,o1}
    \fmf{fermion}{o2,v1}
    \fmfdot{v1}
    \fmfv{label=$Z$}{i1}
    \fmfv{label=$f$, label.angle=0}{o1}
    \fmfv{label=$\bar f$, label.angle=0}{o2}
  \end{fmfgraph*}\end{fmffile}}\hspace{2cm}
  &  $i \frac{g_w}{c_w} \Big[ v_f \gamma^\mu + a_f\, \gamma^\mu \gamma_5\Big]$ & \hspace{.4cm}
  \hspace{0.35cm}\parbox{10\unitlength}{\begin{fmffile}{FFV2}\vspace{0.35cm}
    \begin{fmfgraph*}(52,39)
    \fmfleft{i1}
    \fmfright{o1,o2}
    \fmf{wiggly,tension=2}{i1,v1}
    \fmf{fermion}{v1,o1}
    \fmf{fermion}{o2,v1}
    \fmfdot{v1}
    \fmfv{label=$Z_{LR}$}{i1}
    \fmfv{label=$f$, label.angle=0}{o1}
    \fmfv{label=$\bar f$, label.angle=0}{o2}
  \end{fmfgraph*}\end{fmffile}}\hspace{2cm}
  & $i \frac{g_w}{c_w} \Big[ {\mathfrak v}_f \gamma^\mu + {\mathfrak a}_f\, \gamma^\mu \gamma_5\Big]$\\[0.75cm]
 \end{tabular}}

\vspace{1.5ex}
\hspace{-1.5ex}with 
\vspace{1ex}
 
\begin{tabular}{c | c c | c c}
 $f$   & $v_f$ & $a_f$ & $\mathfrak{v}_f$ & $\mathfrak{a}_f$\\ \hline
 $u$   & $\frac{1-4 c_{2 w}}{12}$ & $ \frac{1}{4}$ & $\frac{1-4 c_{2 w}}{12 \sqrt{c_{2 w}}}$ & $-\frac{1}{4} \sqrt{c_{2 w}}$\\
 $d$   & $\frac{1+2 c_{2 w}}{12}$  & $-\frac{1}{4}$ & $\frac{1+2 c_{2 w}}{12 \sqrt{c_{2 w}}}$ & $\frac{1}{4}\sqrt{c_{2 w}}$ \\
 $e$   & $\frac{1-4 s^2_w}{4}$    & $-\frac{1}{4}$ & $\frac{1-4s_w^2}{4 \sqrt{c_{2 w}}}$    & $\frac{1}{4}\sqrt{c_{2 w}}$\\
 $\nu$ & $0$ & $\frac{1}{2} \delta_{ij}$ & 0 & $\frac{s^2_w}{2 \sqrt{c_{2 w}}} \delta_{ij}$\\
 $N$   & $0$ & $\frac{1}{8} \e^2 \left(t_w^2-1\right) \delta_{ij}$ & 0 & $-\frac{c_w^2}{2
   \sqrt{c_{2 w}}} \delta_{ij}$\\
\end{tabular}
 
 \caption{Neutral-current fermion interaction vertices involving the $Z$ and $Z_{LR}$ bosons, given in terms of their vector and axial components. In the limit of negligible Dirac heavy-light mixings, the $\nu\nu$ and $NN$ couplings are flavour diagonal, and the $Z\nu N$ vertices vanish. \label{tab:zff}}
 \end{table}

\begin{table}\renewcommand{\arraystretch}{1.4}
  \centering \resizebox{.8\linewidth}{!}{\begin{tabular}{c c || c c}
  Diagram & Feynman rule & Diagram & Feynman rule\\ \hline %
  \hspace{0.35cm}\parbox{10\unitlength}{\begin{fmffile}{FFV3}\vspace{0.35cm}
    \begin{fmfgraph*}(52,39)
    \fmfleft{i1}
    \fmfright{o1,o2}
    \fmf{wiggly,tension=2}{i1,v1}
    \fmf{fermion}{v1,o1}
    \fmf{fermion}{o2,v1}
    \fmfdot{v1}
    \fmfv{label=$W^-$}{i1}
    \fmfv{label=$\nu_j$, label.angle=0}{o1}
    \fmfv{label=$\bar \ell_i$, label.angle=0}{o2}
  \end{fmfgraph*}\end{fmffile}}\hspace{2cm}
  &  $\frac{ig_w}{\sqrt{2}}\gamma^\mu P_L \,c_{\xi } \, U_{\nu,ij} $ & \hspace{.4cm}
  \hspace{0.35cm}\parbox{10\unitlength}{\begin{fmffile}{FFV4}\vspace{0.35cm}
    \begin{fmfgraph*}(52,39)
    \fmfleft{i1}
    \fmfright{o1,o2}
    \fmf{wiggly,tension=2}{i1,v1}
    \fmf{fermion}{v1,o1}
    \fmf{fermion}{o2,v1}
    \fmfdot{v1}
    \fmfv{label=$W^-_R$}{i1}
    \fmfv{label=$\nu_j$, label.angle=0}{o1}
    \fmfv{label=$\bar \ell_i$, label.angle=0}{o2}
  \end{fmfgraph*}\end{fmffile}}\hspace{2cm}
  & $\frac{ig_w}{\sqrt{2}}\gamma^\mu P_R\, e^{i \alpha }  s_{\xi }\,U_{\nu,ij} $\\[0.75cm]
  \hspace{0.35cm}\parbox{10\unitlength}{\begin{fmffile}{FFV5}\vspace{0.35cm}
    \begin{fmfgraph*}(52,39)
    \fmfleft{i1}
    \fmfright{o1,o2}
    \fmf{wiggly,tension=2}{i1,v1}
    \fmf{fermion}{v1,o1}
    \fmf{fermion}{o2,v1}
    \fmfdot{v1}
    \fmfv{label=$W^-$}{i1}
    \fmfv{label=$N_j$, label.angle=0}{o1}
    \fmfv{label=$\bar \ell_i$, label.angle=0}{o2}
  \end{fmfgraph*}\end{fmffile}}\hspace{2cm}
  &  $-\frac{ig_w}{\sqrt{2}}\gamma^\mu P_L\, e^{-i \alpha }  s_{\xi }\, U_{N,ij}^*$ & \hspace{.4cm}
  \hspace{0.35cm}\parbox{10\unitlength}{\begin{fmffile}{FFV6}\vspace{0.35cm}
    \begin{fmfgraph*}(52,39)
    \fmfleft{i1}
    \fmfright{o1,o2}
    \fmf{wiggly,tension=2}{i1,v1}
    \fmf{fermion}{v1,o1}
    \fmf{fermion}{o2,v1}
    \fmfdot{v1}
    \fmfv{label=$W^-_R$}{i1}
    \fmfv{label=$N_j$, label.angle=0}{o1}
    \fmfv{label=$\bar \ell_i$, label.angle=0}{o2}
  \end{fmfgraph*}\end{fmffile}}\hspace{2cm}
  & $\frac{ig_w}{\sqrt{2}}\gamma^\mu P_R \,c_{\xi }\, U_{N,ij}^*$\\[0.75cm]
 \end{tabular}}
 \caption{Leptonic charged–current interactions mediated by the charged gauge boson mass eigenstates, $P_L$ and $P_R$ respectively indicating the left-handed and right-handed chirality projectors.\label{tab:WLL}}
 \end{table}

%
%
\section{Decay rates and cross sections} \label{sec:DRatXSec}

In this section, we begin by reviewing the dominant decay rates and branching ratios 
of the LRSM gauge bosons, scalars and heavy neutrinos $N$. 
Our main focus is on the heavy-neutrino sector and its LNV decay channels that may 
give rise to displaced signatures. 
The magnitude of the displacement is determined by the heavy neutrino lifetime in 
its rest frame, together with the production-dependent boost factors that we discuss 
in Section~\ref{subsec:DecaysBrs}.
Subsequently, we provide in Section~\ref{subsec:Xsecs} analytic expressions for the 
cross sections of several heavy neutrino production channels relevant for future 
FCC-ee collisions in which pairs of heavy neutrinos are produced through 
scalar-mediated or gauge boson mediated processes. 
We consider both relevant $2\to2$ processes and $2\to4$ vector and scalar boson 
fusion channels.

%
\subsection{Decay rates, branching ratios and lifetimes} \label{subsec:DecaysBrs}
\subsubsection{\texorpdfstring{$Z$ and $Z_{LR}$ decays to heavy neutrino pairs} ~}
Proposed future electron-positron $Z$-pole runs at a centre-of-mass energy 
$\sqrt{s}\simeq M_Z$ would produce an enormous sample of up to $\mathcal O(10^{12})$ $Z$ 
bosons, as shown in Tables~\ref{tab:FCCee} and~\ref{tab:CEPC} in 
Appendix~\ref{app:FCC_CEPC_benchmarks}. 
This will subsequently open the possibility to search for extremely rare exotic 
$Z$ decays, including those into heavy neutrinos. 
In this context, we consider the gauge coupling of the heavy $N$ to the $Z$ boson, 
as well as that to the $Z_{LR}$ boson for completeness. 
Both of these couplings are flavour diagonal in our approximation (see the Feynman 
rules in Table~\ref{tab:zff}), so we drop all flavour indices in the expressions 
provided below, and we recall that for a Majorana fermion the neutral vector current vanishes. 
Therefore, only the axial couplings remain, and the $Z_{LR}NN$ one turns out to 
be parametrically unsuppressed, while the $ZNN$ one is suppressed by the 
neutral-gauge mixing and thus first appears at ${\cal O}(\e^2)$.

The corresponding partial widths read
\begin{equation} \label{eq:GamZNN}
  \Gamma \left(Z \to NN \right) = \frac{\alpha_w a_N^2}{6 c_w^2} M_Z \beta_{NZ}^3 \, ,
  \qquad\text{and}\qquad
  \Gamma \left(Z_{LR} \to NN \right) = \frac{\alpha_w \mathfrak a_N^2}{6 c_w^2} 
  M_{Z_{LR}} \beta_{NZ_{LR}}^3\, ,
\end{equation}
where $\alpha_w=g_w^2/(4\pi)$ and $a_N,\mathfrak a_N$ are the axial couplings given 
in Table~\ref{tab:zff}. 
These formul\ae\ follow from squaring the amplitudes 
$\mathcal A_Z=(g_w/c_w)\,a_N\,\bar u(p_1)\slashed{\varepsilon}\gamma_5 v(p_2)$ and 
$\mathcal A_{Z_{LR}}=(g_w/c_w)\,\mathfrak{a}_N\,\bar u(p_1)\slashed{\varepsilon}\gamma_5 v(p_2)$, 
hence retaining (through the form of the axial couplings) only the leading terms 
in the neutral-gauge rotation by construction. 
In particular, contributions that convert $Z \to \nu \nu$ into $Z \to NN$ decays 
via two heavy-light insertions are doubly suppressed by small Dirac mixings and are 
neglected here. 
The phase-space factor $\beta_{NV}^3 = (1-4 m_N^2/M_V^2)^{3/2}$ is the usual velocity 
suppression for decays into Majorana fermion pairs, such that the $Z \to NN$ decay 
is kinematically allowed only for $m_N < M_Z/2$ and is induced by the 
${\cal O}(\epsilon^2)$ mixing.
The associated rate hence scales parametrically as $\e^4\sim (M_W/M_{W_R})^4$, 
as expected on general grounds from the decoupling limit.

A simple order-of-magnitude estimate illustrates the potential reach of future 
colliders. 
With $N_Z \sim 10^{12}$ produced $Z$ bosons, a background-free search with 
$\mathcal O(1)$ efficiency would probe branching ratios down to $\mathcal O(10^{-12})$. 
As the $Z\to NN$ branching fraction scales like $M_{W_R}^{-4}$, this na\"ive 
estimate gives
\begin{equation}
  M_{W_R}\sim M_W\,N_Z^{1/4}\sim M_W\times 10^3\sim \mathcal O(10^2)\ \mathrm{TeV}\,,
\end{equation}
\textit{i.e.}\ a parametric sensitivity at the $10-100\,\mathrm{TeV}$ scale.
In practice, realistic detector acceptance, trigger/selection efficiencies and 
vertexing requirements for displaced decays reduce this reach substantially.
We therefore refine the realistic sensitivity in the next sections by including 
selection cuts, tracker geometry and estimated vertexing performance. 
Finally, away from the $Z$ pole ($\sqrt{s}\neq M_Z$), $NN$ pair production proceeds through the off-shell $e^+e^-\to Z, Z_{LR}\to NN$ contributions discussed in Section~\ref{subsec:Xsecs}. 
On-shell $Z$ production, as considered here, indeed remains relevant only very 
close to the $Z$ pole.

\subsubsection{\texorpdfstring{$\Delta$ and Higgs decays to heavy neutrino pairs } ~}
As discussed above, the SM Higgs $h$ and the triplet scalar $\Delta$ mix, opening 
two scalar decay channels into heavy-neutrino pairs with decay rates  
\begin{align} \label{eq:GamD2NN}
  \!\!\! 
  \Gamma\left(\Delta \to NN \right) &= \frac{c_\theta^2 \alpha_w m_\Delta}{8}
  \left( \frac{m_N}{M_{W_R}} \right)^2 \beta_{N\Delta}^3 \, ,
  & 
  \Gamma\left(h \to NN \right) &= \frac{s_\theta^2 \alpha_w m_h}{8}
  \left( \frac{m_N}{M_{W_R}} \right)^2 \beta_{Nh}^3 \, ,
\end{align}
where $\beta_{Ni}^2 = 1 - (2 m_N / m_i)^2$.
These couplings are flavour diagonal (see Table~\ref{tab:SNN}), so each 
decay produces a pair of identical heavy neutrinos.
As can be seen in Eq.~\eqref{eq:GamD2NN}, the decay rates are directly proportional to the heavy neutrino mass $m_N^2$, which is the essence of the spontaneous mass generation (as for charged fermions in the SM).
This contrasts with the $Z \to NN$ decay, where the rate depends on the gauge structure
(couplings, charge assignments and mixings) and is not proportional to the neutrino mass which only enters in the phase space factor.
The dependence on the mixing angle $\theta$ reflects the fact that, in the limit
where the relevant part of the neutral scalar sector effectively reduces to the 
two lightest states, the triplet-like decay width scales by $c_\theta^2$ and the 
SM-like one by $s_\theta^2$. 
Moreover, the decay rates into SM final states follow the standard Higgs 
expressions (see \textit{e.g.}~\cite{Djouadi:2005gi}), with the $h$ channels 
suppressed by $c_\theta^2$ and the $\Delta$ channels~\cite{Nemevsek:2016enw}
suppressed by $s_\theta^2$.
The detailed analytic expressions for the decays of $\Delta$ into SM final states 
(two-body and three-body decays) can be found in~\cite{Fuks:2025jrn}.

\subsubsection{Heavy neutrino decays}
Finally, a heavy neutrino $N_k$ (of flavour $k$) generally decays via a three-body 
channel $N_k \to \ell_\alpha^\pm q_i \bar q_j$, where $\alpha$ denotes the lepton 
flavour and $i$ and $j$ denote the quark flavours. 
In this case, the associated partial decay width can be approximated
by~\cite{Nemevsek:2023hwx}
\begin{align}
  \Gamma(N_k \to \ell_\alpha^\pm q_i \bar q_j) \simeq 
  \frac{\alpha_w^2 m_{N_k}^5}{64 \pi M_{W_R}^4}
  \left|V_{R,\, ij}^\text{CKM} \right|^2\left|U_{N,\, \alpha k} \right|^2
  \left(1 - 8 x + 8 x^2 - x^4 - 12x^2 \log x \right) \, ,
\end{align}
with $x = m_q^2/m_{N_k}^2$ and $m_q$ being the mass of the heavier of the two 
final-state quarks.
If $m_{N_k} \gtrsim M_W$, two-body decay modes $N_k \to \ell_\alpha^\pm W^\mp$ 
open with partial widths given by
\begin{align}
  \Gamma(N_k \to \ell_\alpha^\pm W^\mp) &= \frac{\alpha_w m_{N_k}}{8} s_\xi^2 
  \left(\frac{1}{y} - y \right) \, , \qquad\text{with}\quad
  y = \frac{M_W^2}{m_{N_k}^2} \, .
\end{align}
Such channels may dominate in some parts of the parameter space, as discussed 
in~\cite{Nemevsek:2023hwx}.

%
%
\subsection{Production cross sections} \label{subsec:Xsecs}

\begin{figure}[t]
  \centering
  \includegraphics[width=.73 \columnwidth]{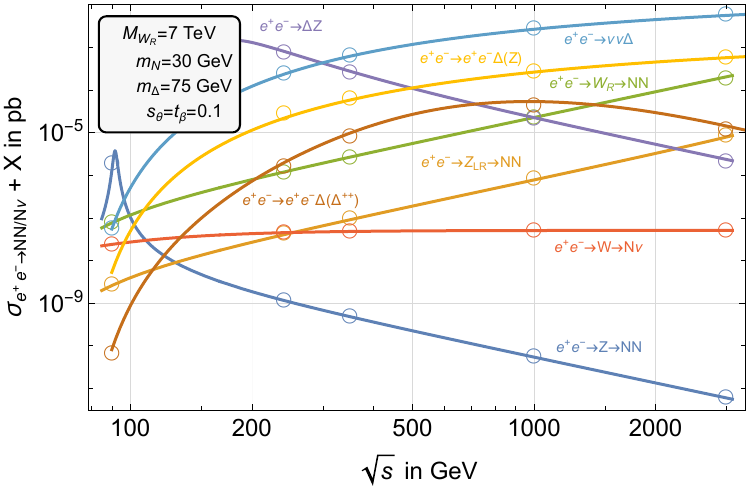}
  \vspace*{-2ex}
  \caption{Cross sections for various heavy neutrino ($N$) and scalar ($\Delta$) 
  production channels, possibly together with SM neutrinos or a $Z$ boson. 
  The solid lines show the analytic results discussed in the text, while the circles correspond to \textsc{MadGraph5\_aMC@NLO} simulations using the LRSM model file of~\cite{Kriewald:2024cgr} for the benchmark runs listed in Tables~\ref{tab:FCCee} 
  and~\ref{tab:CEPC}.  
  \label{fig:Xsecs}}
\end{figure}

In this section, we compute cross sections and several kinematic distributions for
$e^+ e^-$ collisions, focusing on the dominant and several sub-dominant production
channels of the LRSM new states.
In particular, we evaluate the boost factors $\gamma$ and the distributions in the
transverse momentum $p_T$ and polar angle $\cos\theta$ for processes yielding the
production of both heavy neutrinos $N$ and scalars $\Delta$.
A global overview of the total cross sections summarising all calculations achieved 
in this section is presented for a specific benchmark scenario and various 
centre-of-mass energies $\sqrt{s}$ in Figure~\ref{fig:Xsecs}.

\subsubsection{Heavy neutrino production via gauge boson exchanges}
\begin{figure}
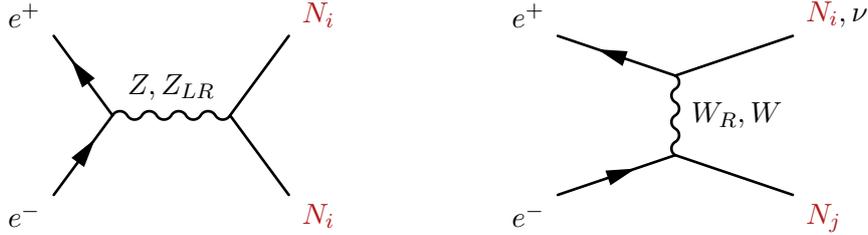

  \vspace{1 em}
  \centering
  \input{figures/fig_epem_ZZLR_NN.tex}
  \hspace{2 cm}
  \input{figures/fig_epem_WR_NNnu.tex}%
  \vspace{1.5ex}
  \caption{Feynman diagrams illustrating the pair production of heavy neutrinos 
  $N$, and their associated production with a light neutrino $\nu$.
  \label{fig:PairAssocN}}
\end{figure}

The simplest processes that produce a heavy neutrino pair $N N$ or an 
associated $N \nu$ pair are $2 \to 2$ scatterings mediated either by the 
$s$-channel exchange of a neutral gauge boson $V^0 = Z, Z_{LR}$ or via the 
$t$- and $u$-channel exchanges of a charged gauge boson $V^\pm = W, W_R$, 
as illustrated in the Feynman diagrams of Figure~\ref{fig:PairAssocN}.  
The neutral $V^0$ bosons only couple diagonally in flavour space, as shown 
in Table~\ref{tab:zff}, whereas the charged bosons can mediate the production 
of $N_i N_j$ pairs with different flavours.  
In the region of interest and for the production of neutrinos of the same 
flavour, the cross sections are well approximated by
\vspace*{-.5ex}
\begin{align} \label{eq:sig_epem_V_NN}
  \sigma_{e^+ e^- \to V^0 \to N N} &= \frac{4 \pi \alpha_w}{c_w^2} 
  \left( v_e^2 + a_e^2 \right) \frac{\sqrt{s}}{(s - M^2)^2 + \Gamma^2 M^2} 
  \frac{\alpha_w a_N^2}{6 c_w^2} \sqrt{s} \beta_{Ns}^3 \, ,
  \\ \label{eq:sig_epem_WR_NN}
  \sigma_{e^+ e^- \to W_R \to N N}  &\simeq \frac{\pi \alpha_w^2 |V_{eN}|^4}{24} 
  \frac{s}{M_{W_R}^4} \beta_{Ns}^3 \, ,
  \\[-.5ex] \label{eq:sig_epem_W_Nnu}
  \sigma_{e^+ e^- \to W \to N \nu}  &\simeq 
  \frac{\pi  \alpha_w^2 s_\xi^2 \left|V_{eN}\right|^2}{2 M_W^2}
  \frac{s \beta_{\nu s}^4}{M_W^2 + s \beta_{\nu s}^2} \, ,
\end{align}
where $\beta_{Ns}^2 = 1 - 4 m_N^2/s$, $\beta_{\nu s}^2 = 1 - m_N^2/s$. 
Moreover, $\Gamma$ is the total width of the exchanged neutral boson $V^0$ and 
$V_{eN} \equiv U_{N,eN}$ in the notation of Table~\ref{tab:WLL}.
Close to the $Z$ pole ($\sqrt s \sim M_Z$), $s$-channel $Z$ exchange dominates 
the production of heavy neutrino pairs (blue line in Figure~\ref{fig:Xsecs}).  
However, moving away from the pole, $W_R$ exchanges take over (green line), 
while $Z_{LR}$ exchanges (orange line) remain sub-dominant due to 
$M_{Z_{LR}} \sim 1.7 M_{W_R}$ in the minimal version of the LRSM.  
Nonetheless, at very high energies $\sqrt s \gtrsim 2~\text{TeV}$, the 
$Z_{LR}$-induced cross section becomes comparable to that at the $Z$ pole, 
reaching the same order of magnitude, though still slightly suppressed relative 
to $W_R$ mediation.

The kinematics of both the $NN$ and $N\nu$ production channels are fixed by the 
centre-of-mass energy $\sqrt s$. 
For the pair production mode $e^+ e^- \to NN$, the invariant mass of the heavy 
neutrino system is given by $m^{\rm inv}_{NN} = \sqrt{s}$.
In the associated production channel $e^+ e^- \to N\nu$ the invariant mass of the
$(N,\slashed{E})$ system is similar and reads $m^{\rm inv}(p_N, p_{\slashed E}) =
\sqrt{s}$, with $p_{\slashed E}$ and $p_N$ being the four-momentum of the escaping 
SM neutrino and that of the heavy neutrino respectively.
These kinematic relations determine the boost of the produced heavy neutrinos.
For $NN$ production, both neutrinos share the energy symmetrically, yielding a boost-factor of
\begin{equation}
  \gamma_{NN} = \frac{\sqrt{s}}{2 m_N} \, .
\end{equation}
For associated $N\nu$ production, the heavy neutrino receives instead a larger 
share of the available energy, and its boost is then given by
\begin{equation}
  \gamma_{N\nu} = \frac{s + m_N^2}{2 \sqrt{s}\, m_N}\, .
\end{equation}
Subsequently, these boost factors directly set the displacement of the 
heavy-neutrino decay products in the laboratory frame.

\begin{figure}
  \centering
  \includegraphics[width=.48\columnwidth]{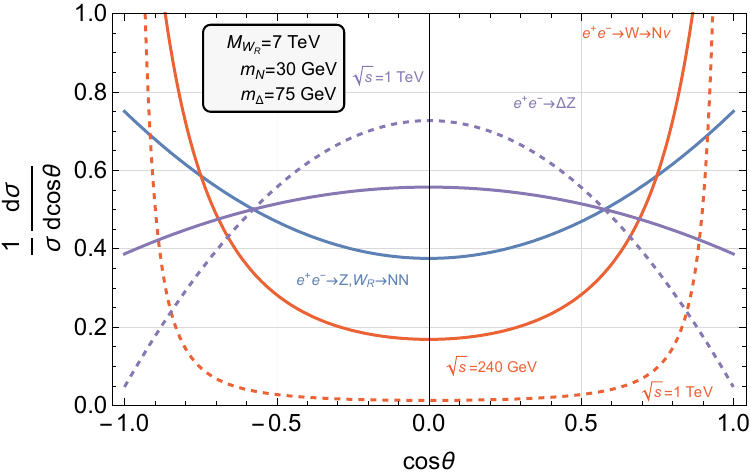}\hfill
  \includegraphics[width=.48\columnwidth]{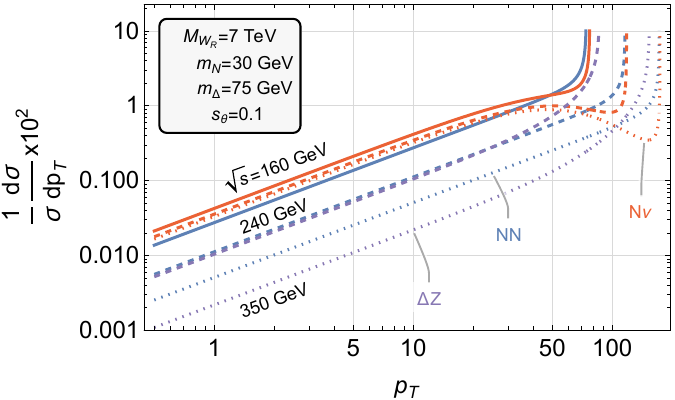}
  \vspace*{-2ex}%
  \caption{{\it Left} -- Polar angle distribution of the heavy neutrino in $NN$
  production via $Z$, $Z_{LR}$ and $W_R$ exchange in blue, as well as in $N\nu$ 
  associated production in red. 
  Results are shown for $\sqrt{s}=240\,\GeV$ (solid) and $1\,\TeV$ (dashed). 
  We additionally display the polar-angle distribution of the $\Delta$ state 
  produced in association with a $Z$ boson ($e^+e^-\!\to\Delta Z$, purple), 
  as discussed in the following subsection. 
  {\it Right} -- Normalised transverse momentum distributions for the 
  same processes. \label{fig:DsigDcosDpT}}
\end{figure}

The angular dependence of the production amplitudes mediated by $Z$, $Z_{LR}$ 
or $W_R$ exchange is independent of $\sqrt{s}$ and $m_N$. 
The normalised differential cross section therefore takes the universal form
\begin{align} \label{eq:dsigdc_Z}
  \frac{1}{\sigma} \frac{\text{d} \sigma}{\text{d} c_\theta} &=
  \frac{3}{8} \left( 1 + c_\theta^2 \right) \,,
\end{align}
where $c_\theta$ denotes the production polar angle with respect to the 
electron beam.
Furthermore, the transverse momentum of the heavy neutrino is bounded by 
$p_T^{\max} = \sqrt{s}\,\beta/2$, and its normalised spectrum for $s$-channel 
vector exchange is well approximated by
\begin{align} \label{eq:dsigdpT_Z}
  \frac{1}{\sigma} \frac{\text{d} \sigma}{\text{d} p_T} &= 
  \frac{1}{p_T^{\max}} \frac{3 x (2 - x^2)}{4\sqrt{1 - x^2}} \, ,
\end{align}
where $x = p_T/p_T^{\max}$. 
This follows directly from the angular distribution in~\eqref{eq:dsigdc_Z}.
In contrast, processes involving $t$-channel and $u$-channel exchange generally 
feature a more forward-peaked (Rutherford-like) angular dependence, 
and their $p_T$ spectra deviate from the compact form above, especially in the 
high-momentum tail.
The corresponding angular and $p_T$ spectra are displayed in
Figure~\ref{fig:DsigDcosDpT} for a benchmark scenario with $M_{W_R}=7\,\TeV$,
$m_N=30\,\GeV$, $m_\Delta=75\,\GeV$ and $s_\theta=0.1$.

In the associated-production channel $e^+ e^- \to N\nu$, the dominant contribution
arises from $t$- and $u$-channel $W$-boson exchanges. 
The corresponding chiral structure differs from that expected from purely 
right-handed $W_R$ exchange, and thus leads to a markedly different angular
behaviour.  
The corresponding normalised distribution is given by
\begin{equation}
  \frac{1}{\sigma} \frac{\text{d} \sigma}{\text{d} c_\theta} = M_W^2 
  \left(M_W^2 + s \beta^2 \right) \left( 
  \frac{1}{\left( 2 M_W^2 + s \beta^2 (1 - c_\theta) \right)^2} + 
  \frac{1}{\left( 2 M_W^2 + s \beta^2 (1 + c_\theta) \right)^2}\right) ,
\end{equation}
where the two terms correspond respectively to the $t$- and $u$-channel 
propagators.  
For large centre-of-mass energies or in the limit $M_W \to 0$, we obtain the
characteristic forward-backward enhancement, which reflects the expected 
Rutherford-like behaviour of $t/u$-channel exchange
\begin{equation}\
  \frac{1}{\sigma} \frac{\text{d} \sigma}{\text{d} c_\theta} 
  \quad \longrightarrow \quad \frac{M_W^2}{\beta^2 s} 
  \frac{3 + c_{2 \theta}}{s_\theta^4} \, , \qquad\text{for} \quad \sqrt s \gg 
  M_W \ \ \text{or}\ \  M_W \to 0\,.
\end{equation}
The propagators produce a strong enhancement at $c_\theta\to \pm 1$, 
\textit{i.e.}\ when the heavy neutrino is emitted nearly co-linearly with 
the beam. 
This sharply contrasts the mild $(1+c_\theta^2)$ dependence of the 
$s$-channel vector case in~\eqref{eq:dsigdc_Z}. 
The resulting distributions are shown in Figure~\ref{fig:DsigDcosDpT} (left)
with the red lines.

The transverse-momentum distribution is likewise modified.  
For $t/u$-channel exchange, the mapping $p_T = (\sqrt{s}/2)\beta\sin\theta$ 
must be folded with the strongly forward-peaked angular distribution, resulting in  
\begin{align}
\begin{split}
  \frac{1}{\sigma} \frac{\text{d} \sigma}{\text{d} p_T} &= \frac{2 x M_W^2 
  \left(M_W^2 + s \beta_{\nu s}^2 \right)}{\sqrt{1-x^2}}  
  \left[ \frac{1}{\left( 2 M_W^2 + s \beta_{\nu s}^2 \left(1 - \sqrt{1-x^2}
  \right) \right)^2} \right. \\
  &+ 
  \left.\frac{1}{\left( 2 M_W^2 + s \beta_{\nu s}^2 \left(1 + \sqrt{1 - x^2} 
  \right) \right)^2} \right] \,.
\end{split}  
\end{align}
The corresponding curves are plotted as red lines in Figure~\ref{fig:DsigDcosDpT} (right).
As expected from the underlying Rutherford-like behaviour, the spectrum deviates
significantly from the compact $s$-channel expression in~\eqref{eq:dsigdpT_Z}, 
especially at large $p_T$.

%
\subsubsection{Higgs and \texorpdfstring{$\Delta$}{Delta} associated production 
with a \texorpdfstring{$Z$}{Z} boson} \label{subsec:xsHiggsDelta}

\begin{figure}
  \centering \vspace{1 em}  
  \input{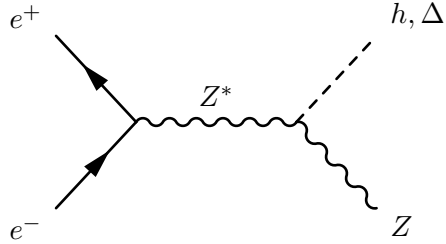}\vspace{.2cm}
  \caption{Illustrative Feynman diagram for associated $Zh$ and $Z\Delta$
  production.  
  \label{fig:DeltaProd}}
\end{figure}

The amplitude $\mathcal{A}_{Z\Delta}$ for the associated production process 
$e^+ e^- \to Z\Delta$, where the triplet $\Delta$ is radiated from an intermediate
$s$-channel $Z$ boson (see Figure~\ref{fig:DeltaProd}), is directly related 
to the SM $e^+ e^- \to Zh$ amplitude $\mathcal{A}_{Zh}$ by the scalar mixing.
Working in the approximation where only the SM-like $ZZ\Delta$ vertex is 
relevant and including the $h$-$\Delta$ mixing angle, we may write
\begin{align} \label{eq:Amp_epem_Zd0}
  \mathcal A_{Z \Delta} = \left(\frac{g_w}{c_w} \right)^2 \frac{M_Z s_\theta}{
  s - M_Z^2}
  \, \overline v_1 \gamma^\mu \left(v_e + a_e \gamma_5 \right) u_2
  \left( \eta_{\mu \nu} - \frac{q_\mu q_\nu}{M_Z^2} \right) 
  \varepsilon^\nu_Z(p_4, \lambda) \, ,
\end{align}
where $\bar v_1$ and $u_2$ are the initial-state spinors, $q$ is the 
four-momentum of the intermediate $Z$ boson and $\varepsilon_Z^\nu$ is the 
final-state $Z$ polarisation vector. 
For clarity, we have omitted the imaginary $i M_Z\Gamma_Z$ term in the $Z$
propagator, which should be reinstated near the $Z$ pole, and the obtained
expression directly stems from the Feynman rule shown in Table~\ref{tab:zff}.
The final-state kinematics involve two unequal masses, thus with $s + t + u = 
m_\Delta^2 + M_Z^2$, $p_3^2 = m_\Delta^2$ and $p_4^2 = M_Z^2$, so that after
averaging over the initial spins and summing over the final polarisations, we 
obtain the following compact expression for the associated squared matrix element 
\begin{align}
  \overline{|\mathcal M_{Z\Delta}|}^2 = \frac{1}{2} \left(\frac{g_w}{c_w} \right)^4 
  \frac{M_Z^2 s_\theta^2}{(s - M_Z^2)^2}
  \left( 2 s + t - m_\Delta^2 + \frac{t}{M_Z^2} 
  \left(m_\Delta^2 - s - t \right) \right) .
\end{align}
The total cross section is obtained by integrating  the matrix element in $t$ over the domain
\begin{equation}
  t_{\min, \max} = - \frac{s - m_\Delta^2 - M_Z^2 \pm \sqrt{\lambda_{Z\Delta}}}{2} \, ,
\end{equation}
with the Källén function
\begin{equation}
  \lambda_{Z\Delta} \equiv \lambda(s,m_\Delta^2,M_Z^2)
  = s^2 + m_\Delta^4 + M_Z^4 - 2 s m_\Delta^2 - 2 s M_Z^2 - 2 m_\Delta^2 M_Z^2\,.
\end{equation}
Performing the integration then yields a compact form for the total cross section, 
\begin{align}
  \sigma_{e^+ e^- \to Z \to \Delta Z} = \frac{\pi \alpha_w^2 s_\theta^2 
  \left(v_e^2 + a_e^2 \right)}{12 c_w^4 s^2} 
  \frac{\sqrt{\lambda_{Z\Delta}} \left(\lambda_{Z\Delta} + 12 s M_Z^2 \right)}{
  \left(s - M_Z^2 \right)^2} \,,
\end{align}
which reduces to the well-known $e^+ e^-\to Zh$ formula upon setting $s_\theta\to 1$ and $m_\Delta\to m_h$.

The behaviour of the cross section for this channel can be understood from this 
last feature: at tree level, the $e^+ e^-\to Z\Delta$ production cross
section is simply the SM $Zh$ one multiplied by the mixing factor $s_\theta^2$
and evaluated at the mass $m_\Delta$. 
Consequently, its size is directly controlled by the scalar mixing. 
In typical LRSM realisations, the mixing decreases as the LR breaking scale is 
raised, so that the $e^+ e^-\to Z\Delta$ rate effectively decouples in the limit 
$M_{W_R} \to \infty$. 
Subsequently, the kinematics reflect the standard features of associated scalar-vector production.
For $s \gg M_Z^2, m_\Delta^2$, the phase-space suppression dominates and leads 
to an overall $\sim \beta_{\Delta s}^6/s$ dependence, which for a light $\Delta$
reduces to the familiar $1/s$ decline visible in Figure~\ref{fig:Xsecs}. 
However, the longitudinal polarisation of the $Z$ boson introduces an additional
contribution proportional to $(\sqrt{s}/M_Z)^2$, as implied by the Goldstone-boson
equivalence theorem. 
This partially compensates the $1/s$ fall-off of the transverse piece and leads 
to a noticeably softer decrease of the full cross section with energy, 
consistent with the behaviour observed in Figure~\ref{fig:Xsecs}.

From a phenomenological perspective, this process is particularly relevant when 
$\Delta \to N N$ has a sizeable branching ratio. 
For mixing values near the current limits (\textit{e.g.}\ $s_\theta = 0.1$), 
the $Z\Delta$ contribution to heavy-neutrino pair production can dominate over 
the $W_R$-mediated channel at lower centre-of-mass energies and actually 
exceeds it for $\sqrt{s} \lesssim 1~\mathrm{TeV}$. 
A further point of interest arises when the final-state $Z$ decays invisibly: 
the process $e^+e^- \to Z\Delta$ with $Z\to\nu\bar\nu$ shared the same final state 
as the weak-boson-fusion topology $e^+ e^- \to \Delta \, \nu \bar\nu$ that 
we address in Section~\ref{subsec:VBF_SBF}.
Around $\sqrt{s}\sim 300~\mathrm{GeV}$ the magnitudes of these two contributions
are comparable, and interference effects between them can become significant. 
A dedicated analysis including the full kinematic dependence is therefore 
required in this region to obtain reliable predictions.

The boost of the triplet scalar $\Delta$ in the associated-production 
process follows directly from its energy,
\begin{align}
  E_\Delta &= \frac{m_\Delta^2 - M_Z^2 + s}{2 \sqrt s} \, ,
  &
  \gamma_\Delta &= \frac{E_\Delta}{m_\Delta} = \frac{m_\Delta^2 - M_Z^2 + s}{
  2 \sqrt s m_\Delta} 
  \xrightarrow{s \gg m_\Delta, M_Z} \frac{\sqrt s}{2 m_\Delta}\,.
\end{align}
This matches the intuitive expectation that, far above threshold, the 
centre-of-mass energy is shared approximately equally between the two comparatively
light final states, giving $E_\Delta \simeq \sqrt{s}/2$ up to small mass
corrections.
The invariant mass of the heavy-neutrino pair produced through the 
$\Delta\to NN$ decay is thus fixed by the resonance condition, 
$m_{NN}^{\rm inv} = m_\Delta$. 

The momentum of the produced $\Delta$ state and the relation between the 
Mandelstam variable $t$ and the production angle $\theta$ follow as
\begin{align} \label{eq:ptDeltaZ}
  p_\Delta &= \frac{1}{2} \sqrt{\frac{\lambda_{Z\Delta}}{s}} \, ,
  &
  t &= -\frac{1}{2} \left( s - m_\Delta^2 - M_Z^2 - \sqrt{\lambda_{Z\Delta}} 
  c_\theta \right)  ,
  &
  \frac{\text{d} t}{\text{d} c_\theta} &= \frac{\sqrt \lambda_{Z\Delta}}{2} \, .
\end{align}
Using these relations, the normalised differential cross section becomes
\begin{align}
\begin{split}
  \frac{1}{\sigma} \frac{\text{d} \sigma}{\text{d} c_\theta} =&{} \frac{3}{4} 
  \frac{ \lambda_{Z\Delta}(1 - c^2_\theta) + M_Z^4 + 6 M_Z^2 s - (M_Z^2 - s)^2 + s^2}{
  \lambda_{Z\Delta} + 12 M_Z^2 s}
  \\
  &\xrightarrow{s \gg m_\Delta, M_Z} \frac{3}{4} \left(1 - c_\theta^2 \right) + 
  \mathcal O\left(\frac{M_Z^2}{s \beta^4} \right),
\end{split}
\end{align}
which exhibits the characteristic $(1 - c_\theta^2)$ dependence expected for 
the production of a scalar together with a vector boson from an $s$-channel current.
This pattern reflects the dominance of the transverse $Z$ component at large $s$,
with deviations at moderate energies arising from the longitudinal mode of the $Z$
boson. 
The resulting $\cos \theta$ distribution is shown in the left panel of 
Figure~\ref{fig:DsigDcosDpT} in purple.

The maximal transverse momentum carried by $\Delta$ is determined
from \eqref{eq:ptDeltaZ}, $p_T^{\max} =\sqrt {\lambda_{Z\Delta}/s}/2$. 
Using $x = p_T/p_T^{\rm max}$ as in the previous section and using the 
chain rule ${\rm d}c_\theta/{\rm d}p_T = -(2x)/(p_T^{\rm max}\sqrt{1-x^2})$, 
we then obtain
\begin{align}
  \frac{1}{\sigma} \frac{\text{d} \sigma}{\text{d} p_T} &= \frac{3}{4} 
  \frac{ \lambda_{Z\Delta} x^2 + M_Z^4 + 6 M_Z^2 s - (M_Z^2 - s)^2 + s^2}
  {\lambda_{Z\Delta} + 12 M_Z^2 s} \frac{2 x}{p_T^{\max} \sqrt{1 - x^2}} \, .
\end{align}
This distribution, displayed in purple in the right panel of 
Figure~\ref{fig:DsigDcosDpT} for several centre-of-mass energies, shows the 
characteristic enhancement near $x\rightarrow1$ (\textit{i.e.}\ $p_T \rightarrow
p_T^{\rm max}$) that is also seen in the $NN$ and $N\nu$ channels. 
This behaviour originates from the Jacobian peak associated with the two-body
kinematics and is only mildly distorted at lower energies by the 
longitudinal-$Z$ contribution.

%
\subsubsection{Vector and scalar boson fusion} \label{subsec:VBF_SBF}

\begin{figure}
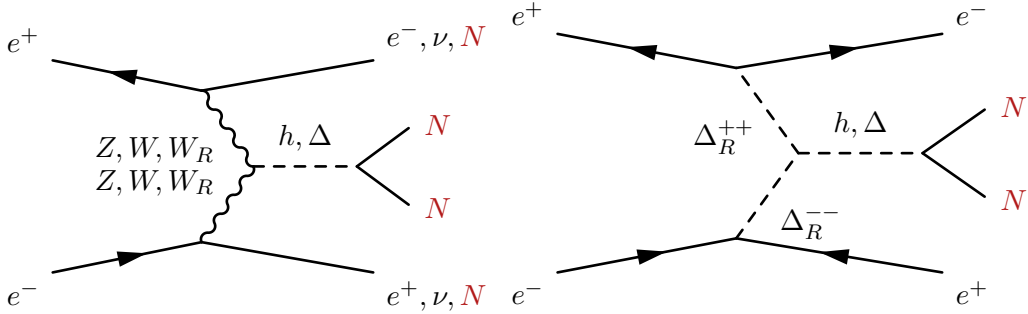

  \centering
  \vspace{1 em}
  \input{figures/fig_epem_VBF_llNN.tex}
  \hspace{.5cm}
  \input{figures/fig_epem_Yuk_llNN.tex}
  \vspace{.5 em}
  \caption{Representative Feynman diagram for heavy neutrino production via vector-boson fusion (left) and doubly-charged scalar-boson fusion (right) in the LRSM.}
  \label{fig:VBFandSBF}
\end{figure}

Fusion topologies, illustrated in Figure~\ref{fig:VBFandSBF}, become increasingly
important at high centre-of-mass energies and can dominate the heavy neutrino 
pair-production signal for $\sqrt{s} \gtrsim\mathcal O(100~\mathrm{GeV})$. 
In the LRSM, two classes of contributions are relevant. 
The first consists of the vector-boson fusion (VBF) channel, in which a neutral
scalar is produced through the fusion of two SM $W$ or $Z$ bosons or their 
$SU(2)_R$ counterparts. 
While $W_R$ fusion can be sizeable, the much larger $Z_{LR}$ mass strongly 
suppresses the associated channel which we therefore neglect in the following. 
The second class of processes consists of the scalar-boson fusion (SBF) 
channel mediated by doubly-charged scalars $\Delta^{\pm\pm}$.

Both the VBF and SBF topologies yield a neutral final-state scalar (either an 
SM-like $h$ or a triplet $\Delta$ state) in the central region, with the scattered
electrons, positrons or neutrinos emitted predominantly at small angles. 
However, the SBF contribution becomes sizeable only when the doubly-charged 
scalars are relatively light and possess enhanced Yukawa couplings to electrons.
Finally, the relative importance of the $h$-mediated and $\Delta$-mediated 
channels is comparable: the $hNN$ interaction is suppressed by the tiny neutrino
Yukawa coupling (see Table~\ref{tab:SNN}), whereas the $\Delta$ channel is 
limited by the trilinear scalar coupling shown in Table~\ref{tab:SSS}.

The calculation of the double-differential VBF cross section for the production 
of a scalar state in $e^+e^-$ collisions follows the classic treatment 
of~\cite{Altarelli:1987ue},\footnote{There is a typo in~\cite{Altarelli:1987ue}; 
the $C_1$ and $C_2$ coefficients in their Eq.(3) should be interchanged.} together 
with the interference with $hZ$ production discussed in~\cite{Kilian:1995tr}.
Starting from the general amplitude with vector and axial fermion currents, 
the spin-averaged squared amplitude can be written in the compact form
\vspace*{-.3ex}
\begin{align} \label{eq:avgM2VBF}
  \overline{\left| \mathcal M \right|}^2 &= 8 g_{VVS}^2 \, \frac{
  C_1 \left( p_A \!\cdot\! p_2 \right)\left( p_B\!\cdot\! p_1 \right) + 
  C_2 \left( p_A \!\cdot\!p_B \right) \left( p_1\!\cdot\! p_2 \right)}
  {\left(q_1^2 - M^2\right)^2\left(q_2^2 - M^2\right)^2} \, ,
\end{align}
where $p_A$ and $p_B$ denote the incoming electron/positron momenta, $p_1$ and $p_2$ 
the outgoing scattered fermion momenta, and $q_1 = p_A - p_1$ and $q_2 = p_B - p_2$ 
the corresponding $t$-channel momentum transfers. 
Moreover, the parameter $M$ is the mediator mass ($M_W$ or $M_Z$ for SM VBF 
contributions, $M_{W_R}$ for the LRSM case), and the $g_{VVS}$ quantity encodes the 
relevant gauge-scalar coupling (for example $g_{ZZh}=g_w M_Z/c_w$, 
$g_{WWh} = g_w M_W$ or any analogous expression derived from Table~\ref{tab:SVV}). 
The coefficients $C_{1,2}$ collect the combinations of vector and axial couplings 
of the exchanged boson to electrons, and are given by
\begin{equation}\label{eq:C12_def}
  C_{1,2} = \left( g_v^2 + g_a^2 \right)^2 \pm 4 g_v^2 g_a^2 \, ,
\end{equation}
with $g_v$ and $g_a$ being the vector and axial couplings appropriate for the chosen
mediator. 
These follow directly from the Feynman rules listed in Tables~\ref{tab:zff} 
and~\ref{tab:WLL}. 
In particular, for the SM $W$ and $Z$ bosons we find
\begin{align}
  &W\ \text{boson}:\  & g_v &= - g_a = \frac{g_w}{\sqrt{2}} \, ,
  \\
  &Z\ \text{boson}:\  & g_v &= \frac{g_w}{c_w} \left(\frac{T_{3L}}{2} - Q s_w^2 \right) ,
  & g_a &= -\frac{g_w}{c_w} \frac{T_{3L}}{2} \, ,
\end{align}
where $T_{3L}$ is the weak isospin quantum number of the relevant fermion.

Following the phase-space integration procedure outlined in Appendix~\ref{app:PhaseSpace},
the triple-differential distribution in the scalar four-momentum $P^\mu$ takes the form
\begin{align} \label{eq:Edsigd3P}
\begin{split}
  E \frac{\text{d} \sigma}{\text{d}^3 P} &= \frac{g_{VVS}^2}{4 s\, s_A\, s_B\, r 
  \left( 2 \pi \right)^4}
  \biggl \{ \bigg[ 2 C_2 \left(1 - c_\chi \right) + C_1 \left(1 + h_A \right) 
  \left(1 + h_B \right) \bigg] 
  \\
  &\times \left[ \frac{2}{h_A^2 - 1} + \frac{2}{h_B^2 - 1} + \frac{1}{\sqrt r} 
  \left( \mathcal L \left( \frac{3 t_A t_B}{r} - c_\chi \right) - \frac{6 s_\chi^2}{\sqrt r}
  \right) \right]
  \\
  &- C_1 \left[ \frac{2 t_A}{h_B - 1} + \frac{2 t_B}{h_A - 1} +  \frac{\mathcal L}{\sqrt r} 
  \left( t_A + t_B + s_\chi^2 \right) \right]  \biggr \} \, ,
\end{split}   
\end{align}
where the shorthand variables $s_{A,B}$, $h_{A,B}$, $t_{A,B}$, $c_\chi$, $s_\chi$, $r$,
$\mathcal L$ are all defined in Appendix~\ref{app:PhaseSpace}. 
This specific choice of notation is the one of~\cite{Altarelli:1987ue} after correcting 
the $C_1\leftrightarrow C_2$ typo in that reference. 
Once the differential cross section $E\,\mathrm d\sigma/\mathrm d^3P$ is known, we 
may obtain distributions in any of the kinematic property of the produced scalar 
field (energy $E$, boost factor $\gamma$, polar angle $c_\theta$, rapidity $y$ and 
transverse momentum $p_T$) by using usual Jacobian transformations and integrations.
We obtain the corresponding total cross sections by integrating $E\,\mathrm d\sigma/
\mathrm d^3P$ over the scalar energy and polar angle,
\begin{align}
  \sigma &= 2\pi \int_{-1}^1 \text{d} c_\alpha \int_m^{E_{\max}} \text{d} E \, P 
  \left( E \frac{\text{d} \sigma}{\text{d}^3 P} \right) , \quad\text{with}\quad  &
  E_{\max} &= \frac{\sqrt s}{2} \left( 1 + \frac{m^2}{s} \right) ,
\end{align}
where $m$ generically stands for the mass of the final-state scalar. 
Equivalently, this expression can be rewritten in terms of the transverse-momentum 
and rapidity variables,
\begin{align}
  \sigma &= 2 \pi \int_{y_-}^{y_+} \text{d} y \int_0^{p_T^{\max}} \text{d} p_T \, p_T 
  \left( E \frac{\text{d} \sigma}{\text{d}^3 P} \right) ,
\end{align}  
with
\begin{align}
  y_{\pm} &= \pm \log \frac{\sqrt s}{m} \, , 
  &
  p_T^{\max 2} &= \left( \frac{s + m^2}{2 \sqrt{s} \cosh y} \right)^2 - m^2  \, .
\end{align}
The formul\ae\ above apply directly to $W$ and $Z$ fusion. 
They can be extended to $W_R$ fusion by replacing the involved couplings 
and masses, as well as the final-state SM neutrinos with the heavy ones. 
SBF contributions are determined analogously by inserting scalar propagators 
and the Yukawa couplings of the $\Delta^{\pm\pm}$ state. Finally, we recall that 
the processes $e^+ e^- \to \Delta \nu \nu$ and $e^+ e^- \to \Delta e^+ e^-$ are 
related to $e^+ e^- \to h \nu \nu$ and $e^+ e^- \to h e^+ e^-$ simply by rescaling 
the $g_{VVh}$ coupling by the appropriate power of the $h$-$\Delta$ mixing angle
$\sin \theta^2$.

We evaluate the double integrals numerically with the limits mentioned above to 
obtain the results shown in Figure~\ref{fig:Xsecs} (yellow and purple lines). 
To achieve agreement with the simulations, one must remove all the
parton level cuts in~\textsc{MG5\_aMC@NLO} on the forward electrons that 
are of no interest to us.
As can be seen in the figure, the VBF cross sections increase with the 
centre-of-mass energy with a dependence stemming from the large flux of 
quasi on-shell, collinear vector bosons radiated off the initial leptons. 
In the high-energy limit, the effective-vector-boson approximation provides useful 
intuition: the initial-state electron and positron radiate 
quasi-real $W/Z$ bosons whose subsequent fusion produces the final-state scalar. 
At the amplitude level, the longitudinal components of the vector bosons whose 
couplings are related to the Goldstone modes are enhanced by factors of order
$\sqrt{s}/M_V$ due to the equivalence theorem.
This enhancement compensates for the propagator suppression and explains the growth 
captured in the expression~\eqref{eq:Edsigd3P}. 
For the SM $W$-boson and $Z$-boson modes, the mediators are light, and the VBF 
rates can become sizeable already at moderate $\sqrt{s}$ values. 
For the fusion of $W_R$ bosons, the heavy mediator mass suppresses the amplitude 
by powers of $1/M_{W_R}^2$ in the propagators, and thus the VBF contribution is
parametrically smaller unless $\sqrt{s}$ is large enough to overcome the mass 
suppression.

\begin{figure}[t]
  \centering
  \includegraphics[width=.49\columnwidth]{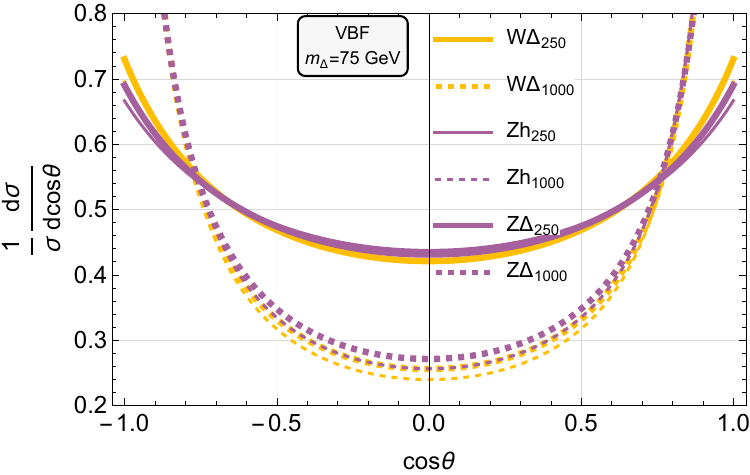}
  \includegraphics[width=.49\columnwidth]{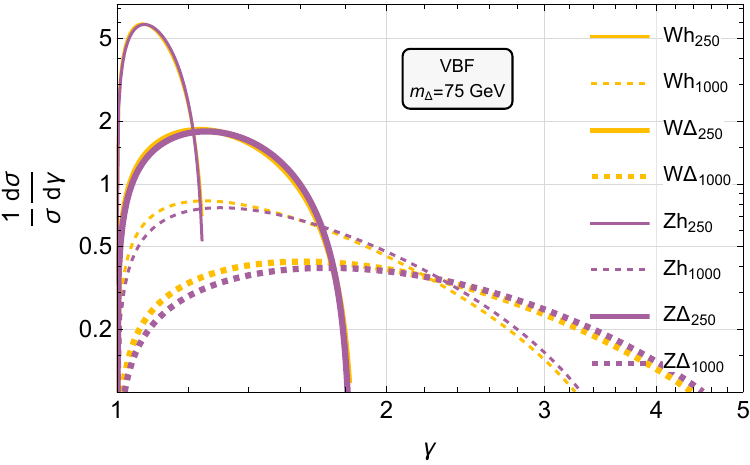}
  \vspace*{-2ex}%
  
  \caption{%
  \textit{Left} -- Normalised polar-angle distributions of the scalar state produced via vector-boson fusion, shown for $W$-boson fusion (thick lines) and $Z$-boson fusion (thin lines). Results are displayed for the SM Higgs boson $h$ and for a triplet scalar $\Delta$ with $m_\Delta = 75\,\GeV$. 
  \textit{Right} -- Normalised distributions of the scalar boost factor $\gamma_h$ and $\gamma_\Delta$ for the vector-boson fusion mode.\label{fig:dsigdcdga_WBF}}
\end{figure}

The polar-angle distributions of the scalar produced via vector-boson fusion are obtained from a numerical integration of the associated matrix element over the scattered fermion phase space, and  are shown in the left panel of Figure~\ref{fig:dsigdcdga_WBF} both for the SM Higgs state $h$ and the LRSM $\Delta$ scalar with a mass that we fix to $75\,\GeV$ (the normalised distributions being independent of the scalar mixing angle). As expected for fusion processes, the distributions are forward-backward symmetric and reflect the dominance of the $t$-channel gauge-boson exchange contributions.

In contrast to the two-body production channels discussed earlier, the scalar energy is not fixed in this case, and the resulting boost-factor distribution follows directly from the scalar energy spectrum via $\mathrm{d}\sigma/\mathrm{d}\gamma = m\,\mathrm{d}\sigma/\mathrm{d}E$. The corresponding distributions are displayed in the right panel of figure~\ref{fig:dsigdcdga_WBF}. At $\sqrt{s}=250\,\GeV$, the spectrum is strongly peaked at $\gamma\simeq 1$, indicating that the scalar is produced nearly at rest in the laboratory frame. At higher centre-of-mass energies, illustrated here for $\sqrt{s}=1\,\TeV$, the available phase space increases and the boost distribution broadens, extending up to $\gamma$ of about 3. This effect is particularly pronounced for the lighter triplet scalar, which can acquire a larger boost for fixed $\sqrt{s}$ due to its smaller mass compared to the SM Higgs boson.

%
%
\section{Signal sensitivities} \label{sec:SigSens}

The displaced decays of the heavy neutrinos provide a particularly powerful probe of the LRSM, as the decay length depends simultaneously on the production kinematics and on the small parameters that control the $N$ interactions. As a result, the different production mechanisms discussed in Section~\ref{sec:DRatXSec} give access to complementary regions of the LRSM parameter space, and displaced signatures can remain observable even when prompt searches lose sensitivity.

The relevant production channels can be grouped into three broad classes which will be analysed separately in the following subsections. The first class consists of gauge-mediated processes, including the $Z\to NN$ decays at the $Z$ pole and heavy neutrino production via $t$-channel exchanges. These channels both rely on the exchange of the LR gauge bosons and are additionally sensitive to gauge-boson mixing effects (see Section~\ref{sec:sensigauge}). The second class comprises the scalar-mediated modes where the heavy neutrinos are produced through the decay of neutral scalars, ($h,\Delta\to NN$), with rates thus controlled by the mixing between the SM Higgs and the triplet scalar $\Delta$ (see Section~\ref{sec:sensiscalar}). Finally, the third class corresponds to a no-mixing scenario and includes the SBF production of the $\Delta$ state mediated by $\Delta_R^{++}$ exchange, which remains active even in the limit where both gauge and scalar mixings are negligible (see Section~\ref{sec:sensiSBF}).

To quantify the sensitivity of future lepton colliders to these displaced signals, we begin with a `purely theoretical' estimate of the expected number of signal events. This quantity is defined as the product of the production cross section, decay branching ratios, the geometric acceptance $\mathcal A_\text{geo}$ to enforce the decays to occur inside the tracking volume and the integrated luminosity $\mathcal L$ (see Table~\ref{tab:FCCee} for the benchmark FCC-ee configurations),
\begin{equation}
  N_\text{signal} = \sigma\times \mathrm{BR}\times \mathcal A_\text{geo} \times 
  \mathcal L \, .
\end{equation}
While the cross section and branching ratios can be straightforwardly computed 
following the discussion in Section~\ref{sec:DRatXSec}, the estimate of the geometric acceptance requires more care.
We need to compute the decay distribution of the heavy $N$ in the laboratory frame on an event-by-event basis by convoluting the phase space of the $N$-production cross section with the decay probability within a certain detector volume,
\begin{equation}
  P_N(d_T) = \frac{1}{\langle d_T\rangle}\exp\left(-\frac{d_T}{
  \langle d_T\rangle}\right) .
\end{equation}
We focus here on the transverse displacement $d_T$ which we use as a proxy for the decay position, and whose average value is given by
\begin{equation}
    \langle d_T \rangle = \frac{p_T^\text{lab}(N)}{m_N} \tau_N \, ,
\end{equation}
where $\tau_N$ is the proper lifetime of the heavy neutrino $N$ and $p_T^\text{lab}(N)$ is its transverse momentum in the laboratory frame.

For the direct production modes (\textit{e.g.}\ $e^+e^-\to NN$), the transverse momentum in the centre-of-mass frame coincides with that in the laboratory frame, and the decay distribution can therefore be evaluated straightforwardly.
For scalar-mediated channels such as $e^+e^- \to Z\Delta$ followed by the decay $\Delta\to NN$, the heavy-neutrino momenta must be boosted from the $\Delta$ rest frame, where $p_N^\Delta=(E_N^\Delta,\vec p_N^\Delta)$, into the laboratory frame along the $\Delta$ momentum direction. The resulting spatial momentum is 
\begin{equation}
  \vec p_N^\text{lab} = \vec p_N^\Delta + \gamma_\Delta \vec\beta_\Delta\left(\frac{\gamma_\Delta
  }{1 + \gamma_\Delta}\vec \beta_\Delta\!\cdot\!\vec p_N^\Delta - E_N^\Delta\right) ,
\end{equation}
with $\vec\beta_\Delta = {\vec p}_\Delta^\text{lab}/E_\Delta^\text{lab}$ and 
$\gamma_\Delta = (1 - |\vec\beta_\Delta|^2)^{-1/2}$.
The expected number of signal events in which one heavy neutrino decays within a transverse displacement interval $[d_T^{\text{min}},\,d_T^{\text{max}}]$ can then be estimated schematically as
\begin{equation}
  N_\text{signal} = \int_{d_T^\text{min}}^{d_T^\text{max}} \mathrm{d}d_T 
  \frac{\mathrm{d}N_\text{signal}}{\mathrm{d}d_T} = \int \mathrm{d} \Phi 
  \frac{\mathrm{d}(\sigma\times \mathrm{BR})}{\mathrm{d}\Phi}\left(
  e^{-\frac{d_T^\text{min}}{\langle d_T \rangle}} -
  e^{-\frac{d_T^\text{max}}{\langle d_T \rangle}}\right) .
\end{equation}
This can be straightforwardly extended to the second heavy neutrino by multiplying with the corresponding integrated decay probability distribution and sampling the phase space accordingly.
\begin{figure}
  \centering
  \includegraphics[width=0.48\linewidth]{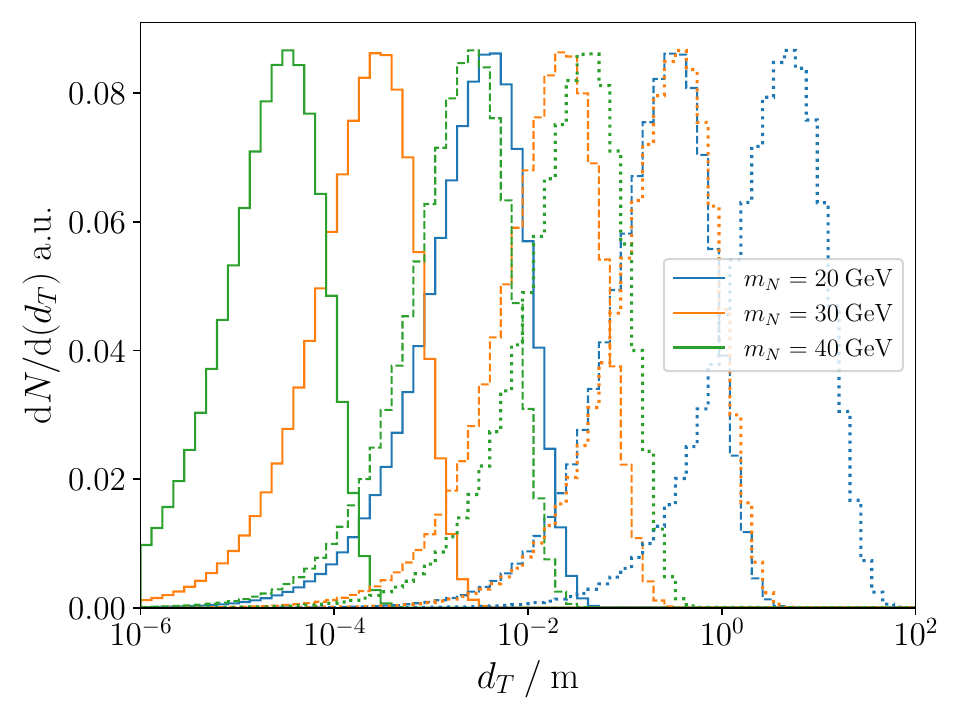}
  \includegraphics[width=0.48\linewidth]{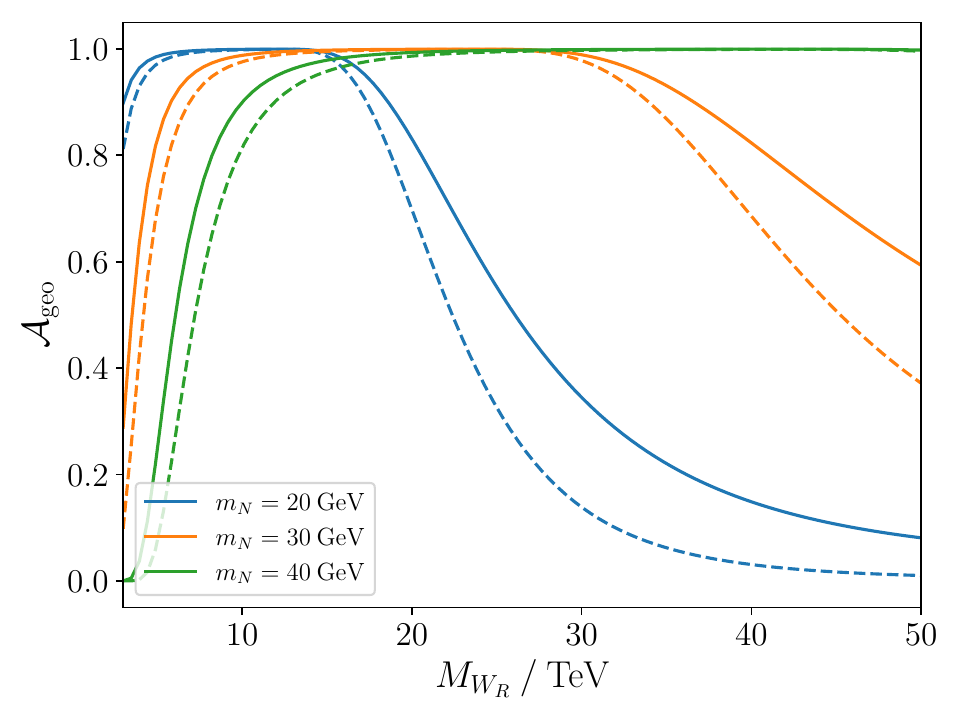}
  \vspace*{-2ex}%
  \caption{%
    \textit{Left} -- Differential event distribution for the process $Z\to NN$ at the $Z$ pole as a function of the transverse displacement $d_T(N)$ of a single heavy neutrino in the laboratory frame. Solid, dashed and dotted curves correspond to $M_{W_R}=5$, $15$ and $30~\TeV$, respectively.
    \textit{Right} -- Geometric acceptance $\mathcal A_\text{geo}$ for heavy-neutrino decays occurring within a transverse displacement window $d_T\in[50~\mu\mathrm{m},\,2~\mathrm{m}]$ that is considered as representative of the FCC-ee tracking volume.}
  \label{fig:disp}
\end{figure}

In Figure~\ref{fig:disp} we show the transverse-displacement distribution of a single heavy neutrino produced in $Z\to NN$ decays at the $Z$ pole for several values of $m_N$ and $M_{W_R}$ (left panel), together with the corresponding geometric acceptance $\mathcal A_\text{geo}$ for decays occurring within $d_T\in[50~\mu\mathrm{m},\,2~\mathrm{m}]$ (right panel).
For relatively light $N$ masses, the decay length can be sizeable even for $M_{W_R}\simeq5$--$15~\TeV$, in which case a significant fraction of decays occurs beyond the typical FCC-ee tracking volume ($\sim2~\mathrm{m}$).
Conversely, for heavier $m_N$, the decay displacement for $M_{W_R}\simeq5$--$15~\TeV$ tends to be too small to yield an appreciable number of displaced vertices.
Owing to the exponential nature of the decay-length distribution, however, long tails are always present, such that even proper lifetimes of $\mathcal O(10~\mathrm{m})$ can still result in a sizeable number of events within the geometric acceptance.

\subsection{Gauge modes: 
\texorpdfstring{$Z\to NN$}{Z -> NN}, 
\texorpdfstring{$e^+e^-\to NN,N\nu$}{e+ e- -> NN, N nu} }\label{sec:sensigauge}
\begin{figure}
  \centering
  \includegraphics[width=0.48\linewidth]{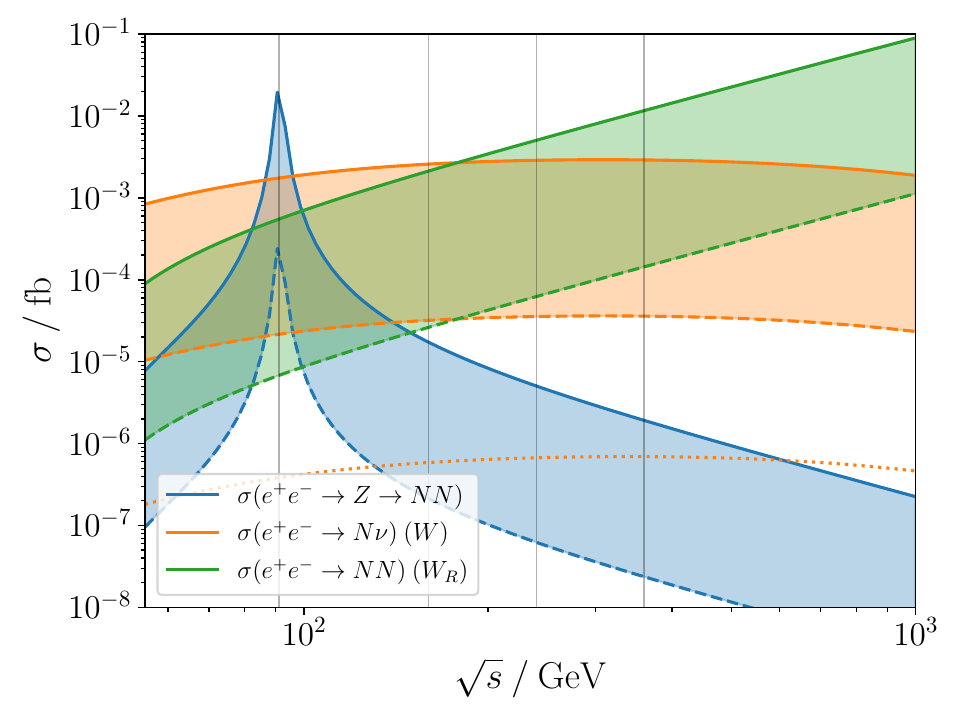}
  \includegraphics[width=0.48\linewidth]{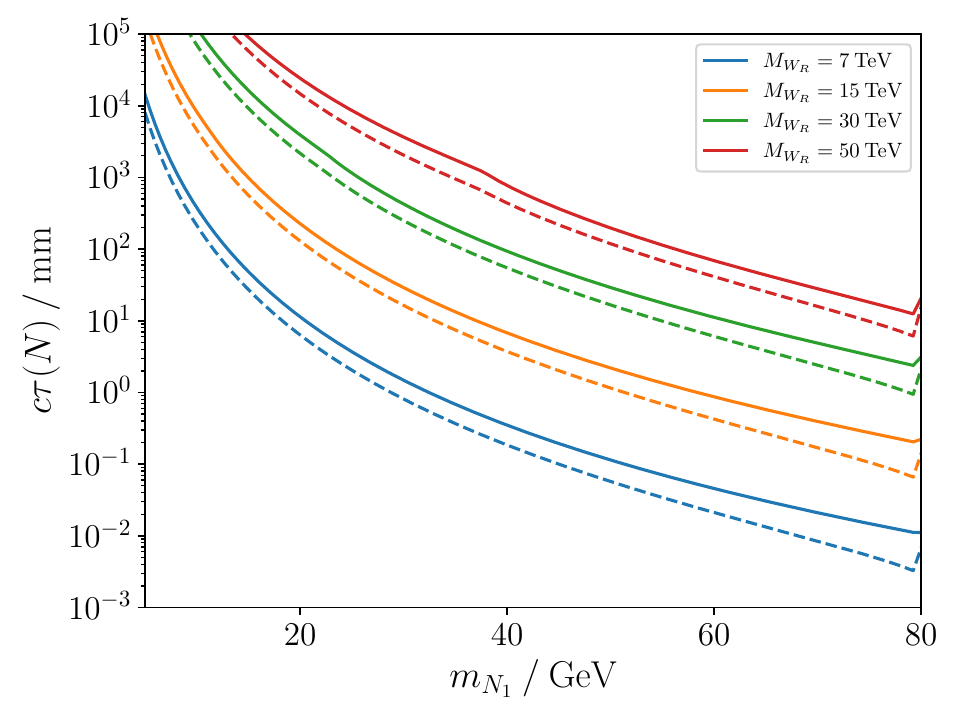}
  \vspace*{-2ex}%
  \caption{%
    \textit{Left} -- Production cross sections for the gauge-mediated channels $Z\to NN$ at the $Z$-pole (blue), $e^+e^-\to NN$ (orange) and $e^+e^-\to N\nu$ (green) as a function of the centre-of-mass energy $\sqrt{s}$. The solid (dashed) lines correspond to $M_{W_R}=5~\TeV$ ($15~\TeV$) with $\tan\beta=0.5$, whereas the dotted orange curve illustrates the behaviour of the $e^+e^-\to N\nu$ cross section for $M_{W_R}=15~\TeV$ and a smaller $\tan\beta=0.05$. In all cases, the heavy-neutrino mass is fixed to $m_N=20~\GeV$, and the vertical grey lines indicate the centre-of-mass energies of the proposed FCC-ee runs. 
    \textit{Right} -- Proper lifetime of the heavy neutrino $N$ as a function of $m_N$ for different values of $M_{W_R}$. The solid (dashed) curves correspond to $\tan\beta=0$ ($0.5$).}
  \label{fig:gauge_xsec}
\end{figure}
We first consider the gauge-mediated production channels, namely heavy neutrino production from $Z\to NN$ decays at the $Z$ pole and the $e^+e^-\to NN,\,N\nu$ processes via $t$-channel $W$ and $W_R$ exchange.
We present the dependence of the corresponding production rate in the left panel of Figure~\ref{fig:gauge_xsec} for a few benchmark choices. In these modes, the relevant parameters are  $M_{W_R}$,  $m_N$ and the $W$-$W_R$ mixing angle  parametrised in terms of $\tan\beta$. Moreover, the $Z\to NN$ decay is controlled by the $Z$-$Z_{LR}$ mixing, which scales as $(M_W/M_{W_R})^2$,  from the dependence of the corresponding partial width on the coupling $a_N$ (see Eq.~\eqref{eq:GamZNN} and  Section~\ref{sec:LRSM}). As a result, this channel is only relevant at the $Z$ pole where $Z$-boson production is enhanced. Additional off-shell contributions from $Z_{LR}$  are  strongly suppressed by the large $Z_{LR}$ mass and remain  negligible at FCC-ee energies.

In contrast, the $W$-$W_R$ mixing plays a central role in the $t$-channel process $e^+e^-\to N\nu$, which is predominantly mediated by the light mostly SM-like $W$ state. This channel is therefore highly sensitive to $\tan\beta$, as highlighted through the orange curves in the left panel of Figure~\ref{fig:gauge_xsec}, and thus only relevant for sizeable values of $\tan \beta \gtrsim 0.2$. On the contrary, the $e^+e^-\to NN$ pair-production mode depends only weakly on this parameter. The heavy-neutrino lifetime is likewise sensitive to $\tan\beta$: for small values of $\tan\beta$, heavy neutrino decays are dominated (up to small Dirac mixing effects) by $W_R$-mediated processes, leading to an approximately unity semi-leptonic branching ratio, $\mathrm{BR}(N\to\ell^\pm q\bar q')\simeq1$. For larger values of $\tan\beta$, $W$-mediated contributions become instead non-negligible, inducing interference effects in the semi-leptonic channel and opening purely leptonic decays such as $N\to\ell^+\ell^-\nu$. This thereby reduces both the potential displaced signal yield and the decay length, the latter being illustrated in the right panel of Figure~\ref{fig:gauge_xsec}.

\begin{figure}
  \centering
  \includegraphics[width=0.65\linewidth]{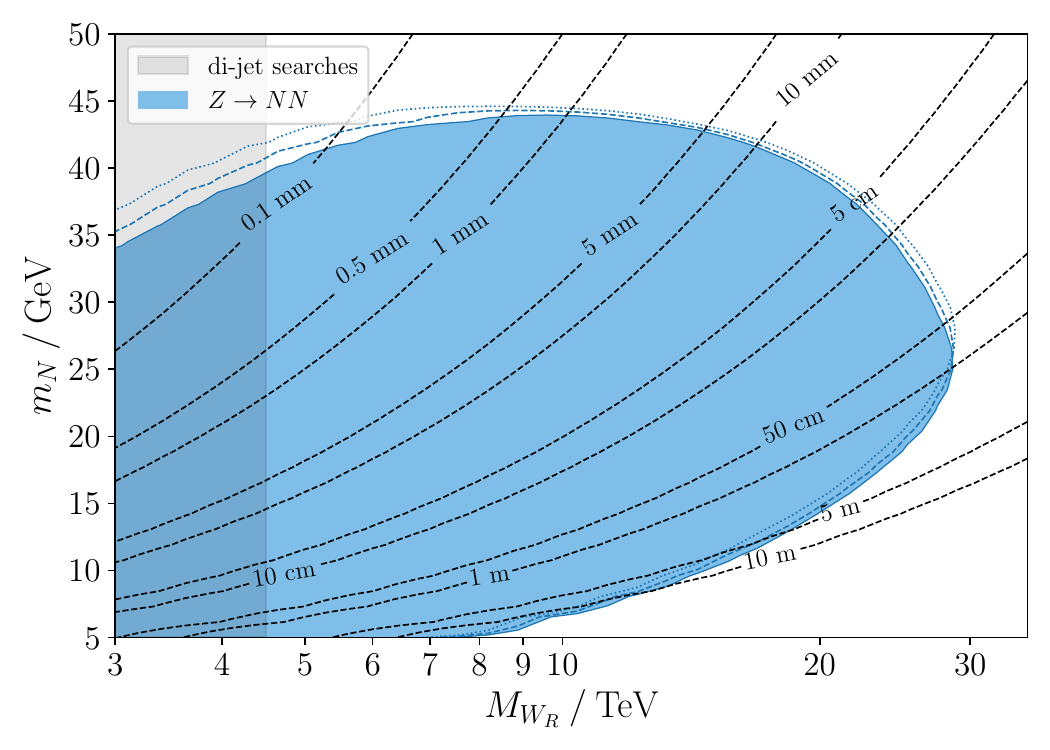}
  \vspace*{-2ex}%
  \caption{%
    Theoretical sensitivity reach of the FCC-ee to heavy neutrinos produced via $Z\to NN$ decays at the $Z$ pole, expressed in the $(M_{W_R}, m_N)$ plane. The shaded region corresponds to $\tan\beta=0.5$, while the dashed and dotted contours refer to $\tan\beta=0.3$ and $\tan\beta=0$ respectively. The sensitivity is defined by requiring at least three signal events with both produced heavy neutrinos decaying within the fiducial transverse displacement range, assuming an integrated luminosity of $205~\mathrm{pb}^{-1}$, and the labelled contours indicate the proper lifetime of the heavy neutrino $N$.
  }
  \label{fig:sens_ZNN}
\end{figure}

To assess the theoretical sensitivity reach of FCC-ee to heavy neutrinos produced in gauge-mediated channels, we perform a scan over the $(m_N,\,M_{W_R})$ parameter space, considering several representative values of the gauge-boson mixing parameter $\tan\beta$.  We begin by the $Z\to NN$ channel at the $Z$ pole which benefits from the large statistics achievable at $\sqrt{s}=m_Z$. In Figure~\ref{fig:sens_ZNN}, we present the maximal parameter-space reach obtained by imposing only a requirement on the transverse decay displacement, namely that both produced heavy neutrinos decay within the fiducial tracking volume $d_T\in[50\,\mu\mathrm{m},\,2\,\mathrm{m}]$ with no additional selection efficiencies or reconstruction effects being implemented. The resulting contours should therefore be interpreted as an optimistic, theory-level upper bound on the sensitivity, and they are defined by requiring at least $N_\text{signal}\ge 3$ expected signal events assuming an integrated luminosity of $205~\mathrm{pb}^{-1}$ at the $Z$ pole. The reach is controlled by the interplay between the $Z\to NN$ rate, which is governed by the $Z$-$Z_{LR}$ mixing and hence suppressed as $(M_W/M_{W_R})^2$, and the heavy-neutrino lifetime, which determines whether the decays occur within the tracking volume. As a result, the sensitivity exhibits only a mild dependence on $\tan\beta$, reflecting the fact that the $Z\to NN$ production mechanism is largely insensitive to any $W_L$-$W_R$ mixing effects.

\begin{figure}
  \centering
  \includegraphics[width=0.48\linewidth]{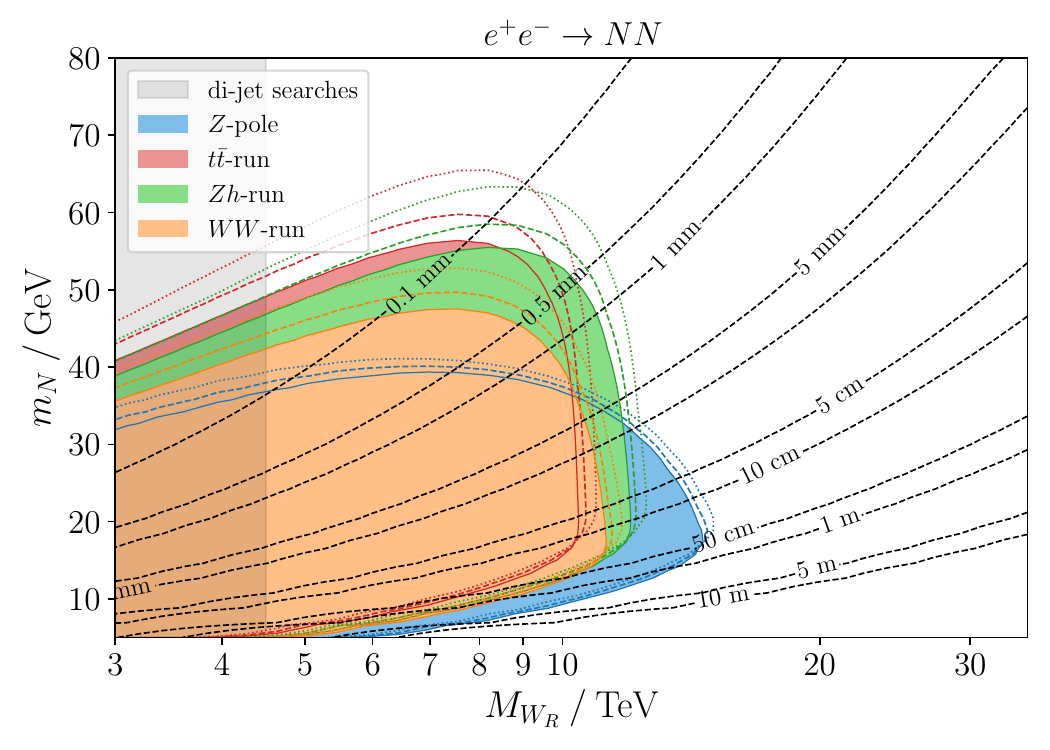}
  \includegraphics[width=0.48\linewidth]{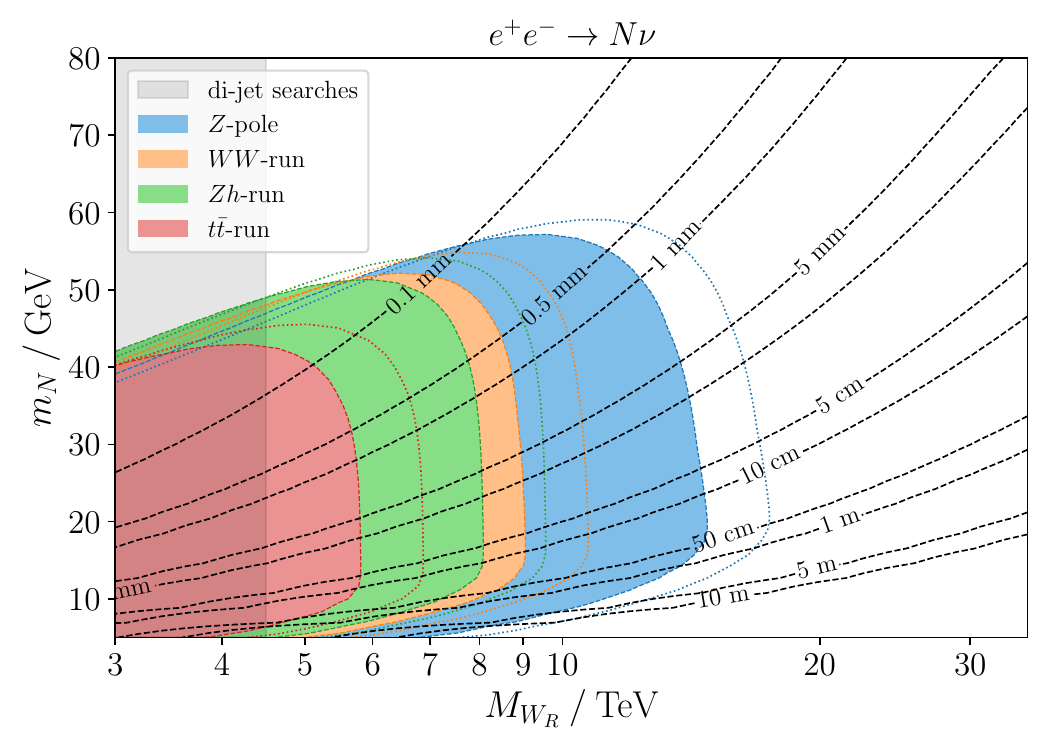}
  \vspace*{-2ex}%
  \caption{%
    Expected sensitivity of the FCC-ee to the $t$-channel production of heavy neutrinos in the processes $e^+e^- \to NN$ (left) and $e^+e^- \to N\nu$ (right), shown in the $(M_{W_R}, m_N)$ plane. The colour code (see the legend) refers to the different FCC-ee operation points. \textit{Left} -- The shaded regions correspond to $\tan\beta = 0.5$, while the dashed and dotted lines indicate $\tan\beta = 0.3$ and $0$ respectively. \textit{Right} -- The shaded regions and dashed lines correspond to $\tan\beta = 0.3$, while the dotted lines indicate $\tan\beta = 0.5$.
  } \label{fig:sens_NNtchannel_Nnu}
\end{figure}
In Figure~\ref{fig:sens_NNtchannel_Nnu}, we present the expected sensitivities to heavy neutrino production when the signal arises from the $t$-channel production of $NN$ pairs mediated by $W_R$-boson exchanges (left panel), as well as from its single production in association with a SM-like neutrino mediated by $W$-boson exchanges (right panel). The expected sensitivity to the $e^+e^-\to NN$ signal extends to progressively larger values of $m_N$ as the centre-of-mass energy increases, reflecting the improved kinematic accessibility of heavier final states at higher $\sqrt{s}$. In contrast, the reach in $M_{W_R}$ remains relatively stable across the different energy runs as the gain in centre-of-mass energy is largely compensated by the corresponding reduced integrated luminosities (see Table~\ref{tab:FCCee}). For $e^+e^-\to N\nu$ production, the sensitivity instead deteriorates with increasing  $\sqrt{s}$. This behaviour is primarily driven by the laboratory-frame angular distribution of the produced heavy neutrino $N$ which becomes increasingly forward-peaked at higher centre-of-mass energies (see Figure~\ref{fig:DsigDcosDpT}). As a consequence, the typical transverse displacement of the $N$ decay products is reduced, leading to a lower efficiency for the selection of displaced vertices. In addition, the total production cross section exhibits only a mild dependence on $\sqrt{s}$ (see Figure~\ref{fig:gauge_xsec}), such that the reduced integrated  luminosities at higher energies further suppress the expected signal yields. 

\medskip

Finally, our results show that the single-production channel exhibits a pronounced dependence on $\tan\beta$ (with even a loss of sensitivity when $\tan\beta$ is too small) as this parameter controls both the production rate and the $N$ decay width. On the other hand, for the $e^+e^-\to NN$ channel this dependence is milder as $\tan\beta$ only affects the $N$ lifetime and its semi-leptonic branching fraction. However, within the regions of parameter space covered in the $(M_{W_R}, m_N)$ plane, the precise value of $\tan\beta$ does not significantly alter the sensitivity contours, rendering the projected bounds robust against its variations. Conversely, an observation of the single-production channel alone would provide a strong indication of sizeable $\tan\beta$ mixing.

\medskip

\subsection{Scalar mixing modes: 
\texorpdfstring{$e^+ e^-\to Z h(\Delta)$}{e+ e- -> Z h(Delta)} and 
\texorpdfstring{$e^+e^-\to \nu\nu h(\Delta)$}{e+ e- -> nu nu h (Delta)}}\label{sec:sensiscalar}

If sizeable mixing is present in the scalar sector, and in particular an $h$-$\Delta$ mixing with $\sin\theta \gtrsim 0.05$, additional production channels for long-lived heavy neutrinos become accessible. In this case, the SM-like Higgs boson can decay as $h\to NN$, and if the triplet scalar $\Delta$ is light enough to be produced on-shell, the decay $\Delta\to NN$ is also potentially relevant if kinematically allowed. Before turning to the corresponding sensitivity reach, we first discuss the parametric dependence of the relevant production and decay rates on the scalar mixing angle $\sin\theta$, the masses $m_\Delta$ and $m_N$, as well as on $M_{W_R}$ which enters the $\Delta NN$ Yukawa coupling (see Table~\ref{tab:SNN}). On the other hand, the parameter $\tan\beta$ plays only a subdominant role here since it affects mainly the $N$ lifetime (as discussed in Section~\ref{sec:sensigauge}).

The size of the scalar mixing angle $\theta$ is constrained by measurements of the effective Higgs couplings to electroweak gauge bosons and SM fermions. At leading order, these SM-like couplings are rescaled by a factor of $\cos\theta$, leading to an approximate bound of \mbox{$\sin\theta \lesssim$ 0.3} from measurements of the $hW^+W^-$ coupling strength. If the triplet scalar is light enough for the decay $h\to\Delta\Delta$ to be kinematically open, then the scalar mixing is further constrained by Higgs width measurements and exotic decay searches, typically requiring $\sin\theta \lesssim 0.2$. 

\begin{figure}
  \centering
  \includegraphics[width=0.48\linewidth]{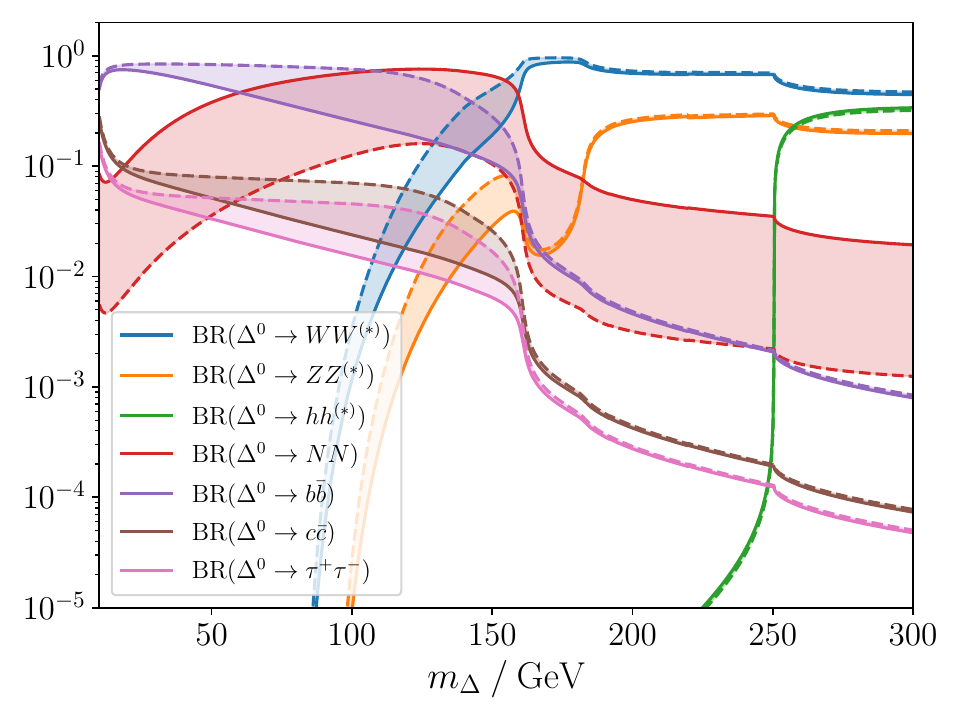}\hfill
  \includegraphics[width=0.48\linewidth]{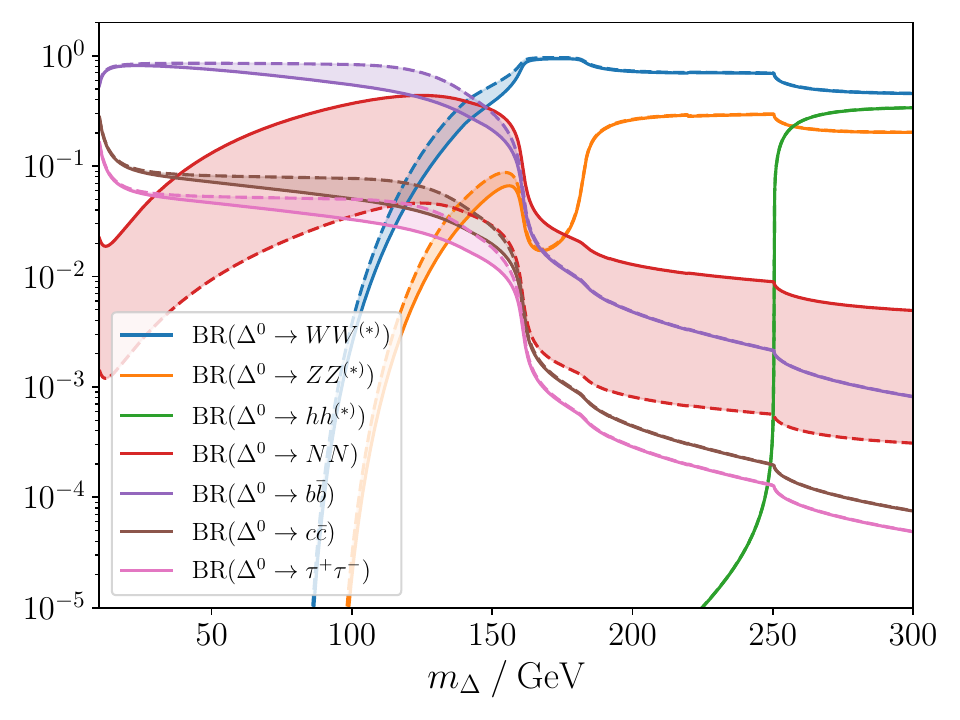}
  \vspace*{-2ex}%
  \caption{%
    Branching ratios of the triplet scalar $\Delta$ into heavy neutrinos and SM final states as a function of $m_\Delta$, with the heavy-neutrino mass fixed to $m_N = m_\Delta/3$. We highlight the dependence on the scalar mixing angle by considering $\sin\theta = 0.05$ (solid) and $\sin\theta = 0.2$ (dashed) for $M_{W_R}=5~\mathrm{TeV}$ (left), as well as that on the right-handed gauge-boson mass with $M_{W_R}=5~\mathrm{TeV}$ (solid) and $20~\mathrm{TeV}$ (dashed) for $\sin\theta=0.1$ (right).
  }\label{fig:Delta_BRs}\vspace{.2cm}
  \end{figure}
  \begin{figure} \includegraphics[width=0.48\linewidth]{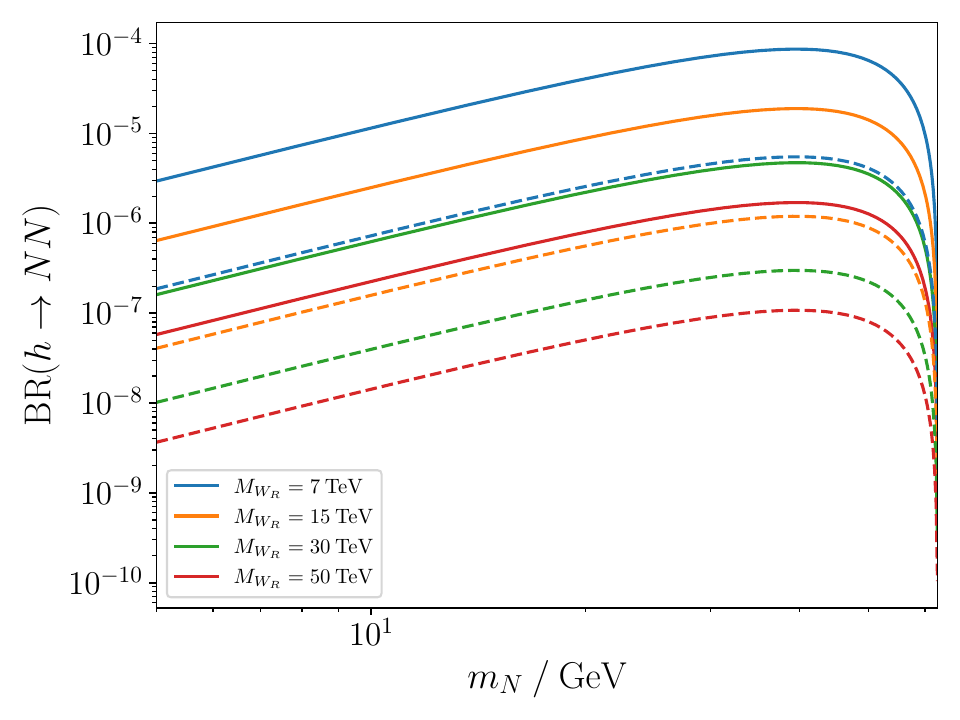}\hfill
  \includegraphics[width=0.48\linewidth]{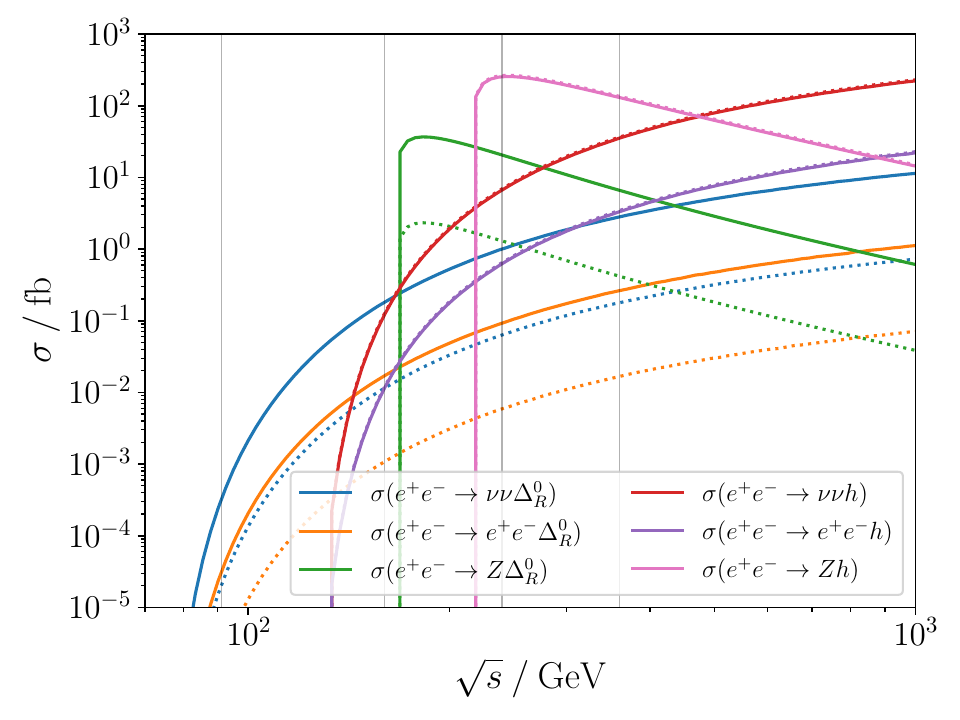}
  \vspace*{-2ex}%
  \caption{%
    Branching ratio and total rates relevant for scalar-mediated heavy neutrino production, for $\sin\theta = 0.2$ (solid) and 0.05 (dashed). \textit{Left} -- Branching ratio $\mathrm{BR}(h\to NN)$ as a function of the heavy neutrino mass $m_N$ for different values of $M_{W_R}$.  \textit{Right} -- Total cross sections for the production of the SM-like Higgs boson $h$ and the triplet scalar $\Delta$ in association with SM states, as a function of the centre-of-mass energy $\sqrt{s}$ and for a fixed triplet mass $m_\Delta = 75\,\mathrm{GeV}$.
  }
  \label{fig:scalar_prod_hNN}
\end{figure}

Since the couplings of the $\Delta$ state to the SM fields are directly proportional to $\sin\theta$, then both the $\Delta$ production rate and decay pattern depend sensitively on the scalar mixing, as illustrated in Figure~\ref{fig:Delta_BRs} for particular benchmark choices. In particular, while the $\Delta$ production cross sections in channels involving SM particles increase with $\sin\theta$, the branching fraction of the phenomenologically relevant decay $\Delta\to NN$ is reduced as larger mixing enhances the competing decay modes into SM states (see also~\cite{Nemevsek:2016enw, Fuks:2025jrn} for analytic expressions). This leads to a non-trivial interplay between production and decay such that the optimal sensitivity may arise for intermediate values of $\sin\theta$. In addition, the Yukawa coupling governing the $\Delta NN$ interaction scales as $m_N/M_{W_R}$ (see Table~\ref{tab:SNN}), implying that larger values of $M_{W_R}$ suppress the $\Delta\to NN$ partial width. As a result, the decay $\Delta\to NN$ can maintain an appreciable branching ratio over a broad range of $\sin\theta$ values in $[0.05,\,0.2]$ for moderate values of $M_{W_R}$, up to the threshold for on-shell decays into electroweak gauge bosons $m_\Delta \simeq 2 M_W$. Above this threshold, decays into SM gauge bosons rapidly dominate and the $\Delta\to NN$ branching ratio drops to the percent level or below.

For Higgs-mediated heavy neutrino production, the situation is somewhat simpler, as illustrated in Figure~\ref{fig:scalar_prod_hNN}. Within the range of $\sin\theta$ values allowed by Higgs coupling measurements, the Higgs production cross sections are only mildly affected by the choice of the LRSM parameters values, while the branching fraction $\mathrm{BR}(h\to NN)$ increases monotonically with $\sin\theta$. Furthermore, as in the case of the $\Delta$ decays, increasing $M_{W_R}$ suppresses the $\mathrm{BR}(h\to NN)$ branching fraction through the same Yukawa coupling dependence, up to an overall scalar mixing factor.

To estimate the theoretical sensitivity of the FCC-ee to scalar-mediated heavy neutrino production, we scan the parameter space in the $(M_{W_R}, m_N)$ plane, fixing $m_\Delta = 3 m_N$ and considering three representative allowed values of the scalar mixing $\sin\theta = 0.05$, 0.1 and 0.2. For all channels, we apply the same displaced-vertex requirement as for the gauge modes, \textit{i.e.}\ we convolute the production cross sections with the probability that both heavy neutrinos from the $h \to NN$ and $\Delta \to NN$ decays lie within the fiducial transverse displacement window $d_T \in [50\,\mu\mathrm{m},\,2\,\mathrm{m}]$.

\begin{figure}
  \centering
  \includegraphics[width=0.48\linewidth]{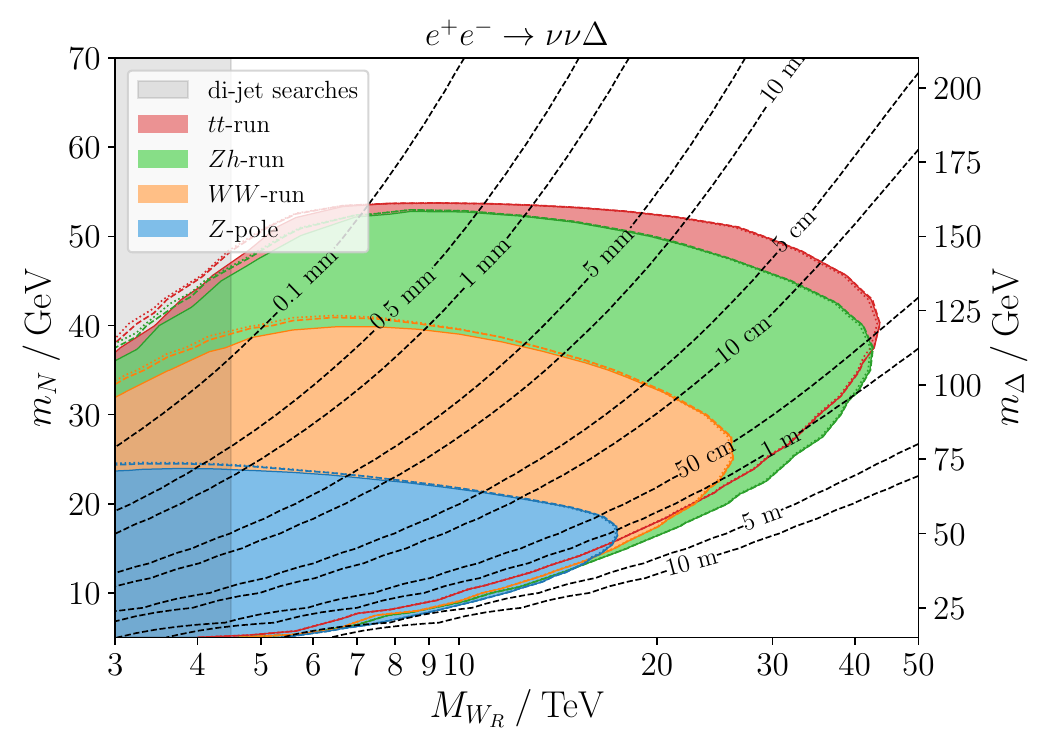}
  \vspace*{-2ex}
  \includegraphics[width=0.48\linewidth]{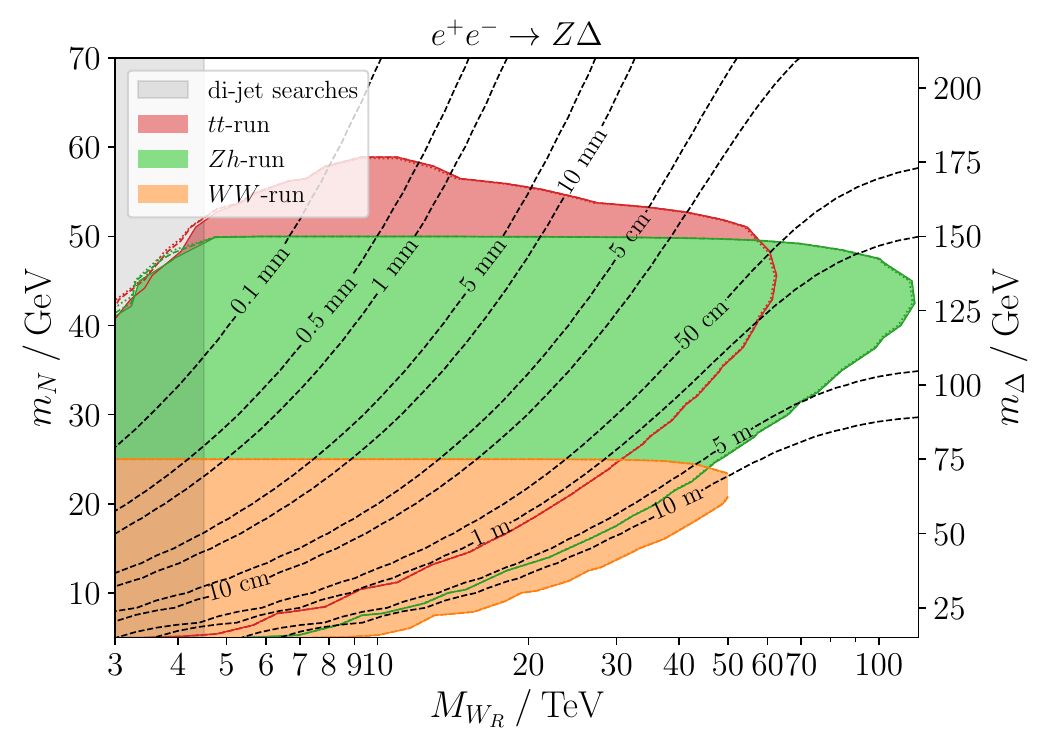}
  \caption{%
    Expected sensitivity of the FCC-ee to the triplet-mediated production of long-lived neutrinos $N$ via the $\Delta\to NN$ decay in the VBF (left) and $\Delta$-strahlung (right) mode, shown in the $(M_{W_R}, m_N)$ plane for fixed $m_\Delta = 3 m_N$. The solid (shaded), dashed and dotted contours correspond to $\sin\theta = 0.05$, 0.1 and 0.2, respectively, while the labelled contours indicate the heavy-neutrino proper decay length.
  } \label{fig:sens_delta}
\end{figure}

We first consider $\Delta$ production via $W$-boson fusion ($e^+e^-\to\nu\bar\nu\Delta$ with $\Delta\to NN$) and the `$\Delta$-strahlung' process ($e^+e^- \to Z\Delta$ with $\Delta\to NN$), and show the corresponding sensitivity contours for the different FCC-ee operation points in the left and right panels of Figure~\ref{fig:sens_delta} respectively. For both production mechanisms, the results exhibit only a mild dependence on the scalar mixing $\sin\theta$ within the considered range. This behaviour reflects the competing effects of increasing $\sin\theta$, which enhances the production cross sections with a simultaneous reduction of $\Delta\to NN$ branching ratio due to the opening of decay channels into SM final states (see Figure~\ref{fig:Delta_BRs}). For the VBF process, the sensitivity improves steadily with increasing centre-of-mass energy. The rise of the fusion cross section with $\sqrt{s}$ largely compensates for the reduced integrated luminosities at higher energies, allowing the FCC-ee to probe right-handed gauge-boson masses up to $M_{W_R} \lesssim 40\,\mathrm{TeV}$ for heavy neutrino masses $m_N \lesssim 55\,\mathrm{GeV}$, corresponding to $m_\Delta \lesssim 160\,\mathrm{GeV}$. The reach is primarily limited by the rapid suppression of the $\mathrm{BR}(\Delta\to NN)$ branching ratio once the $\Delta\to W^+W^-$ channel becomes kinematically accessible, or equivalently for sufficiently large $M_{W_R}$. In contrast, the sensitivity to the $\Delta$-strahlung mode exhibits a markedly different energy dependence. The production cross section peaks close to threshold ($\sqrt{s} \simeq m_\Delta + M_Z$) and then decreases at higher energies (see Figure~\ref{fig:scalar_prod_hNN}). As a consequence, the sensitivity is controlled both by the available phase space for on-shell $Z\Delta$ production and by the reduction of the $\Delta\to NN$ branching ratio for larger $\sin\theta$ and $M_{W_R}$ values. Despite these limitations, the $\Delta$-strahlung mode provides exceptional sensitivity at lower-energy runs. In particular, the $Zh$ run ($\sqrt{s}\simeq240\,\mathrm{GeV}$) has the potential to probe values of $M_{W_R}$ as large as $120\,\mathrm{TeV}$, whereas the $WW$ run ($\sqrt{s}\simeq160\,\mathrm{GeV}$) also offers substantial reach up to $M_{W_R}\simeq60\,\mathrm{TeV}$ for a relatively light triplet mass $m_\Delta \lesssim 70\,\mathrm{GeV}$, where the sensitivity is eventually limited by the heavy-neutrino displacement. Finally, at the $t\bar t$ run ($\sqrt{s}\simeq360\,\mathrm{GeV}$), the larger production cross section allows the sensitivity to extend to $m_\Delta$ values beyond the $\Delta\to W^+W^-$ threshold, partially compensating for the decreasing $\mathrm{BR}(\Delta\to NN)$ values and covering regions up to $m_\Delta \simeq 180\,\mathrm{GeV}$ and $M_{W_R}\simeq70\,\mathrm{TeV}$, \textit{i.e.}\ just below the opening of the $\Delta\to ZZ$ channel.

\begin{figure}
  \centering
  \includegraphics[width=0.48\linewidth]{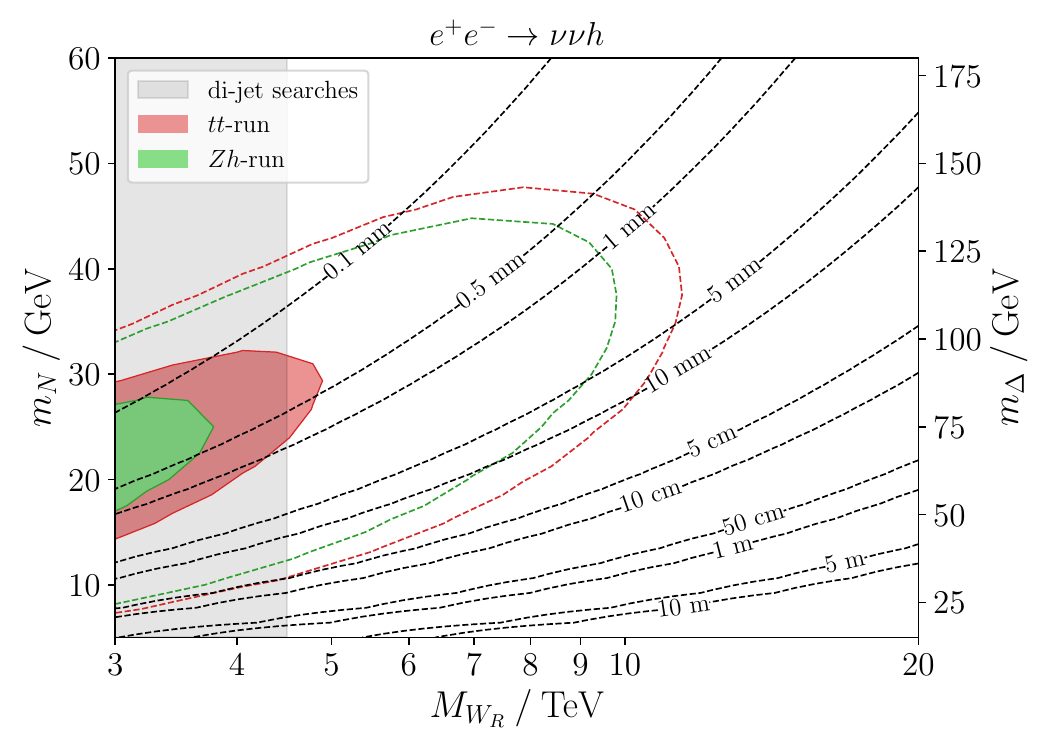}
  \includegraphics[width=0.48\linewidth]{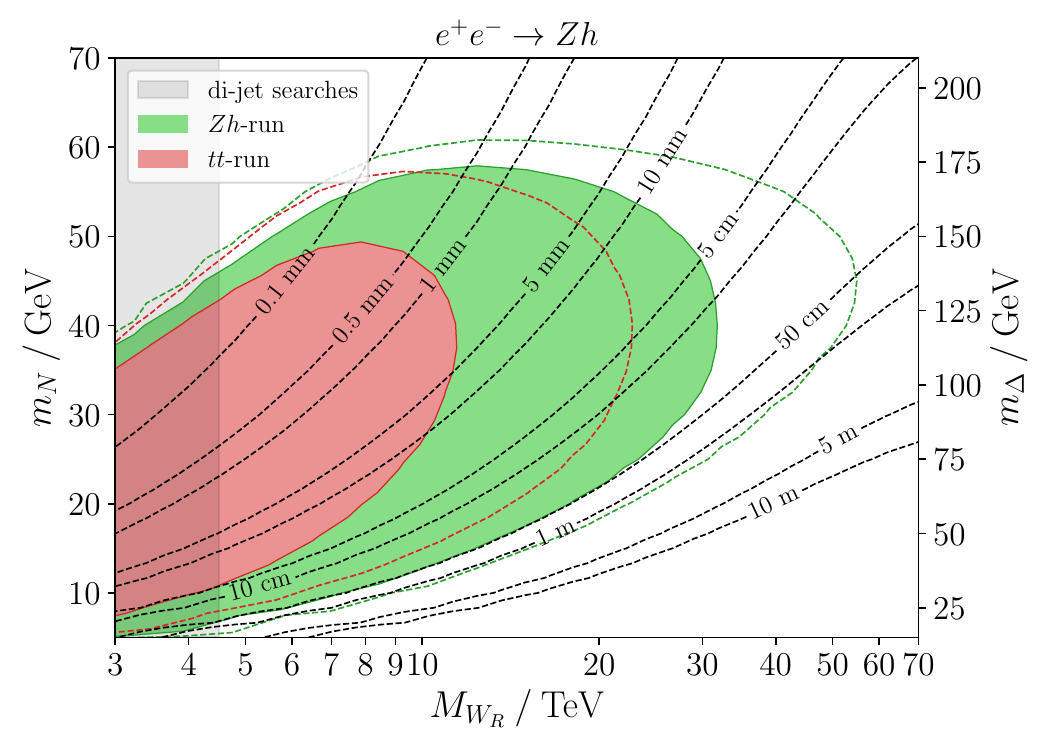}
  \vspace*{-2ex}
  \caption{%
    Expected sensitivity of the FCC-ee to the scalar-mediated production of long-lived neutrinos $N$ via the $h\to NN$ decay in the VBF (left) and Higgs-strahlung (right) mode, shown in the $(M_{W_R}, m_N)$ plane. The filled contours correspond to $\sin\theta = 0.1$, while the dashed ones represent scenarios with $\sin\theta = 0.2$.
  }\label{fig:sens_higgs}
\end{figure}
We now turn to scalar-mediated heavy neutrino production through $h \to NN$ decays. As already mentioned, in this case the situation differs qualitatively from $\Delta$ production. The Higgs production cross sections in both the VBF mode ($e^+e^-\to\nu\bar\nu h$ with $h\to NN$) and through Higgs-strahlung ($e^+e^-\to Zh$ with $h\to NN$) are essentially independent of the scalar mixing within its experimentally allowed range. Instead, the sensitivity is governed by the partial width $\Gamma(h\to NN) \propto \sin^2\theta\, m_N^2/M_{W_R}^2$, which directly controls the branching ratio $\mathrm{BR}(h\to NN)$ (see Figure~\ref{fig:scalar_prod_hNN}). The corresponding sensitivity contours for the VBF and Higgs-strahlung channels are shown in the left and right panels of Figure~\ref{fig:sens_higgs}, respectively. Due to the comparatively small production cross section, the VBF mode yields only limited sensitivity, reaching at most $M_{W_R}\lesssim15\,\mathrm{TeV}$ in the kinematically accessible region defined by $m_N\lesssim m_h/2$, and requires relatively large scalar mixing $\sin\theta\simeq0.2$ to be observable. However, such large mixing values are likely to be challenged by future precision measurements of the effective $hW^+W^-$ coupling. In contrast, the substantially larger Higgs-strahlung cross section allows the FCC-ee to probe $h\to NN$ decays up to $M_{W_R}\simeq50\,\mathrm{TeV}$, even for more moderate scalar mixing values of $\sin\theta\simeq0.1$. An observation of the $h\to NN$ decay at the FCC-ee would therefore provide a strong indication of sizeable scalar mixing in the LRSM. If accompanied by a $\Delta\to NN$ signal, it would then offer a direct probe of the scalar sector responsible for neutrino mass generation, while the absence of the latter would point to a heavier $\Delta$ state lying either above the $\Delta\to W^+W^-,\,ZZ$ thresholds or beyond the kinematic reach of the FCC-ee.

\subsection{Scalar boson fusion: 
\texorpdfstring{$e^+e^-\to e^+ e^-\Delta$}{e+ e- -> e+ e- Delta}} \label{sec:sensiSBF}

The last qualitatively different production mechanism that we consider involves the production of a heavy neutrino pair from the fusion of two doubly-charged scalars into a neutral scalar ($\Delta_R^{++}\Delta_R^{--}\to \Delta$). This corresponds to the full process $e^+e^- \to \ell^+\ell^- \Delta$ with a subsequent $\Delta \to NN$ decay, as illustrated by the Feynman diagram shown in the right panel of Figure~\ref{fig:VBFandSBF}. In contrast to the previously discussed channels, this production mode does not rely on gauge interactions but is instead controlled solely by scalar and Yukawa couplings, thus leading to a markedly different parametric behaviour. First, the Yukawa coupling $Y_\Delta$ of the doubly-charged scalar to the charged leptons can be sizeable if at least one of the heavy neutrinos is sufficiently heavy, with $Y_\Delta \simeq \mathcal{O}(1)$ or even $\mathcal{O}(10)$ with the precise value depending on the structure of the right-handed lepton mixing. Second, the trilinear scalar coupling $\Delta_R^{++}\Delta_R^{--}\Delta$ can receive significant enhancements, depending on the mass spectrum of the triplet scalars and the neutral components of the scalar bi-doublet. The latter effect is particularly important as the neutral scalars of the bi-doublet must be heavy in order to satisfy low-energy flavour constraints, with $m_A \simeq m_H \gtrsim 20~\mathrm{TeV}$~\cite{Bertolini:2014sua, Maiezza:2016bzp}, and the $\Delta_R^{++}\Delta_R^{--}\Delta$ coupling include contributions proportional to $m_A$ at order $\theta\epsilon$ or $\epsilon^2$ (see Table~\ref{tab:SSS}). Although parametrically suppressed, these can nevertheless be sizeable, even if their size is also constrained by perturbative unitarity. Imposing perturbative unitarity to the relevant term of the scalar potential $\mathcal V \supseteq \alpha_3 \,\mathrm{Tr}(\phi^\dagger\phi)\, \Delta_R\Delta_R^\dagger$ indeed leads to the constraint $\alpha_3 < 8\pi$, which imposes, for a given value of $M_{W_R}$, an upper bound on the heavy scalar masses $m_A \simeq m_H$ and therefore limits the maximal enhancement of the $\Delta_R^{++}\Delta_R^{--}\Delta$ coupling. Motivated by this structure, it is instructive to distinguish two regimes for SBF heavy neutrino production. In the first one, the scalar mixing is negligible ($\theta = 0$), such that the dependence of the trilinear coupling on $m_A$ is weak and its strength remains moderate. In the second regime, we consider instead a sizeable scalar mixing (with $\theta \simeq 0.2$), which induces a strong dependence of the trilinear coupling on the heavy scalar masses thereby potentially leading to large enhancements.

\begin{figure}
  \centering
  \includegraphics[width=0.6\linewidth]{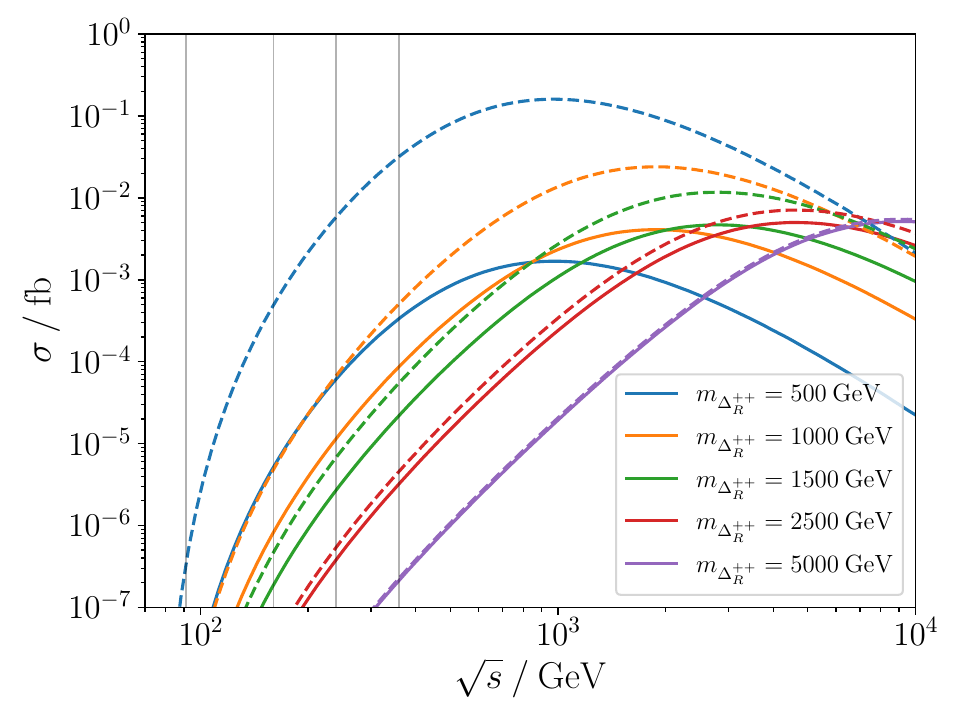}
  \vspace*{-3ex}
  \caption{%
    Production cross section of the SBF process $e^+e^-\to e^+e^-\Delta$ as a function of the centre-of-mass energy $\sqrt{s}$. The solid lines correspond to a vanishing scalar mixing $\theta=0$, the dashed lines refer to scenarios with $\theta=0.2$, while  the vertical grey lines indicate the FCC-ee proposed operation energies.
  }
  \label{fig:SBF_xsec}
\end{figure}

The resulting SBF cross section is presented in Figure~\ref{fig:SBF_xsec} as a function of the centre-of-mass energy for different choices of $m_{\Delta_R^{++}} \in [500,\,5000]~\mathrm{GeV}$ and for both $\theta=0$ and $\theta=0.2$. In all cases, we fix $Y_\Delta^{ee}=1$, $M_{W_R}=7~\mathrm{TeV}$, $m_A \simeq m_H = 20~\mathrm{TeV}$ and $m_\Delta = 75~\mathrm{GeV}$. The SBF production cross section is found sizeable at $\sqrt{s}\sim \mathcal{O}(\mathrm{TeV})$, but remains strongly suppressed at the proposed FCC-ee energies. A scalar mixing of $\theta=0.2$ yields an enhancement of one to two orders of magnitude compared to the $\theta=0$ case, depending on which contribution dominates the $\Delta_R^{++}\Delta_R^{--}\Delta$ coupling, and larger rates are in principle possible for $|Y_\Delta|>1$ since the cross section scales as $|Y_\Delta|^4$. However, constraints on $Y_\Delta$ arise from searches for doubly-charged scalars. Direct searches at the LHC impose a lower bound of $m_{\Delta_R^{++}} \gtrsim 700~\mathrm{GeV}$~\cite{ATLAS:2017xqs}, whereas indirect searches via $e^+e^-\to e^+e^- (\mu^+\mu^-)$ scattering at LEP enforce $m_{\Delta_R^{++}}/Y_\Delta^{ee(e\mu)} \gtrsim 2.5\,(4.9)~\mathrm{TeV}$~\cite{Nomura:2017abh, ALEPH:2013dgf}. As a consequence, no phenomenologically interesting region of the parameter space remains accessible at the FCC-ee energies without being strongly constrained by existing bounds. At higher centre-of-mass energies such as those envisioned at CLIC or at a high-energy muon collider, the SBF process can nevertheless yield a sizeable number of signal events, which should motivate a dedicated sensitivity study that we leave for future work.

%
%
\section{Kinematic reconstruction and experimental sensitivities}
\label{sec:KinRecon}

To precisely assess the experimental sensitivities at a future FCC-ee experiment, we carry out extensive simulations of all relevant signal processes and outline reconstruction strategies that allow us to fully identify the various production and decay modes considered in this work. In the following, we first describe the simulation setup and detector assumptions, and we then detail the reconstruction of the displaced decay vertices. This forms the backbone of all subsequent sensitivity estimates. The event generation is performed using \textsc{MG5\_aMC@NLO}~\cite{Alwall:2014hca}, relying on the UFO~\cite{Degrande:2011ua, Darme:2023jdn} model file developed in~\cite{Kriewald:2024cgr}. The hard-scattering events are subsequently passed to \textsc{Pythia~8.3}~\cite{Bierlich:2022pfr} for parton showers and hadronisation. For the simulation of the response of the detector, we rely on the IDEA detector~\cite{IDEAStudyGroup:2025gbt} card implemented in the fast simulation framework \textsc{Delphes}~3~\cite{deFavereau:2013fsa}. The IDEA detector is a design concept for a future FCC-ee experiment and consists of a low material-budget inner tracker (silicon vertex detector) surrounded by a light gaseous drift chamber. The full tracking volume extends to $r \leq 2\:\mathrm{m}$ radially and $|z| \leq 2\:\mathrm{m}$ in the longitudinal direction, corresponding to a pseudo-rapidity coverage of $|\eta|\lesssim 2.56$. This tracking system is embedded in a $2$--$3\,\mathrm{T}$ solenoidal magnetic field\footnote{Current plans for the $Z$-pole run foresee operations in a $2\,\mathrm{T}$ field, while higher magnetic fields could be envisaged for the other runs. Here we conservatively assume $2\,\mathrm{T}$ for all centre-of-mass energies.} followed by a dual-readout calorimeter system and muon chambers. Due to its large tracking volume, excellent spatial resolution and advanced particle-identification capabilities (thanks to drift chamber and timing measurements), the IDEA detector offers unique conditions for searches for long-lived particles with displaced decays, reaching unprecedented reconstruction capabilities.

The reconstruction of the long-lived heavy neutrinos emerging from our signal processes relies primarily on the identification of their displaced decay vertices. To this end, we employ the vertexing algorithms recently developed in~\cite{Kriewald:2025eiy} and implemented in \textsc{Delphes}, which have been designed to efficiently reconstruct both prompt and displaced vertices within a fast-simulation environment. The reconstruction starts with the classification of tracks as prompt or displaced,  based on the two-dimensional impact-parameter significance $\mathcal S_\mathrm{IP}$ defined as
\begin{equation} \label{eqn:IPsig}
  \mathcal S_\mathrm{IP} = \sqrt{(D_0, Z_0)(\mathrm{Cov}(D_0,Z_0))^{-1}
  \left(D_0, Z_0 \right)^T} \, ,
\end{equation}
where $D_0$ and $Z_0$ denote the (signed) transverse and longitudinal impact parameters, respectively, and $\mathrm{Cov}(D_0,Z_0)$ is their covariance matrix as provided by the track fit in \textsc{Delphes}. Tracks with $\mathcal S_\mathrm{IP} < 5$ are initially classified as prompt, while tracks with $\mathcal S_\mathrm{IP} \geq 5$ are considered displaced. Prompt tracks are fitted to a primary vertex (PV) candidate using the beam-spot position and size as a prior. For each track, the Mahalanobis distance to the PV is then computed. Tracks with a distance $\chi^2_\text{Maha} \geq 9$ are reclassified as displaced, and otherwise as prompt. This procedure enables the reconstruction of displaced vertices with small transverse displacements very close to the PV position, and we adopt a conservative lower cut of $d_T \gtrsim 50\:\mu\mathrm{m}$. Displaced tracks are subsequently grouped into vertex candidates using a graph-based approach in which tracks are paired based on their mutual compatibility, with additional support from track triplets (see~\cite{Kriewald:2025eiy} for details). Vertex candidates are next identified as the connected components of this graph, and we require each displaced vertex to contain at least one charged lepton track and at least three tracks in total.

The resulting displaced vertex (DV) candidates are fitted using a robust, initialisation-independent vertex fitter. For the multi-track decay topologies characteristic of the processes considered in this work, the vertex reconstruction efficiency is close to unity, with the main limitations arising for very low-multiplicity decays (typically for $m_N \lesssim 5\:\mathrm{GeV}$) and from geometric acceptance effects due to very large transverse displacements ($d_T \gtrsim 1.8\:\mathrm{m}$) or highly forward heavy neutrinos with $|\eta| \gtrsim 4$. The achieved vertexing precision ranges from approximately 2 or 3 $\,\mu\mathrm{m}$ for decays occurring within the vertex detector to $\mathcal O(100)\,\mu\mathrm{m}$ for decays reconstructed in the drift chamber. This ensures that even displaced decays with small transverse displacements from 10 to 50 $\mu\mathrm{m}$ can be reliably distinguished from the primary interaction point.

\subsection{Kinematical reconstruction}
\label{sec:kin}
Having reconstructed the displaced decay vertices, the next step consists of reconstructing the four-momenta of the associated long-lived heavy neutrinos. As a starting point, we may define a na\"ive displaced-vertex momentum by summing the fitted momenta of all charged tracks associated with a given DV
\begin{eqnarray}
  p_\text{DV}^\mu = \sum_{i\in \text{tracks}} p_{\text{track}_i}^\mu \, .
\end{eqnarray}
Whereas this implicitly assumes 100\%-efficient particle identification for charged hadrons ($\pi^\pm$, $K^\pm$, $p$ and $\bar p$) and leptons, fast-simulation studies of the IDEA detector indicate that such an assumption is well justified: particle identification based on ionisation cluster counting in the drift chamber combined with time-of-flight measurements can provide a $3\sigma$ separation of $\pi^\pm$, $K^\pm$ and $p/\bar p$ over wide momentum ranges~\cite{Bedeschi:2022rnj, IDEAStudyGroup:2025gbt}, with first test-beam results supporting these expectations~\cite{Elmetenawee:2025ioq}. In addition, the dual-readout electromagnetic calorimeter and the muon system are expected to deliver very high efficiencies for electron and muon identification. Possible sources of particle mis-identification such as in-flight $\pi^\pm$ decays into leptons or fully stopped pions in the calorimeter are difficult to quantify without a full detector simulation. However, for the multi-track displaced decays considered in this work, such effects are expected to be sub-dominant.

While the na\"ive DV momentum provides a useful baseline, it neglects the contributions of the neutral particles (photons and neutral hadrons) produced in the heavy neutrino decays. To address this limitation, we employ a track-based jet-matching procedure that allows for a more complete reconstruction of the heavy neutrino four-momenta. This reconstruction proceeds in two stages. First, all charged and neutral particles are processed by a particle-flow (PFlow) algorithm within \textsc{Delphes}~3, which primarily supplies energy and momentum measurements for neutral particles and provides an additional consistency check for the charged particles.\footnote{Owing to the superior momentum resolution of the tracker, the track-based reconstruction combined with particle identification is used as the primary energy estimator for the charged particles.} Next, isolated charged leptons ($e^\pm$, $\mu^\pm$) are identified as tracks with $p_T \geq 1.0\:\mathrm{GeV}$ and satisfying loose isolation requirements, allowing nearby tracks within a cone of $\Delta R \leq 0.1$ to carry up to $50\%$ of the lepton transverse momentum. Identified isolated leptons are then removed from the PFlow collection to avoid object overlap and double counting in the subsequent jet clustering. The remaining PFlow objects are clustered into jets using \textsc{FastJet-3.5.1}~\cite{Cacciari:2011ma}, employing the Durham $k_T$ algorithm in exclusive mode~\cite{Catani:1991hj}. Depending on the event topology, this procedure reconstructs fully resolved final states with two, four or six jets. In a second step, the reconstructed jets are associated to the displaced vertices based on their shared charged-track content. For this purpose, we use a $p_T$-weighted track-overlap score,
\begin{equation}
  s_\mathrm{overlap} = \frac{2 \sum_{i\in \mathrm{shared}} (p_T^i)^\alpha}{
  \sum_{i\in\mathrm{jet}} (p_T^i)^\alpha + \sum_{i\in\mathrm{DV}}(p_T^i)^\alpha} \, ,
\end{equation}
with $\alpha = 1$. Jets are then assigned to the displaced vertex that maximises this score, provided that $s_\mathrm{overlap} \geq 0.1$. The reconstructed four-momentum $p_\text{N}^\mu$ of the heavy neutrino is finally given by the sum of the four-momenta of all isolated leptons and jets matched with the displaced vertex,
\begin{equation} \label{eqn:pproxy}
  p_\text{N}^\mu = \sum_{i\in \text{leptons}}p_i^\mu + \sum_{i\in \text{jets}} p_i^\mu\,.
\end{equation}
As demonstrated in~\cite{Kriewald:2025eiy} and further illustrated below, this procedure provides an accurate proxy for the true heavy neutrino momentum over a wide range of masses and decay topologies.

\begin{figure}
  \centering
  \includegraphics[width=0.5\linewidth]{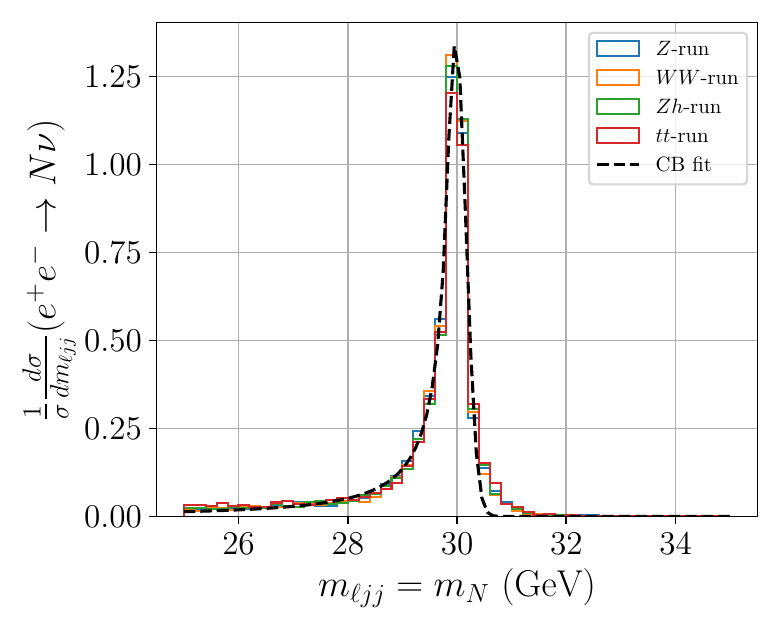}
  \vspace*{-2ex}%
  \caption{%
    Reconstructed invariant mass distribution of the heavy neutrino produced in $e^+e^-\to N\nu$ events for the four FCC-ee run energies and a heavy neutrino mass of 30~GeV. The mass is computed from the proxy momentum defined in Eq.~\eqref{eqn:pproxy}, and a Crystal Ball fit (shown for illustration) demonstrates that the reconstructed peak is very close to the true value $m_N = 30$~GeV.\label{fig:dist_mN}}\vspace{.3cm}
  \end{figure}

To illustrate the performance of the kinematical reconstruction, we consider a benchmark point defined by $M_{W_R} = 7$~TeV, $m_N = 30$~GeV and $m_\Delta = 90$~GeV.\footnote{Since we focus exclusively on normalised distributions, the precise choice of the mixing structure is irrelevant.} For each FCC-ee run energy and signal process, we simulate $10^4$ events, and the decay vertices are reconstructed using the criteria and the jet/lepton-DV matching procedure described above. At this stage, no additional kinematical selections are imposed, such that the distributions that we compute directly reflect the intrinsic reconstruction performance and detector acceptance effects. We begin with the single-$N$ production process $e^+e^-\to N\nu$, and present in Figure~\ref{fig:dist_mN} the invariant mass distribution derived from the reconstructed proxy momentum $p^\mu_\text{N}$ defined in Eq.~\eqref{eqn:pproxy}. This distribution exhibits a narrow peak around the true $m_N$ value, demonstrating that the reconstruction method successfully captures both charged and neutral decay products of the displaced $N$ decay. For illustration, we overlay a fit using a Crystal Ball function, which yields $m_N = 29.98$~GeV. While the fit itself is not used in the FCC-ee sensitivity analysis of Section~\ref{subsec:RecoSens}, the extracted peak position highlights the excellent mass resolution achievable with the proposed reconstruction strategy.

  \begin{figure} \includegraphics[width=0.48\linewidth]{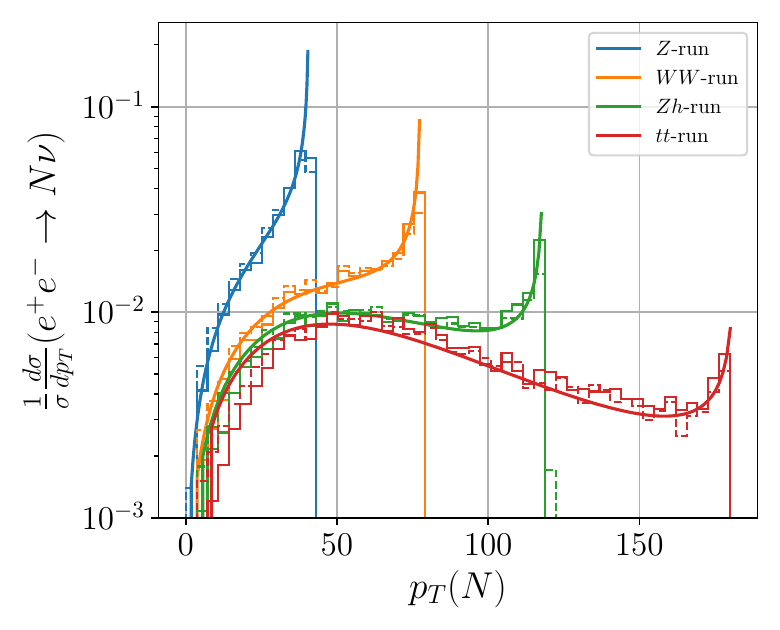}\hfill
  \includegraphics[width=0.48\linewidth]{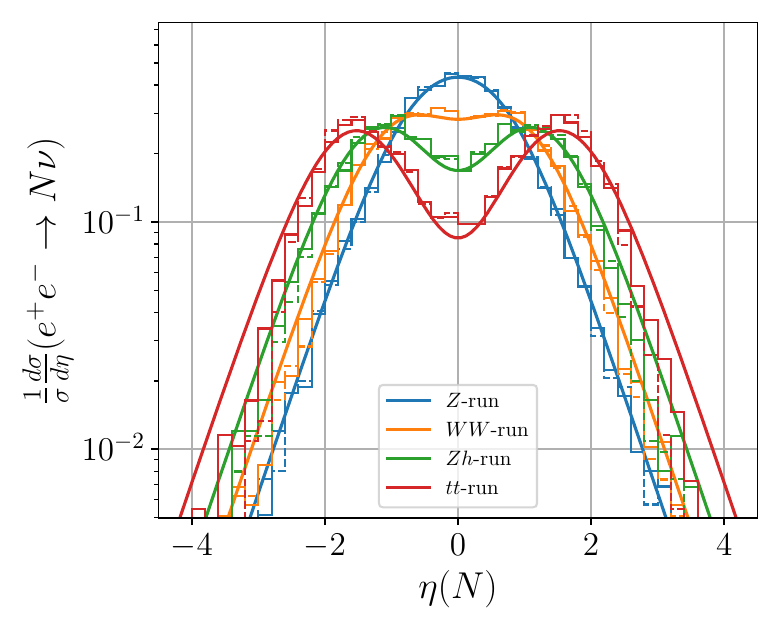}
  \vspace*{2ex}
  \caption{%
    Transverse momentum (left) and pseudo-rapidity (right) distributions of the heavy neutrino originating from $e^+e^-\to N\nu$ events for the four FCC-ee run energies. The solid curves correspond to analytic predictions, the solid histograms to truth-level  Monte Carlo events with a reconstructed displaced vertex, and the dashed histograms to the fully reconstructed distributions after detector simulation.\label{fig:dist_pT_eta}}
\end{figure}

Beyond the mass reconstruction, the transverse momentum and pseudo-rapidity distributions of the heavy neutrino provide a more differential test of the reconstruction procedure. At leading order, $e^+e^-\to N\nu$ is a two-body process such that the scattering angle $\theta$ of the heavy neutrino in the centre-of-mass frame fully determines its kinematics. The pseudo-rapidity distribution follows directly from the differential cross section $\mathrm{d}\sigma/\mathrm{d}\cos\theta$ via the relation $\cos\theta = \tanh\eta$, together with the Jacobian $\mathrm{d}\cos\theta/\mathrm{d}\eta = 1 - \tanh^2\eta$. Similarly, the transverse momentum spectrum reflects the boost of the produced heavy neutrino, which increases with the centre-of-mass energy. The resulting $p_T$ and $\eta$ distributions are shown in Figure~\ref{fig:dist_pT_eta}, where we compare the analytic predictions derived in Section~\ref{subsec:Xsecs} (solid curves) to both the truth-level Monte Carlo distributions after enforcing a restriction to the events in which a displaced vertex can be successfully reconstructed (solid histograms), and the fully reconstructed distributions after detector simulation and vertexing (dashed histograms). Overall, both the truth-level and the reconstructed distributions show excellent agreement with the theoretical expectations, confirming that the reconstruction procedure preserves the essential kinematical features of the signal. A notable deviation is observed for the $t\bar t$ run where a depletion of events appears at large pseudo-rapidities $|\eta|\gtrsim 3$, or equivalently at low transverse momentum. This effect can be traced back to the finite geometric acceptance of the IDEA tracker which provides coverage only up to $|\eta|\lesssim 2.56$. As the centre-of-mass energy increases, $N$ production becomes increasingly forward, causing a growing fraction of events to fall outside the tracking volume. The resulting loss is therefore an expected consequence of detector acceptance rather than a limitation of the reconstruction algorithm itself.

To assess the performance of our reconstruction method in more complex topologies, we now turn to processes featuring the production of two displaced heavy neutrinos. While the reconstruction of a single displaced vertex already provides a stringent test of the vertexing and momentum-proxy strategy, final states with multiple displaced decays pose additional challenges. In particular, they introduce potential combinatorial ambiguities in the association of tracks and jets to the correct vertex, as well as the possibility of asymmetric reconstruction efficiencies between the two $N$ decays. Demonstrating robust reconstruction in such environments is therefore essential for the full exploitation of the signal channels considered in this work.

\begin{figure}
  \centering
  \includegraphics[width=0.48\linewidth]{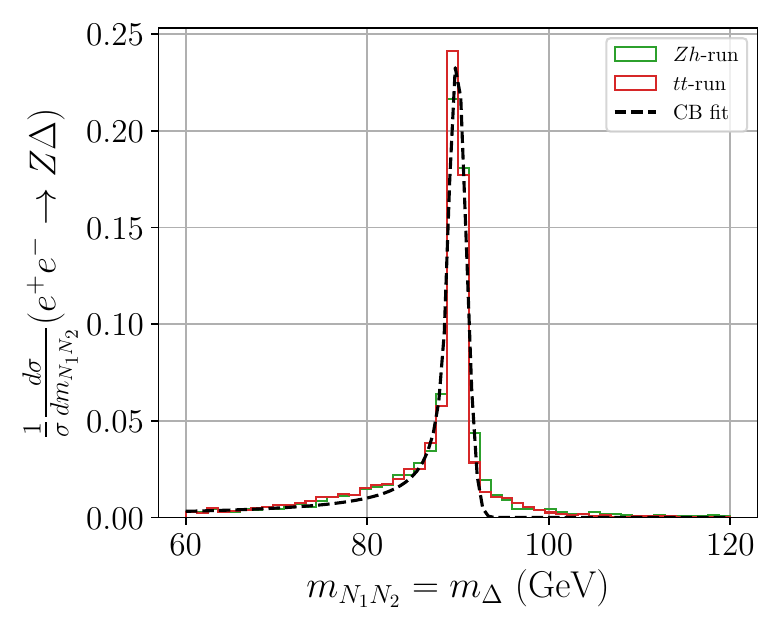}  \hfill
  \includegraphics[width=0.48\linewidth]{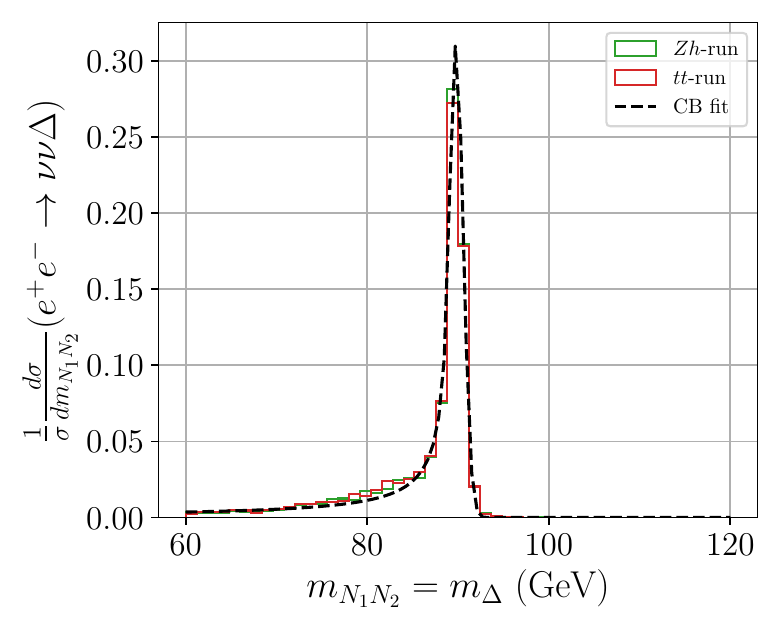}
  \vspace*{-2ex}%
  \caption{%
    Invariant mass distributions of the reconstructed $\Delta$ candidate in the processes $e^+e^-\to Z\Delta$ (left) and $e^+e^-\to\nu\nu\Delta$ via $W$-boson fusion (right), where the triplet scalar decays into a pair of displaced heavy neutrinos. The considered observable is determined from the sum of the two reconstructed heavy neutrino proxy momenta, and we show results for the two relevant FCC-ee run energies. Crystal Ball fits are overlaid for illustration (dashed lines).\label{fig:dist_mDelta}}
\end{figure}
We focus on the production of a scalar $\Delta$ followed by its decay into a pair of neutrinos, $\Delta \to N N$, and reconstruct the $\Delta$ four-momentum as the sum of the two reconstructed heavy neutrino proxy momenta
\begin{equation}
  p_\Delta^\mu = p_{\text{N}_1}^\mu + p_{\text{N}_2}^\mu \, .
\end{equation}
This therefore relies solely on the displaced information and on the jet/lepton-DV matching procedure described above, and does not make use of any prompt objects in the event. In Figure~\ref{fig:dist_mDelta}, we show the invariant mass distributions obtained from the reconstructed $\Delta$ four-momentum for the processes  $e^+e^- \to Z\Delta$ (left panel) and $e^+e^- \to \nu\nu\Delta$ (right panel), where the latter proceeds exclusively via $W$-boson fusion. Results include all relevant FCC-ee run energies, and for illustration we overlay Crystal Ball fits to the reconstructed spectra. Despite the presence of two displaced decays in the final state, the reconstructed mass distributions exhibit a narrow and well-defined peak whose position at $m_\Delta = 89.77$~GeV, which is in excellent agreement with the true input mass of 90~GeV. While the figure highlights residual non-Gaussian tails, these can be attributed to standard detector effects such as finite acceptance for the neutral particles or imperfect jet/lepton-DV associations. They do not compromise the overall mass resolution and our results clearly demonstrate that the momentum-proxy method remains accurate even in multi-vertex topologies.

\begin{figure}
  \centering
  \includegraphics[width=0.48\linewidth]{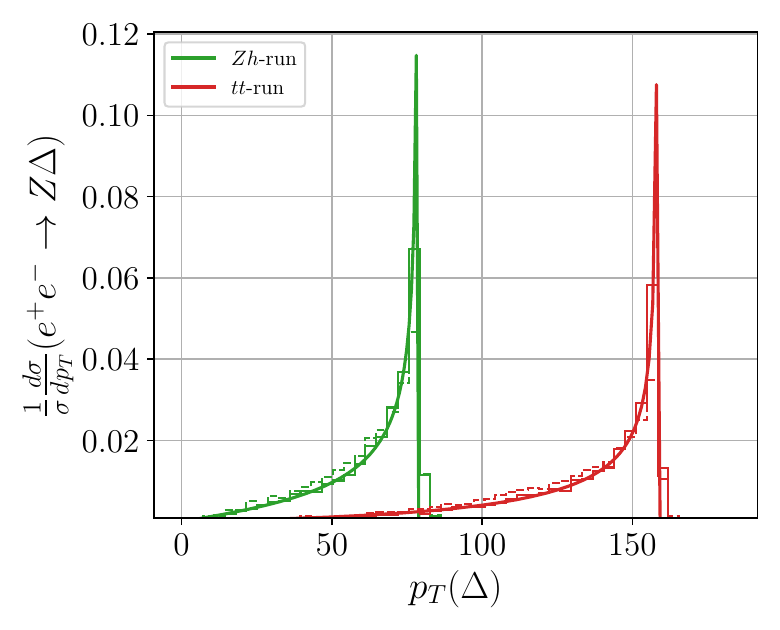}\hfill
  \includegraphics[width=0.48\linewidth]{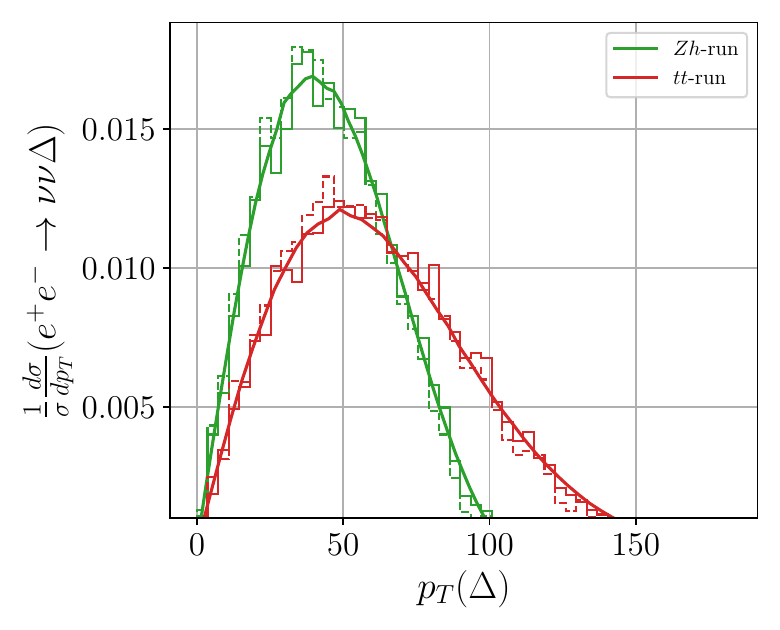}
  \vspace*{-2ex}%
  \caption{Transverse momentum distributions of the reconstructed $\Delta$ candidate in the $e^+e^-\to Z\Delta$ (left) and $e^+e^-\to\nu\nu\Delta$ (right) processes, with the $\Delta$ decaying into a pair of displaced heavy neutrinos. The solid curves show the theoretical predictions, the solid histograms the truth-level  Monte-Carlo distributions and the dashed histograms the reconstructed results after detector simulation and displaced-vertex reconstruction.\label{fig:dist_pT}} \vspace{2em}
  \includegraphics[width=0.48\linewidth]{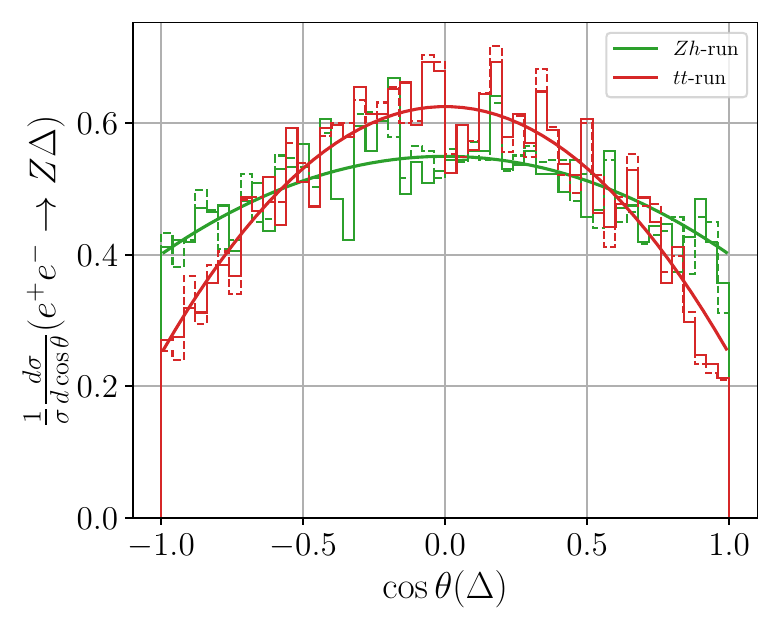}\hfill
  \vspace*{-2ex}%
  \includegraphics[width=0.48\linewidth]{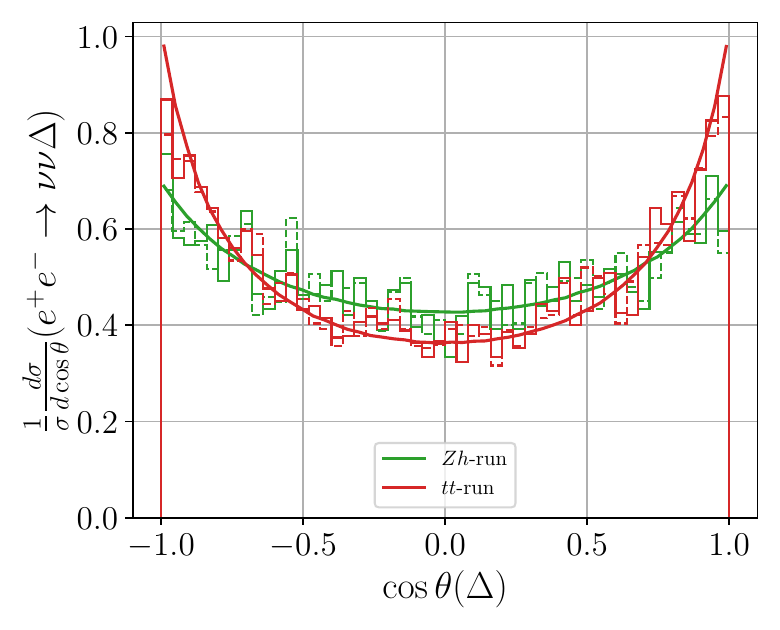}
  \caption{Same as in Figure~\ref{fig:dist_pT} but for the distribution of the cosine of the polar angle of the reconstructed $\Delta$ candidate. \label{fig:dist_cos}}\vspace{2em} 
  \includegraphics[width=0.5\linewidth]{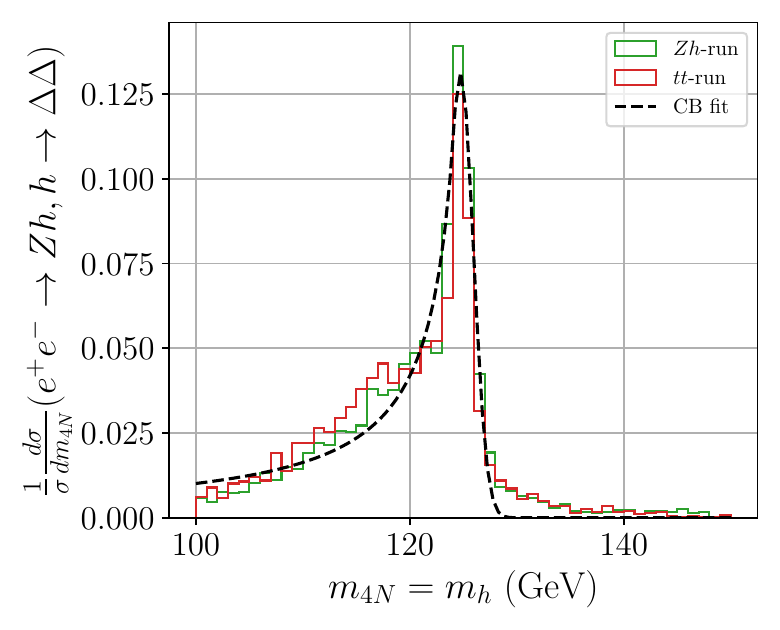}
  \vspace*{-2ex}%
  \caption{Same is in Figure~\ref{fig:dist_mDelta} but for the processes $e^+e^-\to Zh$.\label{fig:mh_dist}}
\end{figure}

In addition, the kinematic distributions of the reconstructed $\Delta$ provide valuable information on the underlying production mechanisms. In Figures~\ref{fig:dist_pT} and~\ref{fig:dist_cos}, we present the transverse momentum $p_T(\Delta)$ and polar-angle $\cos\theta(\Delta)$ distributions, respectively, again for the two relevant FCC-ee operation points at the $Zh$ and $t\bar t$ thresholds. As above, the reconstructed spectra are compared to the corresponding theoretical predictions as well as to truth-level Monte-Carlo distributions restricted to events in which both displaced vertices are successfully reconstructed. In all cases, we observe excellent agreement between the reconstructed distributions and both the truth-level Monte-Carlo and the theoretical predictions, which demonstrates that neither the displaced-vertex reconstruction nor the jet/lepton-DV matching procedure introduces significant kinematic distortions in the parent  $\Delta$ observables. Furthermore, the $e^+e^-\to Z\Delta$ process exhibits a predominantly central production, with events concentrated at small $|\cos\theta|$ and a relatively hard $p_T$ spectrum. This behaviour is characteristic of an $s$-channel production mechanism in which the kinematics are largely fixed by the two-body phase space at a given centre-of-mass energy. In contrast, the $W$-boson fusion process $e^+e^-\to\nu\nu\Delta$ leads to a markedly more forward-peaked angular distribution, accompanied by a broader transverse momentum spectrum that peaks at lower $p_T$. These features reflect the $t$-channel nature of the production mechanism and the enhanced longitudinal boost imparted to the $\Delta$ system. The pronounced differences between the two production modes in both $p_T$ and $\cos\theta$ therefore provide powerful handles for their experimental separation, even in the presence of multiple displaced decays in the final state.

Finally, we consider the cascade decay process  $e^+ e^- \to Zh$, followed by $h \to \Delta \Delta$ with each $\Delta$ decaying into a pair of heavy neutrinos. This channel leads to an exceptionally complex final state, containing four displaced $N$ decays and therefore four final state charged leptons and eight predominantly soft jets. Due to its high multiplicity, relatively soft visible objects, and the presence of multiple displaced vertices, the overall experimental sensitivity of the $4N$ final state is expected to be sub-dominant compared to the simpler signatures discussed above and may thus not be considered a primary discovery channel. Nevertheless, this topology provides a stringent stress test of the displaced-vertex reconstruction and momentum-proxy strategy. Despite the large combinatorial complexity, we find that the long-lived particle reconstruction remains robust and does not introduce significant kinematic distortions, as illustrated by the reconstructed Higgs mass distribution displayed in Figure~\ref{fig:mh_dist}. A clear mass peak is indeed recovered from the sum of the four reconstructed heavy neutrino proxy momenta, a Crystal Ball fit yielding the value $m_{4N} = 124.7$~GeV that is in good agreement with the true Higgs mass.

While the results presented in this section are subject to the intrinsic limitations of fast detector simulation studies, they highlight the reconstruction capabilities envisaged for the IDEA detector at the FCC-ee. Importantly, they demonstrate that displaced vertices can be elevated from mere event tags to a fully fledged kinematic class of objects that enables not only the reconstruction of the long-lived particles themselves but also the identification of the complete underlying production chain, including that of the intermediate and parent states through observables that map directly onto parton-level quantities.

%
\subsection{Reconstruction-level sensitivity estimates} \label{subsec:RecoSens}

In this section, we present the final results of our analysis, culminating in reconstruction-level sensitivity estimates for the various signal channels and centre-of-mass energies considered. We begin by describing the Monte-Carlo event generation procedure, along with the detector-level reconstruction and selection criteria that are common to all channels, and subsequently proceed with the specifics of each production mode.

For each signal process, we generate $10^4$ hard-scattering events for a set of equally spaced heavy neutrino masses, $m_N \in [5,\,10,\,15,\,\dots,\,m_{N_\text{max}}]\:\mathrm{GeV}$, and for a wide range of logarithmically spaced proper decay lengths $c\tau(N) \in [0.02,\,0.04,\,0.1,\,\dots,\,2\times 10^5]\:\mathrm{mm}$. Heavy neutrino decays are then handled by \textsc{Pythia}, together with parton showering and hadronisation, and we restrict the final states to semi-leptonic modes in which $N \to \ell^\pm q \bar q'$ with a flavour-democratic choice for the first two generations of charged leptons ($\ell \in [e,\,\mu]$). Moreover, the different quark flavour contributions ($q,q' \in [u,\,d,\,s,\,c,\,b]$) are weighted according to the CKM matrix, assuming $V_L^\text{CKM} = V_R^\text{CKM}$. In addition to the signal samples, we also simulate representative SM background processes, generating $10^6$ events for each channel. While a comprehensive background study is beyond the scope of this work, we include a set of dominant and illustrative backgrounds at each run energy. Specifically, we simulate $e^+e^- \to Z \to b\bar b$ at the $Z$-pole run, $e^+e^- \to W^+W^-$ with $W^+\to c \bar q, q' \bar b$ and $W^-\to \bar c q, \bar q' b$ decays at the $WW$ run, $e^+e^- \to Zh$ with a $h \to b\bar b$ decay at the $Zh$ run, and $t\bar t$ production at the top threshold run. The proposed centre-of-mass energies and integrated luminosities for the FCC-ee and the CEPC are collected in Tables~\ref{tab:FCCee} and~\ref{tab:CEPC}.

Depending on the signal topology, we apply a set of basic selection cuts designed to suppress the SM backgrounds while maintaining high signal efficiencies. As a common requirement, we demand the reconstruction of exactly two long-lived particle (LLP) candidates,  except for the single-$N$ production process $e^+e^- \to N\nu$, where only one LLP candidate is expected. We further impose an invariant mass window cut on the reconstructed heavy neutrino candidates, $m_N^\text{reco} \in [0.8\,m_N^\text{true},\,1.2\,m_N^\text{true}]$, using the reconstruction techniques and jet definitions introduced in Section~\ref{sec:kin}. To suppress the residual prompt background, we veto events featuring more than two prompt tracks for $m_N^\text{true} \leq 25\:\mathrm{GeV}$ and relax this requirement to at most five prompt tracks for heavier $m_N$. For $NN$ production mediated by a scalar resonance $S = h,\,\Delta$, we additionally require the invariant mass of the system comprising the two reconstructed LLP candidates to satisfy $m_{N_1N_2} \in [0.8\,m_S,\,1.2\,m_S]$. Finally, in the case of direct $e^+e^- \to NN$ production via an $s$-channel $Z$-boson exchange or a $t$-channel $W_R$-boson exchange, we instead require $m_{N_1N_2} \in [0.6\,\sqrt{s},\,\sqrt{s}]$.

These selection criteria suppress all simulated SM background events while yielding signal efficiencies in the range of $70\%-80\%$. Building on the reconstruction performance demonstrated in Section~\ref{sec:kin}, more refined kinematic selections exploiting the reconstructed LLP and parent four-momenta, jet and lepton multiplicities, and dedicated prompt object reconstruction would be possible. However, we leave a detailed optimisation of such strategies for future work, and given the strong background suppression achieved with the basic selections, we proceed with the sensitivity estimates in the background-free approximation. For a Poisson counting experiment with zero expected background, the $95\%$ confidence  level upper limit on the signal yield is $s_{95} = -\ln(0.05) \simeq 3$. Accordingly, we define the sensitivity by requiring at least $N_S \geq 3$ reconstructed signal events.

\begin{figure}[b]
  \centering
  \includegraphics[width=0.48\linewidth]{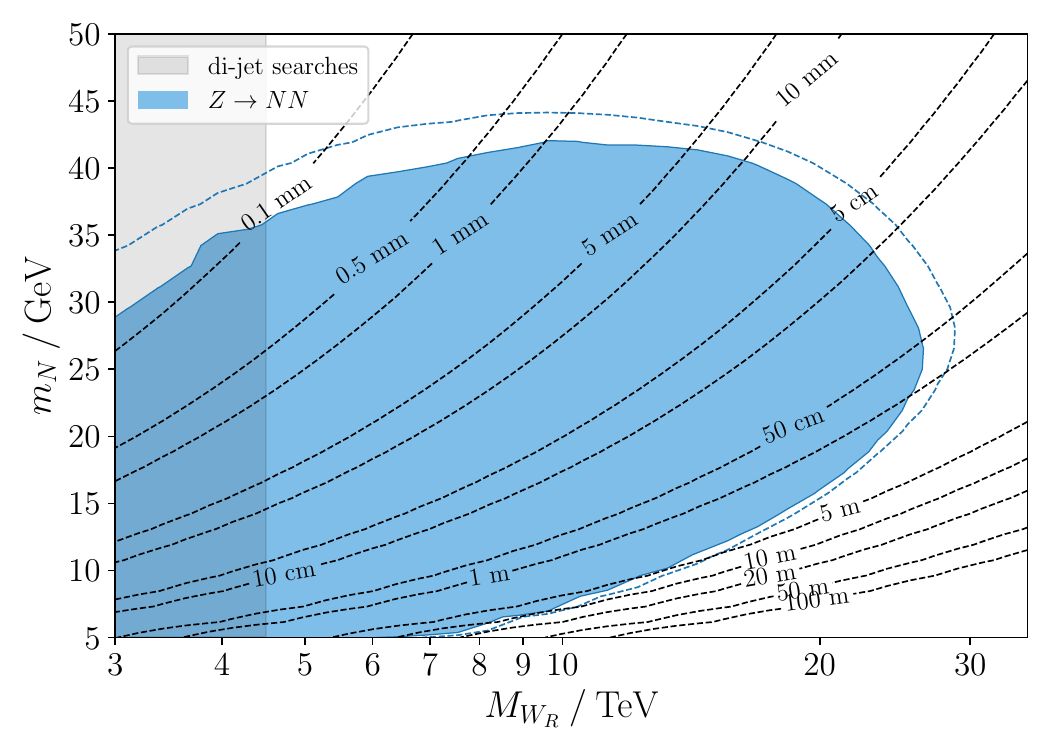}\hfill
  \includegraphics[width=0.48\linewidth]{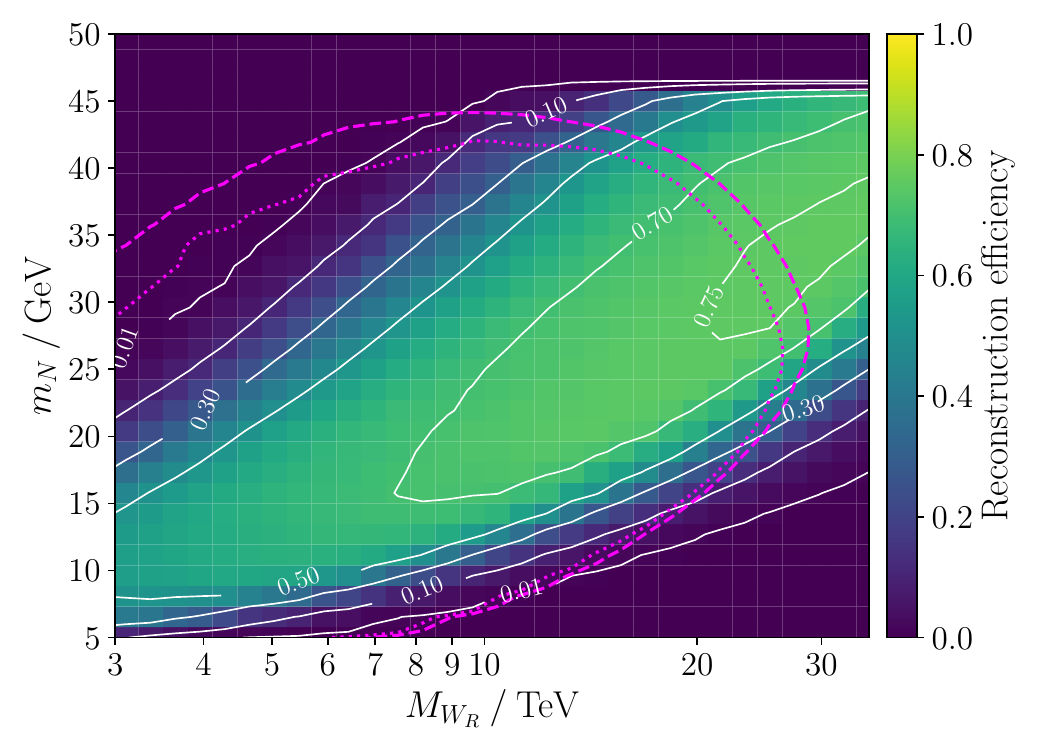}%
  \vspace*{-2ex}
  \caption{%
    \textit{Left} -- FCC-ee sensitivities to heavy neutrino production via $e^+e^- \to Z \to NN$ at the $Z$-pole run, shown in the $(M_{W_R}, m_N)$ plane and for $\tan\beta = 0$. The shaded contour shows the final reconstruction-level sensitivity, while the dashed contour corresponds to the theoretical estimate including the fiducial displacement requirement (see Figure~\ref{fig:sens_ZNN}). \textit{Right} -- Corresponding efficiencies after all kinematic and vertexing selections, with the overlaid magenta dashed (dotted) lines indicating the theoretical (reconstruction-level) sensitivity contours. \label{fig:sens_exp_ZNN}}\vspace{.2cm}
\end{figure}

\paragraph{Gauge modes.}
We begin by discussing the reconstruction-level sensitivities in the gauge-mediated production channels that we obtain using the strategy outlined above. In Figure~\ref{fig:sens_exp_ZNN}, we present the sensitivities for $Z \to NN$ production at the $Z$-pole run in the $(M_{W_R}, m_N)$ plane and for $\tan\beta = 0$. The filled contour shows the final reach, while the dashed contour corresponds to the  theoretical estimate given in Section~\ref{sec:sensigauge}, which includes only the fiducial displacement requirement that both $N$ decay within the tracking volume. The degradation from theoretical to realistic sensitivity is mild, reflecting the high reconstruction efficiencies achieved for this topology, reaching up to about 80\%. The efficiency across the parameter space is shown in the right panel of the figure, where we demonstrate that efficiencies of $\mathcal{O}(75\%)$ are obtained in the parameter space region where the heavy neutrino proper decay length is of $c\tau(N) \sim 10\,\mathrm{cm}$,  corresponding to values relevant for optimal displaced-vertex reconstruction. The strongest loss in efficiency occurs for heavier $N$ masses of $m_N \gtrsim 35\:\mathrm{GeV}$, and for very short lifetimes of $c\tau(N) \lesssim 1\:\mathrm{mm}$. This behaviour can be primarily attributed to the prompt-track veto applied in the event selection. Heavier neutrinos indeed typically produce higher charged-particle multiplicities, increasing the probability that tracks with small displacements are misidentified as prompt. In addition, their decay also produces softer charged particles whose track-parameter resolutions are poorer, leading to smaller impact-parameter significances and a higher chance of misclassification. While relaxing the two-dimensional impact parameter significance requirement $\mathcal S_\text{IP} \geq 5$ (see Eq.~\eqref{eqn:IPsig}) could partially recover the signal efficiency, this would also increase the susceptibility to backgrounds from heavy-flavour decays. Nevertheless, despite this mild degradation, the reconstruction-level reach remains substantial, covering heavy neutrino masses in the range $m_N \in [5,40]\:\mathrm{GeV}$ and extending up to $M_{W_R} \lesssim 25\:\mathrm{TeV}$.

\begin{figure} \includegraphics[width=0.48\linewidth]{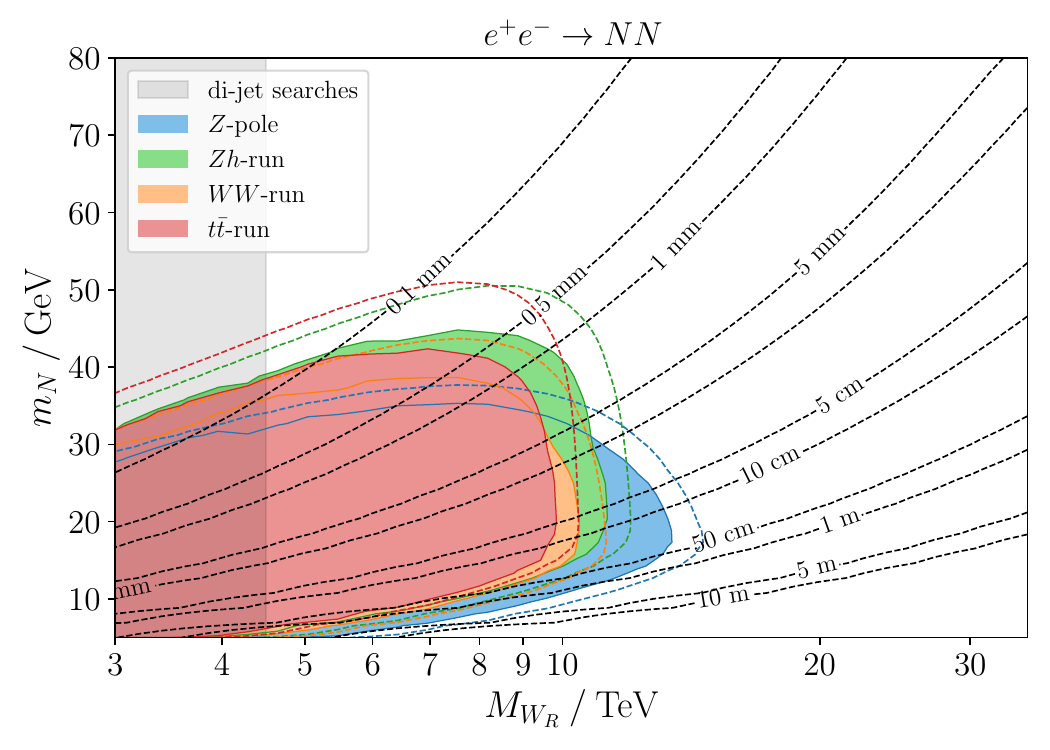}\hfill
  \includegraphics[width=0.48\linewidth]{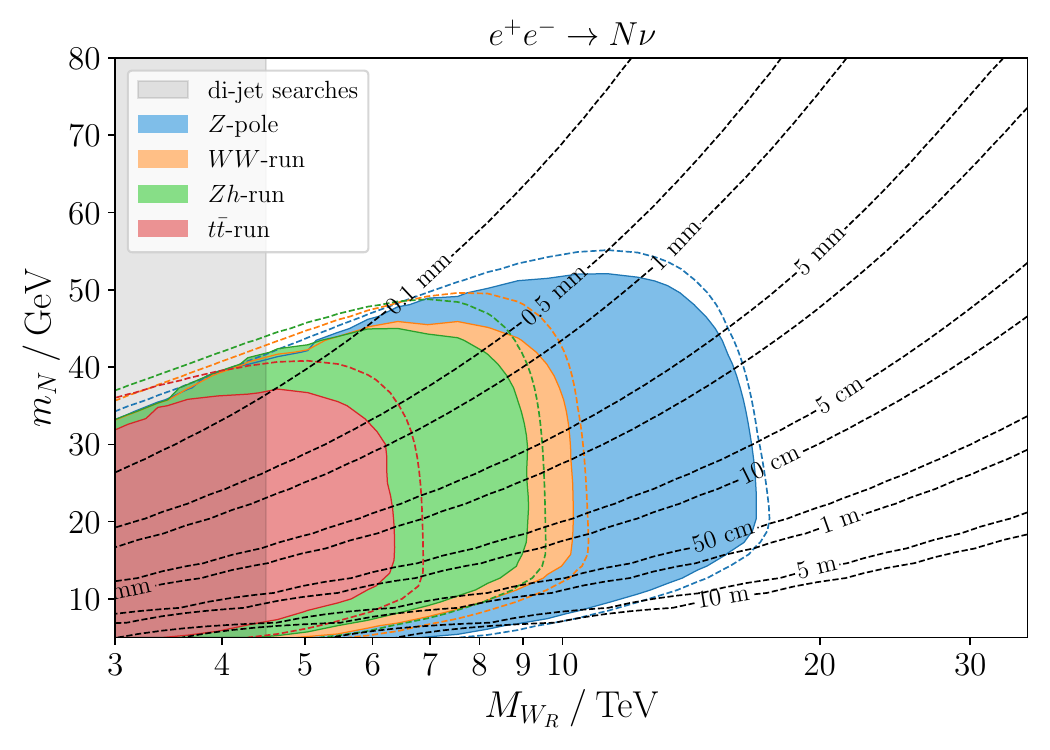}%
  \vspace*{-2ex}
  \caption{%
    FCC-ee sensitivities to heavy neutrino production via $W_R$ exchanges in $e^+e^- \to NN$ (left) and $W$ exchange in $e^+e^- \to N\nu$ (right), shown in the $(M_{W_R}, m_N)$ plane and for $\tan\beta = 0.5$. The filled regions show the reconstruction-level sensitivities, while the dashed contours correspond to theoretical estimates including the fiducial displacement requirement (see Figure~\ref{fig:sens_NNtchannel_Nnu}). The colour code (see the legend) refers to the different FCC-ee operation points. \label{fig:sens_exp_NNtchannel_Nnu}}
\end{figure}

In Figure~\ref{fig:sens_exp_NNtchannel_Nnu}, we show the results for $t$-channel $W_R$-mediated $e^+e^- \to NN$ production, and for the production of a single $N$ in association with a light neutrino via $W$-boson exchange, $e^+e^- \to N\nu$. For the $e^+e^- \to NN$ channel, we again observe a mild degradation relative to the theoretical estimate, although  slightly more pronounced than in resonant $Z \to NN$ production. This is due to the different kinematic structure of the $t$-channel process, which has a more forward-peaked angular distribution. Furthermore, the dominant source of efficiency loss is again the prompt-track veto. In contrast, the degradation observed in the $e^+e^- \to N\nu$ channel is significantly smaller. This can be understood from the lower track multiplicity of the final state, as only a single heavy neutrino decay is present, thereby reducing the probability of misclassifying five displaced tracks as prompt. A mild decrease in efficiency is observed with increasing $\sqrt{s}$,  due to  the increasingly forward-peaked angular structure of the production cross section. This leads to a larger fraction of final-state particles falling outside the tracker acceptance (see also Figure~\ref{fig:dist_pT_eta}) so that some LLP candidates are mis-reconstructed and fail to meet the kinematic selection criteria. However, overall, the reconstruction-level sensitivities closely track the theoretical expectations, with the detector effects and selection cuts introducing only moderate corrections. The resulting reach hence extends up to $M_{W_R} \sim15\:\mathrm{TeV}$ for the $e^+e^- \to NN$ process, and up to $M_{W_R} \sim 17\:\mathrm{TeV}$ for  $e^+e^- \to N\nu$. Thus, both the qualitative and largely the quantitative conclusions of Section~\ref{sec:SigSens} remain valid once  realistic detector effects are taken into account.

\begin{figure}
  \centering
  \includegraphics[width=0.48\linewidth]{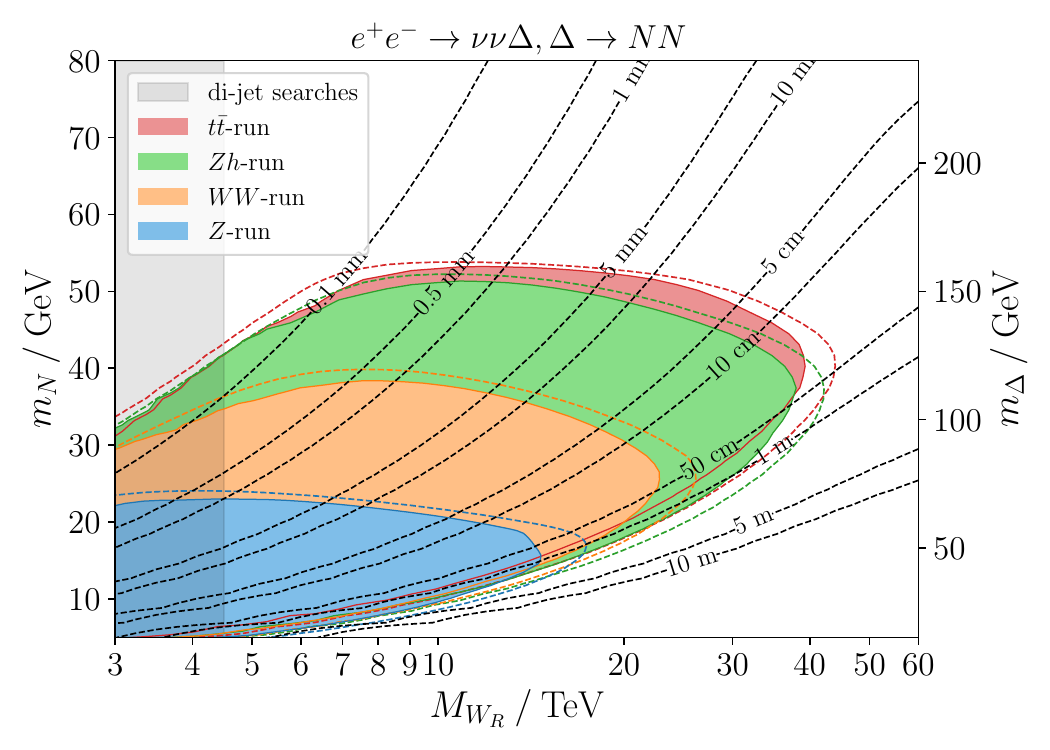}\hfill
  \includegraphics[width=0.48\linewidth]{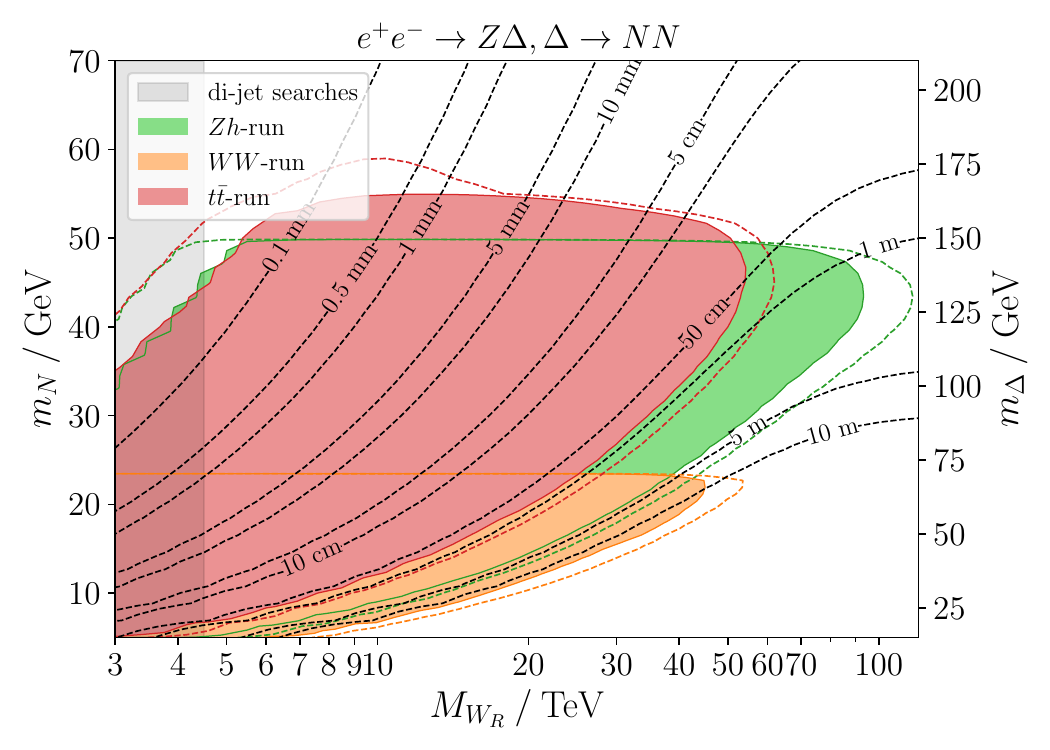}%
  \vspace*{-2ex}
  \caption{%
    Same as Figure~\ref{fig:sens_exp_NNtchannel_Nnu}, but for heavy neutrino originating from $\Delta$ production via weak boson fusion $e^+ e^- \to \nu \nu \Delta$ (left) and $\Delta$-strahlung $e^+ e^- \to Z \Delta$ (right), and for a benchmark choice with $\sin\theta = 0.1$ and $m_\Delta = 3 m_N$. The theory curves are extracted from Figure~\ref{fig:sens_delta}.\label{fig:sens_exp_delta}} 
    \end{figure}
\paragraph{Scalar mixing modes.}
We now turn to scalar-mixing induced production of the $h$ and $\Delta$ states via VBF and strahlung processes, followed by their decays into heavy neutrino pairs $h,\Delta \to NN$. We remind the reader that these channels probe the Yukawa couplings of the neutrinos and the charged leptons and are therefore highly complementary to the gauge-mediated production modes discussed above. The results for the production of the $\Delta$ scalar are shown in Figure~\ref{fig:sens_exp_delta} for both the VBF (left panel) and  $\Delta$-strahlung (right panel) production mode, for a benchmark choice in which $\sin\theta = 0.1$ and $m_\Delta = 3 m_N$. As can be seen, the experimental sensitivity, displayed again in the $(M_{W_R}, m_N)$ plane, closely follows the corresponding theoretical expectations across the explored parameter space. The reach in $M_{W_R}$ is hence mildly reduced with respect to the theoretical estimate, corresponding to an overall signal reconstruction efficiency of approximately $75\%-80\%$. This reduction is largely driven by the detector acceptance and vertex reconstruction effects, and it shows only a weak dependence on $m_\Delta$. The sensitivity in $m_\Delta = 3 m_N$ is instead limited either by the decrease of the $\Delta\to NN$ branching ratio for $m_\Delta \gtrsim 2 M_W$ or by the kinematic threshold for on-shell $Z\Delta$ production.

\begin{figure}  \includegraphics[width=0.48\linewidth]{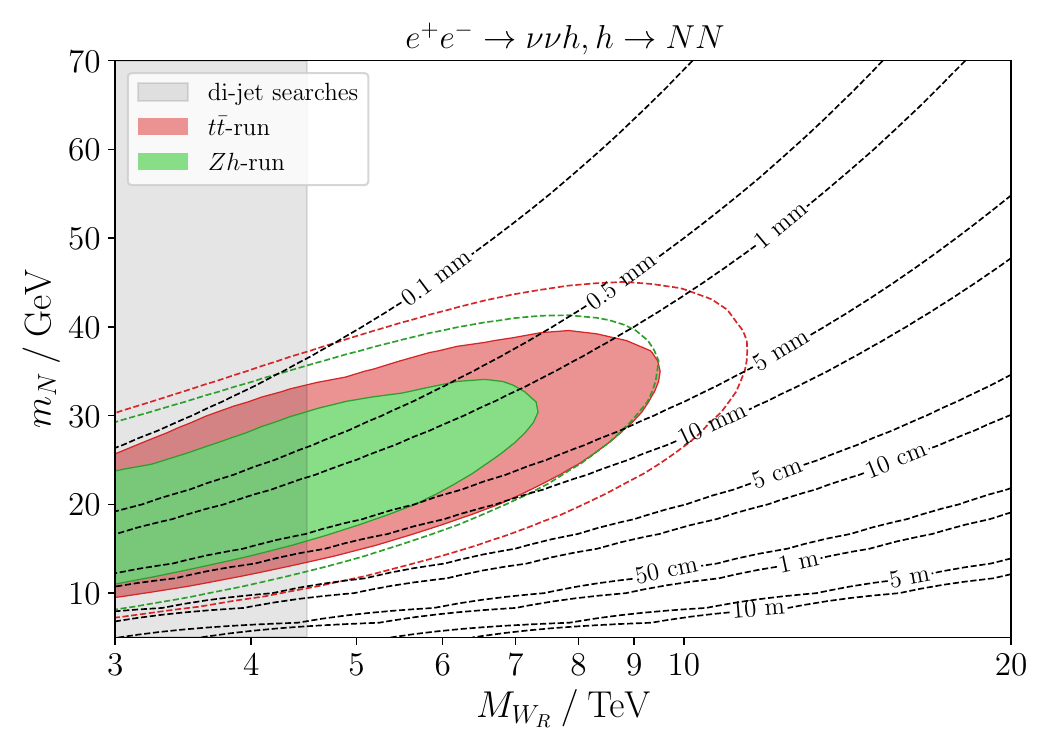}\hfill
  \includegraphics[width=0.48\linewidth]{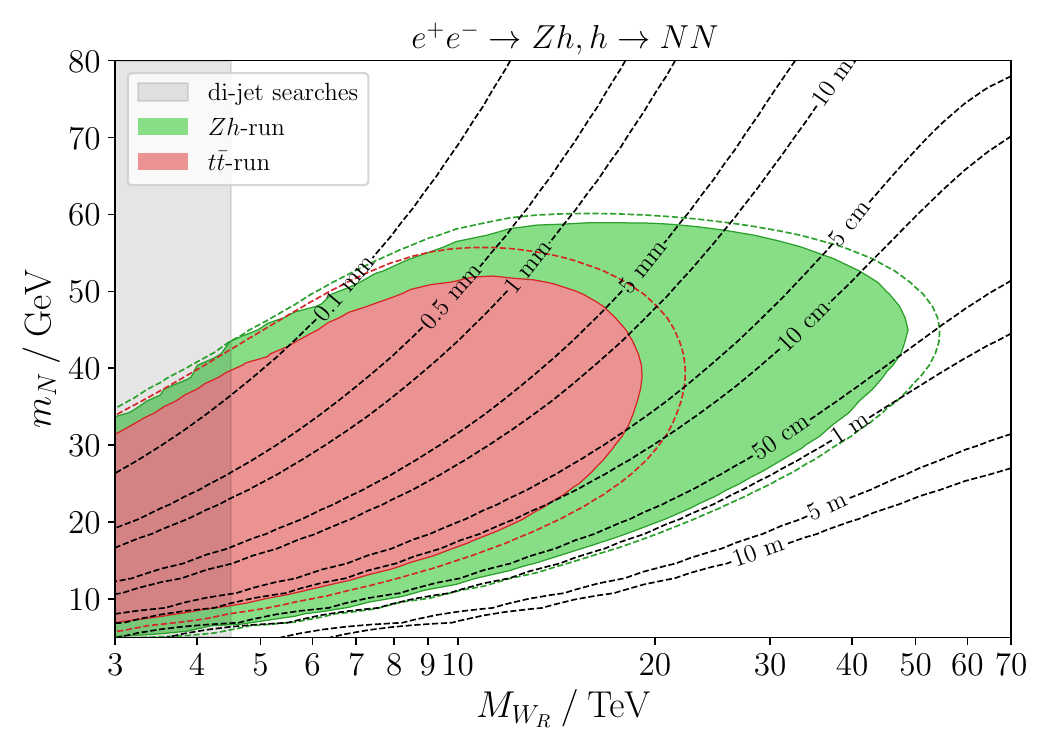}%
  \vspace*{-2ex}
  \caption{%
    Same as Figure~\ref{fig:sens_exp_NNtchannel_Nnu}, but for heavy neutrino originating from Higgs production via weak boson fusion $e^+ e^- \to \nu \nu h$ (left) and Higgs-strahlung $e^+ e^- \to Z h$ (right), and for a benchmark choice with $\sin\theta = 0.2$. The theory curves are extracted from Figure~\ref{fig:sens_higgs}.\label{fig:sens_exp_higgs}}
\end{figure}

In close analogy to $\Delta$ production and decay, we find that similar conclusions apply to Higgs boson production followed by $h \to NN$ decays. The corresponding sensitivities are shown in Figure~\ref{fig:sens_exp_higgs} for  the VBF (left panel) and Higgs-strahlung (right panel) channels, and for a class of LRSM scenarios with $\sin\theta = 0.2$. In the Higgs-strahlung case, the sensitivity in $m_N$ extends essentially up to the kinematic limit $m_N \lesssim m_h/2$, while the VBF channel is ultimately limited by the smaller production cross section. In both cases, the realistic reconstructed sensitivities exhibit the same characteristic shape as the purely theoretical projections, again consistent with reconstruction efficiencies at the level of $75\%$--$ 80\%$.

\begin{figure}
  \centering
  \includegraphics[width=0.6\linewidth]{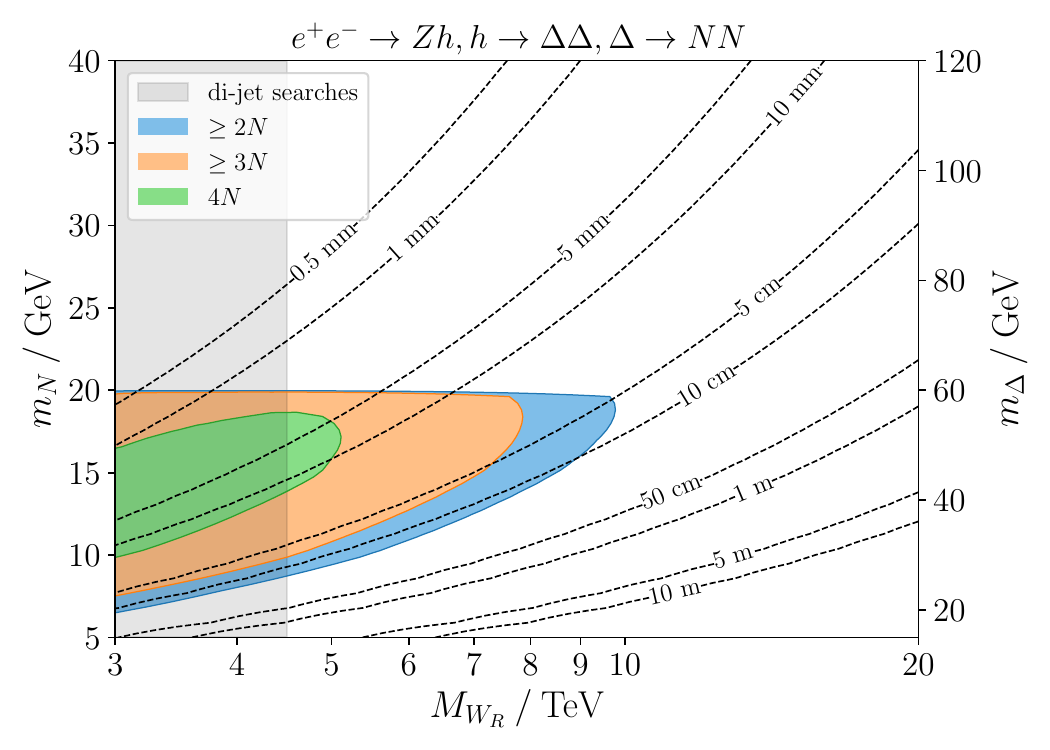}%
  \vspace*{-2ex}
  \caption{%
    FCC-ee sensitivity for  $e^+e^-\to Z h$, $h\to4N$ in the  $Zh$ run,  shown in the $(M_{W_R}, m_N)$ plane for $\sin\theta = 0.2$ and $m_\Delta = 3 m_N$. The filled regions show the reconstruction-level sensitivities when requiring at least two (blue), three (orange) or four (green) reconstructed neutrinos.  \label{fig:sens_exp_4N} }
\end{figure}

Finally, while not considered a primary discovery channel, we investigate the spectacular process $e^+ e^- \to Z h$ followed by the cascade decay $h \to \Delta \Delta \to 4N$. As already mentioned, fully reconstructing this topology requires a more challenging simultaneous identification of multiple displaced vertices associated with four long-lived particles. In Figure~\ref{fig:sens_exp_4N}, we present the sensitivity reach obtained by requiring at least two (blue), three (orange) or four (green) successfully reconstructed LLP  with invariant masses within $m_N^\text{reco} \in [0.8\,m_N^\text{true},\,1.2\,m_N^\text{true}]$. The prompt track veto is applied as in the previous analyses, and we fix $m_\Delta = 3 m_N$ and $\sin\theta = 0.2$ as a benchmark. Although the resulting sensitivity is significantly weaker than for the other scalar production modes, this channel highlights the exceptional capabilities of the IDEA detector for reconstructing complex long-lived particle topologies. Moreover, the $h \to 4N$ final state offers a striking signature of four same-sign leptons, corresponding to $\Delta L = 4$ lepton number violation and providing a smoking-gun signal of the Majorana nature of the neutrinos in left-right symmetric scenarios.

In summary, thanks to the LLP reconstruction strategy developed in this work and the excellent tracking performance envisaged for the IDEA detector, we have demonstrated that the theoretically estimated sensitivities for scalar-mediated heavy neutrino production in the LRSM remain robust once more realistic detector effects and finite reconstruction efficiencies are taken into account.

%
%
\section{Conclusion and Outlook} \label{sec:Conclusion}

%
In this work, we present a comprehensive study of the phenomenology of displaced heavy neutrino signals at future $e^+e^-$ colliders within the Left-Right Symmetric Model. We systematically investigate both gauge-mediated and scalar-mixing-mediated production channels over a wide range of currently proposed centre-of-mass energy and luminosity benchmarks, and we compute the corresponding total cross sections and kinematic distributions. In addition to the well-known $W$, $Z$ and $W_R$ mediated processes in the $s$-, $t$- and $u$-channels, we provide a detailed analysis of the scalar-mixing-mediated production modes and fusion topologies,  significantly extending the coverage of the parameter space.

The analytic predictions for production rates, decay widths, and kinematic distributions are validated through numerical simulations using an LRSM model implementation suitable for collider event simulations. We further perform a realistic simulation of the response of a detector that could be relevant for future electron-positron collider projects. For this task we rely on a newly developed vertexing algorithm. Large portions of the parameter space indeed yield significantly displaced heavy neutrinos, which can thus be efficiently reconstructed using an appropriate and dedicated vertexing algorithm.

The Majorana nature of the heavy neutrinos and their spontaneous mass generation play a central role in the LRSM, and they are known to give rise to striking signatures of lepton number violation. In this work, rather than relying on a conventional same-sign leptons signature for background rejection, as is typically done at hadron colliders, we base our analysis on vertex displacement and reconstructed LLP kinematics, thus retaining all charge combinations in the signal samples. The ultimate sensitivity to the LR scale is therefore primarily driven by the number of reconstructable signal events, which in turn depends on the integrated luminosity and the size and performance of the detector and tracking system. 

We find that the various production channels are largely complementary and dominate in different centre-of-mass energy regimes, especially for fusion topologies. By reconstructing the heavy neutrino momenta from the displaced vertices, we demonstrate that kinematic observables can be accurately reproduced at the detector level with realistic vertexing efficiencies. This enables further discrimination between the different production mechanisms and provides sensitivity to a wide range of LRSM parameters, including $m_N$, $M_{W_R}$, $\xi_{LR}$, $\sin\theta$, and $M_D$. Our study hence highlights the full potential of the FCC-ee, especially for displaced signals involving soft leptons and jets. More precisely, a wealth of signatures exists within the LRSM, and they can be efficiently accessed at the FCC-ee while remaining extremely challenging at hadron colliders due to overwhelming backgrounds. 

Remarkably, for relatively light heavy neutrinos with $m_N < 100~\mathrm{GeV}$, the FCC-ee sensitivity extends well beyond the reach of the LHC, probing LR scales in the deep multi-10~TeV range. The channel with the highest sensitivity is the associated $Z\Delta$ production mode during the FCC-ee $Zh$ run, which approaches an impressive reach of $M_{W_R} \simeq 100~\mathrm{TeV}$ as shown in Figure~\ref{fig:sens_exp_delta}. Interestingly, the Dirac mass mixings predicted by the model contribute to the heavy neutrino decay width, slightly reducing its lifetime and thereby enhancing the overall sensitivity. 

Equally noteworthy is the excellent agreement between the reconstructed-level and truth-level distributions, including $m_{\ell jj}$ (Figure~\ref{fig:dist_mN}), transverse momentum, and pseudo-rapidity spectra (Figures~\ref{fig:dist_pT_eta} and~\ref{fig:dist_pT}), as well as the $m_{NN}$ and $m_{4N}$ reconstructed invariant masses (Figures~\ref{fig:dist_mDelta} and~\ref{fig:mh_dist}). In particular, the reconstruction of the Higgs kinematics from four displaced decay products in this last case elevates displaced vertices from simple discovery handles to fully fledged kinematic objects, even in case of high-multiplicity and soft final states.

%
Looking ahead, several promising directions remain open. Prompt heavy neutrino signatures and their background rejection merit further study, particularly to optimise the sensitivity at the upper end of the accessible $m_N$ range. While the present analysis focuses on electrons and muons that behave similarly, extending the study to final states involving $\tau$ leptons would be highly motivated, albeit more challenging due to additional neutrinos and displaced decay chains. Furthermore, mixed-flavour final states are allowed within the LRSM, even if constrained by lepton flavour violation bounds, and thus represent another interesting avenue for future work. Moreover, the LLP reconstruction techniques developed here could also be exploited to investigate spin correlations, CP violation, and oscillation phenomena, which we leave for future exploration. Finally, beyond heavy neutrino final states, mixed triplet scalar decays into SM particles and $NN$ pairs could further enrich the FCC-ee physics programme and deserve a dedicated study.

Summarising, this work constitutes the first comprehensive analysis of displaced heavy Majorana neutrino signatures in the Left-Right Symmetric Model that consistently bridges the gap between analytic predictions at the Lagrangian level and fully reconstructed displaced objects at the detector level. By combining detailed theoretical control over production and decay mechanisms with realistic simulation and modern displaced-vertex reconstruction, we demonstrate that long-lived particle searches at future lepton colliders can extend well beyond event counting and enable full kinematic reconstruction and direct access to fundamental model parameters. Our results thus establish the FCC-ee as a uniquely powerful probe of left-right symmetry breaking and the Majorana nature of neutrinos, with sensitivity reaching deep into the multi-TeV regime and opening a qualitatively new window onto the origin of neutrino masses.

%
%
\section*{Acknowledgments}
JK and MN are supported by the Slovenian Research Agency under the research core
funding No. P1-0035 and in part by the research grants J1-3013 and N1-0253.
The work of BF has been partly supported by Grant ANR-21-CE31-0013 
(project DMwithLLPatLHC) from the French \emph{Agence Nationale de la Recherche}. 
The work of BF, JK and MN has received further support by the bilateral project 
Proteus PR-12696/50194VC.
MN and JK would like to thank the LPTHE at Sorbonne Université for its hospitality during
their stay, where the majority of this work was carried out. BF is grateful to the IJS for supporting his visits.
MN would like to thank the Institut Français for its support, as well as the 
CA22130 - Comprehensive Multiboson Experiment-Theory Action (COMETA).

\appendix

%
%
\section{General three-body phase space}
\label{app:PhaseSpace}
{\bf Massless scattered particles.}
We first consider the simpler case of $2 \to 3$ kinematics of $A + B \to 1 + 2 + P$, 
where $A,B,1,2$ are massless.
We set up the following four momenta
\begin{align}
  p_A + p_B &= p_1 + p_2 + P \, ,
  &
  p_A^2 = p_B^2 = p_1^2 = p_2^2 &= 0 \, ,
  &
  P^2 &= m^2 \, ,
\end{align}
and $s = (p_A + p_B)^2 = 2 p_A\!\cdot\!p_B$.
We recall that the VBF amplitude in~\eqref{eq:avgM2VBF} is given by
\begin{align}
  \overline{\left| \mathcal M \right|}^2 &= 8 g_{VVS}^2 \frac{
  C_1 \left( p_A\!\cdot\!p_2 \right)\left( p_B\!\cdot\!p_1 \right) + C_2 \left( p_A\!\cdot\!p_B \right) 
  \left( p_1\!\cdot\!p_2 \right)} {\left(q_1^2 - M^2\right)^2\left(q_2^2 - M^2\right)^2} \, ,
\end{align}
with $q_1 = p_A - p_1$, $q_2 = p_B - p_2$.
We evaluate the various scalar products in two different frames: the laboratory frame `LAB' and the rest frame of the produced massless particles 1 and 2 that will be denoted as the `1-2' frame.
The product between the beam momenta $p_{A,B}$ and the momentum $P$ is evaluated in the 
laboratory frame,
\begin{align} \label{eq:LabFrame}
  &\text{LAB:} & p_{A,B} &= \frac{\sqrt s}{2} \left( 1, \pm 1 \right)  ,
  &
  P_\mu &= \left( E, P \, c_\alpha \right)  ,
  &
  2 P\!\cdot\!p_{A, B} &= \sqrt s \left( E \mp P \, c_\alpha \right)  ,
\end{align}
where $P_\mu P^\mu = E^2 - P^2 = m^2$.
The `1-2' rest frame is defined by $\vec p_1 + \vec p_2 = 0$, and we set the 
following frame assignments
\begin{align}
  p_1 &= \left( E_1, \vec p_1 \right)  , &
  p_2 &= \left( E_1, - \vec p_1 \right)  ,
\end{align}
consistent with $p_1^2 = p_2^2 = 0$.
We are interested in the scalar products of $p_{1,2}$ with $p_{A,B}$ and choose to direct $p_A$
along the $z$ axis and $p_B$ in the $x-z$ plane at an angle $\chi$
\begin{align}
  p_A &= E_A \left(1, 0, 0, 1 \right) \,
  &
  p_B &= E_B \left(1, s_\chi, 0, c_\chi \right)  .
\end{align}
The remaining $p_1$ flows along the arbitrary direction set by $\theta$ and $\varphi$,
\begin{align}
  p_1 &= \left( E_1, \vec p_1 \right) , &
  p_2 &= \left( E_1, - \vec p_1 \right)  , &  
  \vec p_1 &= E_1 \left(s_\theta c_\varphi, s_\theta s_\varphi , c_\theta \right)  , &
\end{align}
such that the integration measure in the $p_1$ kinematics is given by $\text{d}^3 p_1 = \text{d} c_\theta  \, \text{d} \varphi \, \text{d} p_1 p_1^2$.
The $P\!\cdot\!p_{A, B}$ scalar products do not depend on the $\theta$ and $\phi$ angles, and we can 
leave them unevaluated in the `1-2' frame. This follows from contracting the four momentum conservation relation with $2 p_A$,
\begin{equation}\begin{split}
  \hspace*{-.3cm}2 p_A\!\cdot\!\Big(p_A \!+\! p_B\Big) &= 2 p_A\!\cdot\!\Big(P + p_1 + p_2\Big) \, ,  \\
  0 + s &= 2 P\!\cdot\!p_A + 2 p_A \!\cdot\!\left( p_1 + p_2 \right) = 2 P\!\cdot\! p_A + 2 p_A \!\cdot\! \left( 2 E_1, \vec 0 \right) 
   = 2 P  \!\cdot\! p_A + 4 E_A E_1 \, ,
\end{split}\end{equation}
and similarly for $p_B$, such that we get the following invariants
\begin{align}
  2 P\!\cdot\! p_{A, B} &= s - 4 E_1 E_{A, B} \, , &
  E_1 E_{A, B} &= \frac{1}{4} \left( s - 2 P\!\cdot\! p_{A, B} \right) = \frac{s_{A, B}}{4}  \, ,
\end{align}
expressed with the shorthand notation 
\begin{align}
  s_{A, B} &= s - 2 P \!\cdot\!p_{A, B} \, .
\end{align}
The $P\!\cdot\!p_{A, B}$ products thus depend only on the energies and are independent of $\theta$ 
and $\varphi$.
Instead, the angular dependence shows up in the $p_{A,B}\!\cdot\!p_1$ products,
\begin{align} \label{eq:PA1}
  2 p_A\!\cdot\! p_1 &= 2 E_1 E_A \left( 1 - c_\theta \right) = \frac{s_A}{2} \left( 
  1 - c_\theta \right) \, , 
  \\ \label{eq:PB1}
  \begin{split}
  2 p_B \!\cdot\!p_1 &= 2 E_1 E_B \left( 1 - s_\chi s_\theta c_\varphi - c_\chi c_\theta \right) 
  = \frac{s_B}{2} \left( 1 - s_\chi s_\theta c_\varphi - c_\chi c_\theta \right)  .
  \end{split}
\end{align}

The $\chi$ angle is also independent of $\theta, \varphi$ and is expressed
with the three invariants $s, s_{A,B}$. We start with the scalar product $p_A \!\cdot\!p_B$ that contains a dependence on $c_\chi$
\begin{align} \label{eq:cchi}
  s &= 2 p_A\!\cdot\! p_B = E_A E_B \left( 1 - c_\chi \right) \, ,
  &
  c_\chi &= 1 - \frac{s}{2 E_A E_B} = 1 - \frac{8 s E_1^2}{ s_A s_B } \, .
\end{align}
We then express $E_1$ in terms of the $s_{A, B}$ invariants from
\begin{align} \label{eq:sp12}
  \left( p_1 + p_2 \right)^2 &= 2 p_1\!\cdot\! p_2 = 4 E_1^2 = \left( p_A + p_B - P \right)^2 
  = m^2 - s + s_A + s_B \, ,
\end{align}
such that $c_\chi$ is fixed by $s, s_{A,B}$, and finally by the massive particle energy $E$ 
and polar angle $c_\alpha$ in~\eqref{eq:LabFrame},
\begin{align}
  c_\chi & = 1 - \frac{2 s \left( m^2 - s + s_A + s_B \right)}{ s_A s_B } 
  \quad\Rightarrow\quad p_1\!\cdot\! p_2 = \frac{s_A s_B}{4 s} \left( 1 - c_\chi \right) .
\end{align}
The angular dependence from propagators in the denominator of~\eqref{eq:avgM2VBF} 
is given by
\begin{align}
  q_1^2 &= \left( p_A - p_1 \right)^2 = - 2 p_A\!\cdot\! p_1 = - \frac{s_A}{2} \left( 
  1 - c_\theta \right) ,
  \\
  \begin{split}
  q_2^2 &= -2 p_B\!\cdot\! p_2 = - 2 p_B \!\cdot\!\left( p_A + p_B - p_1 - P \right) 
  \\
  &= 2 p_B\!\cdot\! p_1 - s_B
  = -\frac{s_B}{2} \left( 1 + s_\chi s_\theta c_\varphi + c_\chi c_\theta \right) .
  \end{split}
\end{align}

The distribution of the cross section over the massive particle momentum is obtained by integrating
the amplitude squared over the $p_1$ angles $\theta$ and $\varphi$
\begin{align} \label{eq:defEdsigd3P}
  E \frac{\text{d} \sigma}{\text{d}^3 P} &= \frac{1}{s \left( 4 \pi \right)^5} 
  \int_{-1}^1 \text{d} c_\theta \int_0^{2 \pi} \text{d} \varphi \, \overline{
  \left| \mathcal M \right|}^2 \, .
\end{align}
The shorthand notation used in Section~\ref{subsec:VBF_SBF} is then defined by
\begin{align} \label{eq:sht}
  s_{A,B} &= s - \sqrt s E \pm \sqrt s P c_\alpha \, , &
  h_{A,B} &= 1 + 2 \frac{M^2}{s_{A,B}} \, , &
  t_{A,B} &= h_{A,B} + c_\chi h_{B,A} \, , 
\end{align}
and
\begin{align} \label{eq:rscL}
  r &= h_A^2 + h_B^2 + 2 c_\chi h_A h_B - s_\chi^2 \, ,
  &
  \mathcal L &= \log \frac{h_A h_B + c_\chi + \sqrt r}{h_A h_B + c_\chi - \sqrt r} \, ,
\end{align}
where $c_\chi = 1 - (2 s s_\nu)/(s_A s_B)$ is the angle between $A$ and $B$ in
the `1-2' frame.

\medskip

For the fusion process where a pair of (pseudo-)scalar bosons $SS$ annihilates into a Higgs boson, the amplitude squared is given by
\begin{align} \label{eq:avgM2SBF}
  \overline{\left| \mathcal M \right|}^2 &= g_{S S h}^2 \frac{
  C_S \left( p_A\!\cdot\! p_1 \right)\left( p_B\!\cdot\! p_2 \right) }
  {\left(q_1^2 - M^2\right)^2\left(q_2^2 - M^2\right)^2} \, ,
\end{align}
where $C_S = 4 (s_e^2 + p_e^2)^2$, and $s_e$ and $p_e$ are the scalar and 
pseudoscalar couplings of $S$ to electrons, while $g_{SSh}$ is the triple scalar 
vertex coupling two scalars $S$ to the Higgs boson $h$. 
We evaluate the scalar products using similar manipulations as in the VBF case
\begin{align}
  p_A\!\cdot\! p_1 &= \frac{s_A}{4} \left(1 - c_\theta \right)  ,
  &
  \begin{split}
  p_B\!\cdot\! p_2 &= p_B \!\cdot\!\left( p_A + p_B - P - p_1 \right) 
  \\
  &= \frac{s_B}{4} \left( 1 + s_\chi s_\theta c_\varphi + c_\chi c_\theta \right)  ,
  \end{split}
  \\[.4cm]
  \left( q_1^2 - M^2 \right)^2 &= \left( \frac{s_A}{2} \right)^2 \left( 
  h_A - c_\theta \right)^2  ,
  &
  \left( q_2^2 - M^2 \right)^2 &= \left(\frac{s_B}{2} \right)^2 \left( 
  h_B + s_\chi s_\theta c_\varphi + c_\chi c_\theta \right)^2  .
\end{align}
We introduce these expressions in~\eqref{eq:avgM2SBF}, then insert the squared matrix element $| \mathcal M |^2$ into~\eqref{eq:defEdsigd3P} and finally integrate over $\varphi$ and $c_\theta$ to get
\begin{align} \label{eq:Edsigd3P_SBF}
  E \frac{\text{d} \sigma}{\text{d}^3 P} = 
  \frac{\pi g_{S S h}^2 C_s}{4 s s_A s_B \left( 4 \pi \right)^5 r^2}  
  \left( \frac{H_s}{\sqrt r} \mathcal L 
  - \frac{G_s}{\left( h_A + 1 \right) \left( h_B + 1 \right)} \right)  .
\end{align}
The logarithmic factor $\mathcal L$ is the same as in \eqref{eq:rscL}, and the two coefficients $H_s$ and $G_s$ are symmetric matrices given by
\begin{align}
  H_s =  
  \begin{pmatrix}
    1 & h_A & h_A^2 & h_A^3
  \end{pmatrix}
  \begin{pmatrix} 
    s_\chi^2 (c_\chi - r) & - s_\chi^2 (1 + 2 c_\chi) & 2 c_\chi & 1 - c_\chi 
    \\
    \cdot & 4 - (1 - c_\chi) s_\chi^2 & c_\chi-s_\chi^2-1 & 2 c_\chi
    \\
    \cdot & \cdot & 4-s_\chi^2 & 0
    \\
    \cdot &  \cdot   & \cdot & 0
    \end{pmatrix}
    \begin{pmatrix}
    1 \\ h_B \\ h_B^2 \\ h_B^3
  \end{pmatrix} ,
  \\[.3cm]
  G_s =  
  \begin{pmatrix}
    1 & h_A & h_A^2 & h_A^3
  \end{pmatrix}
  \begin{pmatrix} 
    2 s_\chi^2 & - \frac{s_\chi^4}{s_{\chi/2}^2} & 1 + 3 c_{2 \chi} & 2 
    \left(1 + c_\chi \right)
    \\
    \cdot & c_{3 \chi} + 7 c_\chi & 2 \left(2 +  c_{2 \chi} + 3 c_\chi  \right) & 4 c_\chi
    \\
    \cdot & \cdot & 7 + c_{2 \chi} & 0
    \\
    \cdot &  \cdot   & \cdot & 0
    \end{pmatrix}
    \begin{pmatrix}
    1 \\ h_B \\ h_B^2 \\ h_B^3
  \end{pmatrix} ,
\end{align}
where the symmetry over the exchange $h_A \leftrightarrow h_B$ was made explicit.

\bigskip

{\bf Massive scattered particles.}
More generally, we allow all particles to be massive, and the phase space for $2\to 3$ 
scattering can be decomposed~\cite{Byckling:1973bk} as
\begin{equation}
  \mathrm{d}\Phi_3(s) = \frac{\pi}{32 s (2\pi)^5\lambda(s, m_a^2, m_b^2)^{\frac{1}{2}}}
  \int\frac{\d s_1 \d s_2\d t_1 \d t_2}{\sqrt{-\Delta_4}} \, .
\end{equation}
We label the particle momenta as $p_a + p_b = p_1 + p_2 + p_3$ and introduce the 
Mandelstam variables as
\begin{align}
  s &= (p_a + p_b)^2\,, & s_1 &= (p_1 + p_2)^2\,, & s_2 &= (p_2 + p_3)^2 \, ,
  \\
  t_1 &= (p_a - p_1)^2\,, & t_2 &= (p_b - p_3)^2 \, , 
\end{align}
where $\lambda(x,y,z) = x^2-2 x y-2 x z+y^2-2 y z+z^2$ denotes the Källén function 
and the Gram determinant is given by
\begin{equation}
  \Delta_4 = \frac{1}{16} 
  \begin{vmatrix}2 m_a^2 & s - m_a^2 - m_b^2 & 
  m_a^2 + m_1^2 - t_1 & s - s_1 + t_2 - m_b^2
  \\ 
  s - m_a^2 - m_b^2 & 2 m_b^2 & s - s_2 + t_1 - m_a^2 & m_b^2 + m_3^2 - t_2
  \\
  m_a^2 + m_1^2 - t_1 & s - s_2 + t_1 - m_a^2 & 2 m_1^2 & s - s_1 - s_2 + m_2^2
  \\ 
  s - s_1 + t_2 - m_b^2 &  m_b^2 + m_3^2 - t_2& s - s_1 - s_2 + m_2^2 & 2 m_3^2
  \end{vmatrix} \, .
\end{equation}
For the $2 \to 3$ scattering cross sections studied here, we can safely neglect the 
external fermion masses and set $m_a = m_b = m_1 = m_3 = 0$, while retaining 
$m_2 = m_h,m_\Delta$.
The derivation of the physical integration region shown in the following is 
still general and can be translated to the fully massive case (with 
considerably lengthier formul\ae).
The physical region in the momenta $s_1,\,s_2,\, t_1\,,t_2$ can be derived 
as follows.
The global condition is set by requiring $\Delta_4 \leq 0$, rendering the phase 
space real.
Setting $\Delta_4 = 0$ for a fixed $s$ leads to a quadratic equation for the 
remaining Mandelstam variables and its solution determines the boundary, \textit{e.g.}
\begin{align}
  t_2^\pm(s_1, s_2, t_1) &= \frac{1}{(s - s_2)^2}\left[(s - s_2)(m_S^2 s - s_1 s_2) + 
  s t_1(2 m_S^2 + s - s_1) - (s + s_1) s_2 t_1 \right.\nonumber
  \\
  &\left. {}\pm2 \sqrt{s(m_S^2 + s - s_1 - s_2)(m_S^2 s - s_1 s_2)
  t_1(s - s_2 + t_1)} \right] .
\end{align}
The discriminant of the $t_2$ boundary sets additional positivity conditions 
on $t_1$, $s_1$, and $s_2$, leading to the boundaries
\begin{align}
  t_1^-(s_2) &= -s + s_2\,, & t_1^+ &= 0\,,
  \\
  s_2^-(s_1) &= \frac{m_S^2 s}{s_1}\,, & s_2^+(s_1) &= m_S^2 + s - s_1\,,
  \\
  s_1^- &= m_S^2\,, & s_1^+ &= s\,.
\end{align}
The phase space can then be numerically integrated using the nested boundaries as
\begin{equation}
  \d \Phi_3(s) = \frac{\pi}{32 s (2\pi)^5\lambda(s, m_a^2, m_b^2)^{\frac{1}{2}}}
  \int_{s_1^-}^{s_1^+}\int_{s_2^-(s_1)}^{s_2^+(s_1)}\int_{t_1^-(s_2)}^{t_1^+(s_2)}
  \int_{t_2^-(s_1, s_2, t_1)}^{t_2^+(s_1, s_2, t_1)}\frac{\d s_1 \d s_2\d t_1 \d t_2}{
  \sqrt{-\Delta_4}} \, ,
\end{equation}
ensuring that only the physical phase space is sampled.

\subsection{Distribution with respect to \texorpdfstring{$\cos\theta_S$}{cos(thetaS)}}
In order to derive the event distributions with respect to $\cos\theta_S$, \textit{i.e.}\ 
the scattering angle of the scalar $S$ with respect to the beam line, we can first 
fix a frame for $S$ as $p_S = (E_S, |\vec p_S|\sin\theta_S, 0, |\vec p_S|\cos\theta_S)$
with 
\begin{equation}
  E_S = \frac{s_1 + s_2}{2\sqrt{s}}\,,\quad|\vec p_S| = \frac{1}{2\sqrt{s}}
  \sqrt{\lambda(s, s - s_1 - s_2 + m_S^2, m_S^2)} \equiv 
  \frac{1}{2\sqrt{s}}\sqrt{\lambda_S} \, .
\end{equation}
From the definition of the Mandelstam variables, we can then easily derive
\begin{equation}
  t_2 = \frac{1}{2}\left(s_1 - s_2 + \cos\theta_S\sqrt{\lambda_S} + 2 t_1\right) ,
\end{equation}
and replace the integration over $t_2$ with the Jacobian
\begin{equation}
  \frac{\mathrm{d} t_2}{\mathrm{d}\cos\theta_S} = \frac{\sqrt{\lambda_S}}{2} \, .
\end{equation}
In both the phase space measure as well as in the squared amplitude, $t_2$ has 
to be replaced with its expression in terms of $\cos\theta_S$.
The integration boundaries are now more restricted for a fixed $\cos\theta_S$, 
and in order to again only sample physical phase space in the Monte Carlo integration 
the boundaries for $t_1$ have to be changed.
Starting from the negativity condition of the Gram determinant we find
\begin{align}
\begin{split}
  t_1^{\pm} &= \frac{1}{8 m_S^2 s - 2(s_1 + s_2)^2}\left(s((s_1 + s_2)^2 + 
  \cos\theta_S (s_1 - s_2)\sqrt{\lambda_S} \right.
  \\
  & - s_2(s_1 + s_2)(s_1 + s_2 - \cos\theta_S\sqrt{\lambda_S}) - 2 m_S^2 s(2 s - 2 s_2 + 
  \cos\theta_S\sqrt{\lambda_S})
  \\
  &\left. {}\pm 2 \sqrt{(1 - \cos\theta_S^2)s (m_S^2 + s - s_1 - s_2)(m_S^2s - s_1 s_2)\lambda_S}\right),
\end{split}    
\end{align}
from which it is evident that the integration boundaries for $s_{1,2}$ remain 
unchanged.
We can now numerically integrate over $s_{1,2}$ and $t_1$ to derive the 
phase space distribution $\mathrm{d}\sigma/\mathrm{d}\cos\theta$.

\subsection{Distribution with respect to \texorpdfstring{$p_T(S)$}{pTS}}
The phase space distribution with respect to the transverse momentum $p_T$ of the produced scalar $S$ can be derived in a similar manner.
With the help of
\begin{equation}
  p_T(S) = \left| \vec p_S \right|\sqrt{1 - \cos\theta_S^2} \, ,
\end{equation}
we can derive 
\begin{eqnarray}
    t_2^\pm(p_T) = \frac{1}{2}\left(2 t_1 + s_1 - s_2 \pm \sqrt{s}\sqrt{\frac{\lambda_S}{s} - 4 p_T^2}\right).
\end{eqnarray}
There are two branches for $t_2^\pm(p_T)$, owed to the non-linear relationship between
$p_T$ and $\cos\theta_S$, and they have to be summed over.
The Jacobian of the transformation is given by
\begin{equation}
  \left|\frac{\mathrm{d}t_2}{\mathrm{d} p_T}\right| = \frac{2 p_T\sqrt{s}}
  {\sqrt{\frac{\lambda_S}{s}-4 p_T^2}} \, .
\end{equation}
As in the case of the $\cos\theta_S$ distribution, the phase space becomes 
more restricted, and we have to derive new integration boundaries.
From the positivity condition of 
\begin{equation} \label{eqn:lambdapTpos}
 \lambda_S - 4 s p_T^2 = (s_1 + s_2)^2 - 4 s(m_S^2 + p_T^2) \geq 0 \, ,
\end{equation}
the boundaries for $s_{1,2}$ also have to be changed.
This can be heuristically understood such that either $s_1$ or $s_2$ have to 
provide enough energy to produce $S$ on-shell with non-vanishing $p_T$.
We choose new boundaries for $s_2$ which can be easily derived from 
Eq.\eqref{eqn:lambdapTpos} as
\begin{equation}
  s_2^-(s_1) =  \mathrm{max}\left(\frac{m_S^2 s}{s_1}, 2\sqrt{s(m_S^2 + p_T^2) 
  } -s_1\right)  ,
\end{equation}
while the upper bound of $s_2$ as well as the boundaries for $s_1$ remain unchanged.
The new boundaries for $t_1$ are much more involved and are again derived from the 
negativity condition of the Gram determinant as
\begin{eqnarray}
  t_1^\pm &=& \frac{1}{8 m_S^2 s - 2(s_1 + s_2)^2}\left[s\left((s_1 + s_2)^2 + (
  s_1 - s_2)\sqrt{\lambda_S - 4 p_T^2 s}\right) \right.
  \\
  &-&\left. 2 m_S^2 s\left(2 s - 2 s_2 + 
  \sqrt{\lambda_S - 4 p_T^2 s}\right)- s_2(s_1 + s_2)\left(s_1 + s_2 - \sqrt{\lambda_S - 4 p_T^2
  s}\right)\right.\nonumber
  \\
  &\pm&\left. 2\sqrt{s(m_S^2 + s - s_1 - s_2)(m_S^2 s - s_1 s_2)(
  4(m_S^2 - p_T^2)s - (s_1 + s_2)^2 + \lambda_S)}\right] . \nonumber
\end{eqnarray}
The maximum $p_T(S)$ that can be externally fixed is given by
\begin{equation}
  p_T^\text{max}(S) = \frac{s - m_S^2}{2 \sqrt{s}} \, .
\end{equation}
Numerically integrating over $s_{1,2}$ and $t_1$ gives us the phase 
space distribution $\mathrm{d}\sigma/\mathrm{d}p_T(S)$.

%
%
\section{Running benchmarks for the FCC and CEPC} \label{app:FCC_CEPC_benchmarks}

For convenience, we collect the currently proposed running benchmarks for
the FCC-ee~\cite{FCC:2025lpp} in Table~\ref{tab:FCCee} and the CEPC~\cite{Antusch:2025lpm} 
in Table~\ref{tab:CEPC}.
 
\begin{table}[b]
\renewcommand{\arraystretch}{1.3}\setlength{\tabcolsep}{16pt}
  \centering
  \resizebox{.99\textwidth}{!}{\begin{tabular}{lccccccc}
    Working point & $Z$ pole & $WW$ thresh.\ & $ZH$ & \multicolumn{2}{c}{$t \overline t$} 
    \\ \hline
    $\sqrt{s}$ {(GeV)} & \!\!\!\!\!88, 91, 94 & 157, 163 & 240 & 340--350 & 365 
    \\ 
    Lumi/IP {($10^{34}$\,cm$^{-2}$s$^{-1}$)} & 140 & 20 & 7.5 & 1.8 & 1.4 
    \\ 
    Lumi/year {(ab$^{-1}$)} & 68 & 9.6 & 3.6 & 0.83 & 0.67 
    \\ 
    Run time {(year)} & 4 & 2 & 3 & 1 & 4 
    \\ 
    Integrated lumi.\ {(ab$^{-1}$)} & 205 & 19.2 & 10.8 & 0.42 & 2.70 
    \\ \hline
    &  &  & \!\!\!$2.2 \times 10^6$ $ZH$ & \multicolumn{2}{c}{$2 \times 10^6$ $t\overline t$} 
    \\
    Event yield &  \!\!\!\!\!$6 \times 10^{12}$ $Z$ & $2.4 \times 10^8$ $WW$ & $+$ & 
    \multicolumn{2}{c}{$+\,370$k $Zh$} 
    \\
    &  &  & \!\!\!\!\!65k $WW \to h$ & \multicolumn{2}{c}{$+\,92$k $WW \to h$}\\ \hline
  \end{tabular}}
  \caption{Parameters for the FCC-ee, adapted from the feasibility study 
  report~\cite{FCC:2025lpp}. \label{tab:FCCee}} \vspace{.7cm}
  \resizebox{.99\textwidth}{!}{\begin{tabular}{lcccc}
    \hline 
    Operation mode & $Z$ factory  & $WW$ thresh.  & Higgs factory & $t \overline{t}$ 
    \\
    \hline 
    $\sqrt{s}$ (GeV) & 91.2 & 160 & 240 & 360 \\
    Lumi/IP  ($10^{34} \text{cm}^{-2}\text{s}^{-1}$) & 191.7 & 26.6 & 8.3 & 0.83 
    \\
    Run time (year) & 2 & 1 & 10 & 5 
    \\
    Integrated lumi. (ab$^{-1}$, 2~IPs) & 100 & 6 & 20 & 1 
    \\
    \hline
    Event yield & $4.1 \times 10^{12}$ & $2 \times 10^{8}$ & $4.3 \times 10^{6}$ & 
    $0.6 \times 10^{6}$ 
    \\
    \hline
  \end{tabular}}
  \caption{Operation parameters for the CEPC, adapted from the study 
  report~\cite{Antusch:2025lpm}.}
  \label{tab:CEPC}
\end{table}

\bibliographystyle{JHEP}

\bibliography{MajoranaHiggsFCCee}

\end{document}